\documentclass[12pt]{article}
\usepackage{ae,aecompl}

\usepackage[T1]{fontenc}
\usepackage[utf8]{inputenc}
\usepackage{geometry}
\usepackage{array}
\usepackage{float}
\usepackage{enumitem}
\usepackage{varwidth}
\usepackage{amsmath}
\usepackage{amsthm}
\usepackage{amssymb}
\usepackage{graphicx}
\usepackage{setspace}
\usepackage[authoryear]{natbib}
\usepackage[pdfusetitle,
 bookmarks=true,bookmarksnumbered=false,bookmarksopen=false,
 breaklinks=false,pdfborder={0 0 1},backref=false,colorlinks=false]
 {hyperref}

\makeatletter

\providecommand{\tabularnewline}{\\}
\newenvironment{cellvarwidth}[1][t]
    {\begin{varwidth}[#1]{\linewidth}}
    {\@finalstrut\@arstrutbox\end{varwidth}}
\floatstyle{ruled}
\newfloat{algorithm}{tbp}{loa}
\providecommand{\algorithmname}{Algorithm}
\floatname{algorithm}{\protect\algorithmname}

\theoremstyle{definition}
\newtheorem*{example*}{\protect\examplename}
\theoremstyle{plain}
\newtheorem{assumption}{\protect\assumptionname}
\theoremstyle{plain}
\newtheorem{lem}{\protect\lemmaname}
\theoremstyle{plain}
\newtheorem{thm}{\protect\theoremname}
\theoremstyle{definition}
\newtheorem{defn}{\protect\definitionname}
\theoremstyle{plain}
\newtheorem{cor}{\protect\corollaryname}
\theoremstyle{plain}
\newtheorem{prop}{\protect\propositionname}

\usepackage{setspace}
\usepackage{pdflscape}
\usepackage{datetime}

\DeclareMathOperator{\Var}{Var}

\DeclareMathOperator{\Cov}{Cov}

\date{}

\makeatother

\providecommand{\assumptionname}{Assumption}
\providecommand{\corollaryname}{Corollary}
\providecommand{\definitionname}{Definition}
\providecommand{\examplename}{Example}
\providecommand{\lemmaname}{Lemma}
\providecommand{\propositionname}{Proposition}
\providecommand{\theoremname}{Theorem}

\makeatletter
\AtBeginDocument{%
	\let\NAT@hyper@cite\@empty
	\let\NAT@hyper@link\@empty
}
\makeatother

\begin{document}
\title{A New Bayesian Bootstrap for \\ Quantitative Trade and Spatial Models}
\author{Bas Sanders, \emph{Harvard University}\textbf{\thanks{E-mail: bas\_sanders@g.harvard.edu. I thank my advisors, Isaiah Andrews,
Pol Antràs, Anna Mikusheva and Jesse Shapiro, for their guidance and
generous support. I also thank Dmitry Arkhangelsky, Marc Melitz, Neil
Shephard, Rahul Singh, Elie Tamer, Davide Viviano and Chris Walker
for helpful discussions. I am also grateful for comments from participants
of the Harvard Graduate Student Workshops in Econometrics and Trade.
All errors are mine.}}}
\date{\monthname\ \number\year}
\maketitle
\begin{abstract}
Economists use quantitative trade and spatial models to make counterfactual
predictions. Because such predictions often inform policy decisions,
it is important to communicate the uncertainty surrounding them. Three
key challenges arise in this setting: the data are dyadic and exhibit
complex dependence; the number of interacting units is typically small;
and counterfactual predictions depend on the data in two distinct
ways—through the estimation of structural parameters and through their
role as inputs into the model’s counterfactual equilibrium. I address
these challenges by proposing a new Bayesian bootstrap procedure tailored
to this context. The method is simple to implement and provides both
finite-sample Bayesian and asymptotic frequentist guarantees. Revisiting
the results in \citet{waugh2010international}, \citet{caliendo2015estimates},
and \citet{artucc2010trade} illustrates the practical advantages
of the approach.
\end{abstract}
\begin{spacing}{1.32}

\section{\protect\label{sec:Introduction}Introduction}

Economists use quantitative trade and spatial models to answer counterfactual
questions. For example, what is the effect on welfare levels and inequality
when trade costs or tariffs between a set of countries change? What
happens to employment shares and wages across sectors after a sudden
liberalization of the manufacturing sector? Since such counterfactual
predictions often inform policy decisions, it is important to communicate
the uncertainty surrounding them. However, in practice, counterfactuals
are often reported without any measure of uncertainty. For instance,
in a survey only 2 out of 36 papers report any uncertainty quantification
for their counterfactual predictions.\footnote{The survey includes all papers published between 2015 and 2024 in
five general-interest economics journals (American Economic Review,
Econometrica, Journal of Political Economy, Quarterly Journal of Economics,
and Review of Economic Studies) that contain the phrase “bilateral
trade flows” or “bilateral flows” and conduct a counterfactual exercise.}

Counterfactual predictions are typically constructed in two steps.
First, the data are used to estimate a finite-dimensional structural
parameter, for example using the generalized method of moments (GMM).
Second, the estimator—together with the observed data—is used to compute
the counterfactual prediction. For instance, in the canonical Armington
model \citep{armington1969theory}, the first step involves estimating
a trade elasticity using observed bilateral trade flows, and the second
step combines the estimated elasticity with the trade flows to predict
welfare changes under a hypothetical shift in trade costs.

Quantifying uncertainty for such a counterfactual raises three main
challenges. First, the data are often dyadic, meaning that each observation
reflects an interaction between two units. This induces a strong dependence
structure across observations. Second, the number of interacting units—such
as countries or sectors—is typically small, making it important to
use methods that retain a clear interpretation in small samples. Third,
the data enter both the estimation of the structural parameter and
the computation of the counterfactual prediction, so that the prediction
depends on the data in two distinct ways. This creates a non-classical
setting for uncertainty quantification \citep{sanders2023counterfactual}.

To address these challenges and support more informed policy decisions,
I propose a Bayesian approach. Specifically, to quantify uncertainty
for the estimator of the structural parameter, I introduce a new Bayesian
bootstrap procedure that is intuitive, easy to implement, and theoretically
grounded. The method amounts to reweighting the data using products
of draws from an exponential distribution. It readily extends to settings
where only a subset of all possible flows is observed (e.g. because
observations that equal zero are dropped), or where the data are polyadic
(i.e., each observation involves more than two units). Moreover, because
the approach is Bayesian, uncertainty quantification for the counterfactual
prediction follows automatically from the posterior distribution of
the structural parameter. 

The key theoretical contribution of this paper is to introduce and
justify a new Bayesian bootstrap procedure tailored to polyadic data
structures. The method extends the classical Bayesian bootstrap \citep{rubin1981bayesian,chamberlain2003nonparametric}
to settings where each observation involves more than a single unit.
A central result is that this procedure admits a finite-sample Bayesian
interpretation. Specifically, under a particular choice of model and
prior, the posterior distribution converges to the Bayesian bootstrap
distribution as a prior informativeness parameter tends to zero. The
model assumes that the polyadic data are generated as functions of
unit-specific latent variables, which are drawn independently from
a common distribution. A Dirichlet process prior is placed on this
distribution, and the Bayesian bootstrap distribution emerges as the
limiting posterior when the prior becomes uninformative. In the main
text, I provide several additional motivations for the model and prior
underlying my results. 

The fact that the Bayesian bootstrap procedure admits a finite-sample
Bayesian interpretation is particularly relevant in applications with
a small number of units. In small samples, posterior distributions
are often non-Gaussian, and asymmetries in their shape can have important
policy implications: for instance, right-skewness could suggest greater
potential for large welfare gains, while left-skewness indicates a
chance of substantial welfare losses. Capturing this asymmetry is
critical for informed decision-making. 

In addition to its finite-sample Bayesian validity, the procedure
is also asymptotically valid in a frequentist sense under mild regularity
conditions. These conditions are generally satisfied, for example,
by the class of GMM estimators, including the Pseudo Poisson Maximum
Likelihood (PPML) estimator of \citet{silva2006log}. This dual validity
makes the procedure competitive with existing methods in the literature—which
are reviewed later in the introduction—that rely exclusively on asymptotic
approximations. For frequentist uncertainty quantification of the
counterfactual prediction, I provide a delta method-type result that
accounts for the fact that counterfactual predictions depend on the
data in two distinct ways.

Throughout the paper, I use the application in \citet{waugh2010international}
as a running example. In this setting, the structural parameter is
a productivity parameter common across countries; the interacting
units are 43 countries; and the estimation method is simple OLS on
dyadic trade flows—a special case of GMM. The posterior variance implied
by my procedure is considerably larger than the heteroskedasticity-robust
variance reported in \citet{waugh2010international}, which does not
account for dependence across dyads. The counterfactual objects of
interest are various inequality statistics under alternative trade
cost schedules, for which I construct credible intervals; these intervals
are narrow, and the economic conclusions are robust.

To further illustrate the flexibility of the method, I also revisit
results in \citet{caliendo2015estimates} and \citet{artucc2010trade}.
In \citet{caliendo2015estimates}, the structural parameters are sector-specific
trade elasticities; the interacting units are countries; and the estimation
method is simple OLS on \textit{triadic flows}. The number of countries
per sector ranges from 12 to 15. The credible intervals for the elasticities
are substantially wider than the heteroskedasticity-robust confidence
intervals reported in the original paper, and they often include regions
of the parameter space where model assumptions are violated. For some
sectors, the posterior distribution of the elasticity is approximately
normal; for others, it is skewed or heavy-tailed. The counterfactual
objects of interest are changes in welfare due to NAFTA, originally
reported in \citet{caliendo2015estimates} without uncertainty quantification.
Skewness in the posterior distributions of the elasticities induces
asymmetry in the posterior distribution of welfare changes, shifting
probability mass away from zero. The credible intervals around welfare
predictions reflect substantial uncertainty and considerable heterogeneity
across countries, although the ranking of welfare effects remains
unchanged.

In the setting of \citet{artucc2010trade}, the structural parameters
are the mean and variance of workers’ switching costs between sectors;
the interacting units are \textit{six} sectors, and the estimation
method is \textit{over-identified GMM} with three instruments. The
posterior distributions for both parameters are non-normal and exhibit
heavy right tails, indicating substantial uncertainty—particularly
regarding the possibility of large switching costs. The counterfactual
objects of interest are changes in various economic outcomes following
a liberalization of the manufacturing sector, and the resulting credible
intervals again reveal substantial uncertainty. Notably, accounting
for this uncertainty reveals that equilibrium wages may plausibly
increase as a result of liberalization with a posterior probability
of 25\%—a finding not visible from point estimates alone. 

The method I propose substantially improves how uncertainty is quantified
for both the structural parameter estimator and the counterfactual
prediction, relative to current practice in quantitative trade and
spatial economics. In the survey mentioned above, 24 out of 36 papers
report a standard error for the estimator of the structural parameter.
However, the most common approach is to compute heteroskedasticity-robust
standard errors, clustering either on dyads or only on the origin
or destination unit. A more flexible alternative, used in several
papers, is two-way clustering on both the origin and destination units.
While this allows for richer dependence, it still fails to capture
key dyadic correlations—for example, between the trade flow from Germany
to the United States and the trade flow from France to Germany. Ideally,
one would allow for dependence between flows that have at least one
unit in common. A small literature proposes methods to account for
such dyadic dependence \citep{fafchamps2007formation,cameron2014robust,aronow2015cluster,graham2020dyadic,graham2020network,davezies2021empirical}.
However, it remains rare for published papers in quantitative trade
and spatial economics to adopt these tools.\footnote{To date, among all papers citing this literature, only two papers
both contain the phrase “bilateral trade flows” or “bilateral flows”
and explicitly account for dyadic dependence: \citet{rosendorf2023alliance}
and \citet{wigton2024lexical}.}

I compare my method to two alternatives from the existing literature.
The closest is the pigeonhole bootstrap introduced by \citet{davezies2021empirical},
which extends the standard resampling bootstrap to polyadic settings.
Both my approach and the pigeonhole bootstrap reweight polyadic observations
using specific weights. One key difference is that the Bayesian bootstrap
assigns continuous and strictly positive weights to all observations,
whereas the pigeonhole bootstrap draws discrete weights and may assign
zero weight to some. Another difference is that theoretical guarantees
for the pigeonhole bootstrap rely on asymptotic approximations that
assume a large number of interacting units. However, the applications
I study involve small numbers of units. In these settings, I show
that the pigeonhole bootstrap can be numerically unstable and tends
to produce wider confidence intervals than the Bayesian bootstrap
procedure. By contrast, in settings with a large number of interacting
units, the approaches are equivalent.

A second alternative for uncertainty quantification is to derive frequentist
standard errors. \citet{graham2020dyadic,graham2020network} build
on earlier work \citep{fafchamps2007formation,cameron2014robust,aronow2015cluster}
to develop consistent variance estimators for maximum likelihood estimators.
I extend these results to Z-estimators—that is, estimators defined
as the solution to a system of estimating equations. As with the pigeonhole
bootstrap, the validity of these frequentist standard errors relies
on asymptotic approximations, which may perform poorly when the number
of interacting units is small. Again, when the number of interacting
units is large, using analytic standard errors is equivalent to using
my approach.

Both \citet{davezies2021empirical} and \citet{graham2020dyadic,graham2020network}
only focus on uncertainty quantification for the estimator of the
structural parameter.\footnote{As mentioned above, only 2 out of 36 papers in the survey report uncertainty
quantification for their counterfactual prediction. \citet{adao2017nonparametric}
samples from the asymptotic distribution of the estimator, while \citet{allen2020universal}
samples uniformly over its confidence interval.} Since the counterfactual prediction depends on this estimator as
an input, it inherits the challenges associated with dyadic data and
a small number of interacting units. As a result, valid uncertainty
quantification for counterfactual predictions using these methods
also relies on asymptotic approximations. In contrast, the Bayesian
bootstrap procedure provides valid finite-sample uncertainty quantification
for counterfactual predictions, supporting better-informed policy
decisions in small-sample settings.

This paper contributes to several literatures. First, it adds to a
growing body of work aimed at improving counterfactual analysis in
quantitative trade and spatial economics \citep{kehoe2017quantitative,adao2017nonparametric,dingel2020spatial,adao2023putting,sanders2023counterfactual,ansari2024quantifying}.
Second, it advances research on bootstrap methods designed for settings
in which standard resampling approaches fail \citep{janssen1994weighted,davezies2021empirical,menzel2021bootstrap}.
Third, it engages with the emerging literature on uncertainty quantification
in polyadic settings \citep{snijders1999non,graham2020dyadic,graham2020network,menzel2021bootstrap,davezies2021empirical,graham2024sparse}.
While the Bayesian bootstrap is briefly mentioned in \citet{graham2020network}
in the context of dyadic regression, it has not been further developed
or applied in polyadic settings.

The rest of the paper is organized as follows. The next section introduces
the setting and the proposed Bayesian bootstrap procedure. Sections
\ref{sec:Theory_FS_Bayesian} and \ref{sec:Theory_Asympotics} present
the main theoretical contributions, focusing on finite-sample Bayesian
results and asymptotic validity, respectively. Section \ref{sec:Extensions}
discusses several extensions of the core framework. The method is
then applied in Section \ref{sec:Applications} to the empirical settings
studied in \citet{caliendo2015estimates} and \citet{artucc2010trade}.
Section \ref{sec:Alt_methods} compares the proposed approach to alternative
methods for uncertainty quantification. Section \ref{sec:Conclusion}
concludes.

\section{\protect\label{sec:Setting_Proposed_method}Setting and Proposed
Method}

In this section I introduce the setting and goal of the paper. I lay
out my proposed method and illustrate it using my running example.
I consider misspecification-robust uncertainty quantification for
over-identified GMM as a special case. Theoretical justifications
are deferred to Sections \ref{sec:Theory_FS_Bayesian} and \ref{sec:Theory_Asympotics}. 

\subsection{Setting and Goal}

\subsubsection{Data Environment}

We observe a sample of bilateral data $\left\{ X_{k\ell}\right\} _{k\neq\ell}\in\mathcal{X}^{n\left(n-1\right)}$
with $\mathcal{X}\subseteq\mathbb{R}^{d_{X}}$, with $n\in\mathbb{N}$
the \textit{number of interacting units}. Since we consider bilateral
data, the effective \textit{sample size} is $n\left(n-1\right)$. 
\begin{example*}[\citealp{waugh2010international}]
In \citet{waugh2010international}, the interacting units are $43$
countries so $n=43$. The data are
\[
X_{k\ell}=\left(\lambda_{k\ell},\lambda_{kk},\tau_{k\ell},p_{k},p_{\ell}\right)\in\mathcal{X}=\left[0,1\right]^{2}\times\left(1,\infty\right]\times\mathbb{R}_{+}^{2},
\]
for $k\neq\ell$.\footnote{The sample size in \citet{waugh2010international} is not actually
$43\cdot42=1806$ but $1373$, because observations with $\lambda_{k\ell}=0$
are dropped. I will come back to this in Section \ref{subsec:MRGMM}.} Here, $\lambda_{k\ell}$ denotes country $\ell$'s expenditure share
on goods from country $k$, $\tau_{k\ell}$ denotes estimated iceberg
trade costs from country $k$ to country $\ell$, and $p_{k}$ denotes
the aggregate price in country $k$. $\triangle$
\end{example*}

\subsubsection{\protect\label{subsec:theta}Structural Estimator and Estimand}

Denote the empirical distribution of the data by 
\begin{equation}
\mathbb{P}_{n,X_{ij}}=\sum_{k\neq\ell}\frac{1}{n\left(n-1\right)}\cdot\delta_{X_{k\ell}},\label{eq:emp_dist}
\end{equation}
for $\delta_{x}$ the Dirac measure at $x$. The Dirac measure at
a single observation $X_{k\ell}$ corresponds to a degenerate probability
distribution which puts a mass of $1$ at that observation. The empirical
distribution assigns mass $\frac{1}{n\left(n-1\right)}$ to each observation
$X_{k\ell}$ and because there are $n\left(n-1\right)$ observations
this is a valid distribution. 

The researcher aims to estimate a structural parameter using the observed
data $\left\{ X_{k\ell}\right\} _{k\neq\ell}$. I assume the estimator
$\hat{\theta}$ is a function of the empirical distribution\footnote{For ease of exposition I assume $\hat{\theta}$ is a scalar, but the
same arguments apply to vector-valued $\hat{\theta}$.}:
\begin{assumption}[Structural estimator]
\label{assu:theta_tilde}We have
\begin{equation}
\hat{\theta}=T\left(\mathbb{P}_{n,X_{ij}}\right),\label{eq:theta_tilde}
\end{equation}
for a known function $T:\Delta\left(\mathcal{X}\right)\rightarrow\Theta\subseteq\mathbb{R}$. 
\end{assumption}
Here, $\Delta\left(\mathcal{X}\right)$ denotes the set of all probability
distributions over $\mathcal{X}$. Assumption \ref{assu:theta_tilde}
covers many common estimators:
\begin{align*}
T_{\mathrm{average}}\left(\mathbb{P}_{n,X_{ij}}\right) & =\mathbb{E}_{\mathbb{P}_{n,X_{ij}}}\left[X_{ij}\right]=\sum_{k\neq\ell}\frac{1}{n\left(n-1\right)}\cdot X_{k\ell}\\
T_{\mathrm{regression}}\left(\mathbb{P}_{n,\left(F_{ij},R_{ij}\right)}\right) & =\underset{\vartheta\in\Theta}{\arg\min}\ \mathbb{E}_{\mathbb{P}_{n,\left(F_{ij},R_{ij}\right)}}\left[\left(F_{ij}-\vartheta R_{ij}\right)^{2}\right]=\frac{\sum_{k\neq\ell}F_{k\ell}R_{k\ell}}{\sum_{k\neq\ell}R_{k\ell}^{2}}\\
T_{\mathrm{GMM}}\left(\mathbb{P}_{n,X_{ij}}\right) & =\underset{\vartheta\in\Theta}{\arg\min}\ \mathbb{E}_{\mathbb{P}_{n,X_{ij}}}\left[\psi\left(X_{ij};\vartheta\right)\right]^{'}\Omega\mathbb{E}_{\mathbb{P}_{n,X_{ij}}}\left[\psi\left(X_{ij};\vartheta\right)\right].
\end{align*}
Estimators that are not covered by Assumption \ref{assu:theta_tilde}
are estimators that involve multiple flows, such as the average flow
over triads of units in a graph.
\begin{example*}[\citealp{waugh2010international}]
The relevant empirical distribution in \citet{waugh2010international}
is
\[
\mathbb{P}_{n,X_{ij}}=\sum_{k\neq\ell}\frac{1}{n\left(n-1\right)}\cdot\delta_{\left(\lambda_{k\ell},\lambda_{kk},\tau_{k\ell},p_{k},p_{\ell}\right)}.
\]
The author aims to estimate a productivity parameter which governs
the dispersion of efficiency levels across countries. An arbitrage
condition motivates the simple linear regression using $\left\{ \log\frac{\lambda_{k\ell}}{\lambda_{kk}}\right\} _{k\ne\ell}$
and $\left\{ \log\left(\tau_{k\ell}\frac{p_{k}}{p_{\ell}}\right)\right\} _{k\neq\ell}$:
\begin{align}
\hat{\theta} & =-T_{\mathrm{Waugh}}\left(\mathbb{P}_{n,X_{ij}}\right)=-\underset{\vartheta\in\Theta}{\arg\min}\ \mathbb{E}_{\mathbb{P}_{n,X_{ij}}}\left[\left(\log\frac{\lambda_{ij}}{\lambda_{ii}}-\vartheta\log\left(\tau_{ij}\frac{p_{i}}{p_{j}}\right)\right)^{2}\right]\nonumber \\
 & =-\frac{\sum_{k\neq\ell}\log\frac{\lambda_{k\ell}}{\lambda_{kk}}\log\left(\tau_{k\ell}\frac{p_{k}}{p_{\ell}}\right)}{\sum_{k\neq\ell}\left(\log\left(\tau_{k\ell}\frac{p_{k}}{p_{\ell}}\right)\right)^{2}}.\label{eq:theta_tilde_Waugh}
\end{align}
The estimator $\hat{\theta}$ indeed satisfies Assumption \ref{assu:theta_tilde}.
$\triangle$
\end{example*}
Note that estimators that can be written as in Equation \eqref{eq:theta_tilde}
are \textit{permutation invariant} with respect to the observed data
$\left\{ X_{k\ell}\right\} _{k\neq\ell}$, because for any permutation
$\sigma:\left\{ 1,...,n\right\} \rightarrow\left\{ 1,...,n\right\} $,
we have
\begin{align*}
\hat{\theta} & =T\left(\sum_{k\neq\ell}\frac{1}{n\left(n-1\right)}\cdot\delta_{X_{k\ell}}\right)=T\left(\sum_{k\neq\ell}\frac{1}{n\left(n-1\right)}\cdot\delta_{X_{\sigma\left(k\right)\sigma\left(\ell\right)}}\right).
\end{align*}
For the purposes of structural estimation, it is then without loss
to assume that the observed data are \textit{jointly exchangeable},
which means that the joint distribution does not change when we relabel
the indices, so that
\[
\left\{ X_{k\ell}\right\} _{k\neq\ell}\overset{d}{=}\left\{ X_{\sigma\left(k\right)\sigma\left(\ell\right)}\right\} _{k\neq\ell},
\]
for any permutation $\sigma:\left\{ 1,...,n\right\} \rightarrow\left\{ 1,...,n\right\} $.\footnote{In other papers concerning dyadic dependence \citep{graham2020dyadic,graham2020network,davezies2021empirical},
joint exchangeability is used as a primitive assumption. I instead
motivate it by focusing on the class of estimators that satisfy Assumption
\ref{assu:theta_tilde}.} Joint exchangeability of the data implies that the elements of $\left\{ X_{k\ell}\right\} _{k\neq\ell}$
have a common \textit{marginal} probability distribution, which I
will denote by $\mathbb{P}_{X_{ij}}$. The \textit{structural estimand
of interest} then is 
\begin{equation}
\theta\equiv T\left(\mathbb{P}_{X_{ij}}\right).\label{eq:estimand}
\end{equation}
Note that this estimand might differ from the structural parameter
of interest if the model is misspecified, as illustrated in the next
example. In such cases, my results deliver valid inference for the
estimand $\theta$. 
\begin{example*}[\citealp{waugh2010international}]
Given that \citet{waugh2010international} considers a simple regression,
for the purposes of estimation, it is without loss to assume that
the observed data $\left\{ X_{k\ell}\right\} _{k\neq\ell}$ are jointly
exchangeable. Here, joint exchangeability means that the joint distribution
of bilateral data remains unchanged if we relabel the countries. Joint
exchangeability implies that there exists some marginal distribution
$\mathbb{P}_{X_{ij}}$ from which all the observations are drawn.
The structural estimand of interest then equals the negative of the
function $T_{\mathrm{Waugh}}$ applied to $\mathbb{P}_{X_{ij}}$,
so that 
\[
\theta\equiv-T_{\mathrm{Waugh}}\left(\mathbb{P}_{X_{ij}}\right)=-\underset{\vartheta\in\Theta}{\arg\min}\ \mathbb{E}_{\mathbb{P}_{X_{ij}}}\left[\left(\log\frac{\lambda_{ij}}{\lambda_{ii}}-\vartheta\log\left(\tau_{ij}\frac{p_{i}}{p_{j}}\right)\right)^{2}\right].
\]
This estimand corresponds to the coefficient in the regression 
\begin{equation}
\log\frac{\lambda_{ij}}{\lambda_{ii}}=-\theta\log\left(\tau_{ij}\frac{p_{i}}{p_{j}}\right)+\varepsilon_{ij},\label{eq:Waugh_regression}
\end{equation}
for an orthogonal error term $\varepsilon_{ij}$. The regression coefficient
in Equation \eqref{eq:theta_tilde_Waugh} was motivated by the model
equation 
\[
\log\frac{\lambda_{ij}}{\lambda_{ii}}=-\theta_{M}\log\left(\tau_{ij}\frac{p_{i}}{p_{j}}\right),
\]
which does not hold exactly in-sample because there is country-level
uncertainty due to country-level productivity shocks. The regression
equation will only recover the true structural parameter $\theta_{M}$
under the assumption that the productivity shocks are exogenous and
follow a Fréchet distribution. Nevertheless, \citet{waugh2010international}
conducts estimation using \eqref{eq:theta_tilde_Waugh}, so going
forward, I focus on the estimand $\theta$ rather than $\theta_{M}$.
$\triangle$
\end{example*}

\subsubsection{\protect\label{subsec:Counterfactual-Predictions}Counterfactual
Predictions}

In quantitative trade and spatial models, researchers are interested
in forming counterfactual predictions. Since these predictions are
relative to some observed factual situation, they are functions of
the realized bilateral data $\left\{ X_{k\ell}\right\} _{k\neq\ell}$
and the structural estimator $\hat{\theta}$:
\begin{assumption}[Counterfactual prediction]
\label{assu:gamma}The reported counterfactual prediction of interest
can be written as
\begin{equation}
\hat{\gamma}=g\left(\left\{ X_{k\ell}\right\} _{k\neq\ell},\hat{\theta}\right),\label{eq:key_eqn}
\end{equation}
for a known function $g:\mathcal{X}^{n\left(n-1\right)}\times\Theta\rightarrow\mathbb{R}$.
\end{assumption}
The corresponding estimand is 
\[
\gamma=g\left(\left\{ X_{k\ell}\right\} _{k\neq\ell},\theta\right)\equiv g\left(\left\{ X_{k\ell}\right\} _{k\neq\ell},T\left(\mathbb{P}_{X_{ij}}\right)\right).
\]
In conventional economic models the estimand of interest is a function
only of the distribution of the data, rather than the actual observations.
The dependence of $\gamma$ on both the realized bilateral data and
their distribution creates a non-classical setting, as was noted in
\citet{sanders2023counterfactual}.
\begin{example*}[\citealp{waugh2010international}]
After obtaining the estimator $\hat{\theta}$, we can use the model
presented in \citet{waugh2010international} to find the equilibrium
wage vector for a given counterfactual trade cost schedule. The relevant
counterfactual mapping is
\begin{equation}
\left\{ X_{k\ell}\right\} _{k\neq\ell},\hat{\theta},\left\{ \tau_{k\ell}^{\mathrm{cf}}\right\} \mapsto\left\{ \hat{w}_{k}^{\mathrm{cf}}\right\} .\label{eq:eql_mapping_Waugh}
\end{equation}
It maps the realized data, the structural estimator and a counterfactual
trade cost schedule to a vector which contains the counterfactual
wage for all 43 countries. Appendix \ref{subsec:W_model_details}
outlines the details of this mapping. From the counterfactual wage
vector we can compute various scalar objects of interest, such as
the wage levels of specific countries or summary statistics across
the wage vector. 

\citet{waugh2010international} considers a series of counterfactuals
that calculate inequality statistics of the equilibrium wage vector
for different trade cost schedules. The inequality statistics are
the variance of log wages, the ratio of the 90th and 10th percentile
of wages, and the mean percentage change in wages. The different counterfactual
trade cost schedules are autarky ($\tau_{ij}^{\mathrm{cf}}=\infty$
for all $i\neq j$), symmetry ($\tau_{ij}^{\mathrm{cf}}=\min\left\{ \tau_{ij},\tau_{ji}\right\} $
for all $i\neq j$) and free trade ($\tau_{ij}^{\mathrm{cf}}=1$ for
all $i\neq j$). Using the equilibrium mapping in Equation \eqref{eq:eql_mapping_Waugh},
it follows that each counterfactual prediction can be written as in
Equation \eqref{eq:key_eqn}. The resulting point estimates are reported
in Table 4 of \citet{waugh2010international} without any uncertainty
quantification. $\triangle$
\end{example*}
The discussion in the previous two sections highlights two distinct
statistical objects of interest: the structural estimator $\hat{\theta}$
and the counterfactual prediction $\hat{\gamma}$. To quantify uncertainty
for each, I proceed in two steps. First, in Section \ref{subsec:UQ_theta},
I present a Bayesian bootstrap procedure to quantify uncertainty for
$\hat{\theta}$. Then, in Section \ref{subsec:UQ_gamma}, I use Assumption
\ref{assu:gamma} to quantify uncertainty for $\hat{\gamma}$. 

\subsection{\protect\label{subsec:UQ_theta}Bayesian Uncertainty Quantification
for $\hat{\theta}$}

To quantify uncertainty for $\hat{\theta}$, I consider a bootstrap
procedure. Specifically, in each bootstrap iteration $b=1,...,B$,
$\hat{\theta}^{*,\left(b\right)}$ is computed by replacing the empirical
distribution in Equation \eqref{eq:emp_dist} by a weighted version
of this empirical distribution,
\[
\mathbb{P}_{n,X_{ij}}^{*,\left(b\right)}=\sum_{k\neq\ell}\omega_{k\ell}^{\left(b\right)}\cdot\delta_{X_{k\ell}}.
\]
The weights $\left\{ \omega_{k\ell}^{\left(b\right)}\right\} _{k\neq\ell}$
are computed using draws from a Dirichlet distribution,
\begin{equation}
\omega_{k\ell}^{\left(b\right)}=\frac{W_{k}^{\left(b\right)}\cdot W_{\ell}^{\left(b\right)}}{\sum_{s\neq t}W_{s}^{\left(b\right)}\cdot W_{t}^{\left(b\right)}},\quad\left(W_{1}^{\left(b\right)},...,W_{n}^{\left(b\right)}\right)\sim\mathrm{Dir}\left(n;1,...,1\right).\label{eq:Dirichlet_weights}
\end{equation}
In practice, it is convenient that the Dirichlet distribution $\mathrm{Dir}\left(n;1,...,1\right)$
can be constructed from i.i.d. draws from an exponential distribution:
\begin{align*}
\left(V_{1}^{\left(b\right)},...,V_{n}^{\left(b\right)}\right) & \overset{\mathrm{iid}}{\sim}\mathrm{Exp}\left(1\right)\\
W_{k}^{\left(b\right)} & =\frac{V_{k}^{\left(b\right)}}{\sum_{s=1}^{n}V_{s}^{\left(b\right)}},\quad k=1,...,n\\
\omega_{k\ell}^{\left(b\right)} & =\frac{V_{k}^{\left(b\right)}\cdot V_{\ell}^{\left(b\right)}}{\sum_{s\neq t}V_{s}^{\left(b\right)}\cdot V_{t}^{\left(b\right)}},\quad k,\ell=1,...,n.
\end{align*}
The procedure to quantify uncertainty for the estimator $\hat{\theta}$
is summarized in Algorithm \ref{alg:BB}.

\begin{algorithm}[h]
\caption{\protect\label{alg:BB}Bayesian bootstrap procedure}

\begin{enumerate}
\item Input: Bilateral data $\left\{ X_{k\ell}\right\} _{k\neq\ell}$ and
estimator function $T:\Delta\left(\mathcal{X}\right)\rightarrow\Theta$.
\item For each bootstrap draw $b=1,...,B$:
\begin{enumerate}
\item Sample $\left(V_{1}^{\left(b\right)},...,V_{n}^{\left(b\right)}\right)\overset{\mathrm{iid}}{\sim}\mathrm{Exp}\left(1\right)$.
\item Compute 
\[
\hat{\theta}^{*,\left(b\right)}=T\left(\sum_{k\neq\ell}\frac{V_{k}^{\left(b\right)}\cdot V_{\ell}^{\left(b\right)}}{\sum_{s\neq t}V_{s}^{\left(b\right)}\cdot V_{t}^{\left(b\right)}}\cdot\delta_{X_{k\ell}}\right).
\]
\end{enumerate}
\item Report the quantiles of interest of $\left\{ \hat{\theta}^{*,\left(1\right)},...,\hat{\theta}^{*,\left(B\right)}\right\} $.
\end{enumerate}
\end{algorithm}
This procedure is a natural generalization of the univariate Bayesian
bootstrap \citep{rubin1981bayesian,chamberlain2003nonparametric}.
It is intuitive and easy to implement, as it just requires drawing
from an exponential distribution and reweighting the data appropriately. 

In Sections \ref{sec:Theory_FS_Bayesian} and \ref{sec:Theory_Asympotics}
I will provide various theoretical motivations for the Bayesian bootstrap
procedure. The key takeaway from Section \ref{sec:Theory_FS_Bayesian}
is that Algorithm \ref{alg:BB} produces draws from a limiting posterior
for $\theta$ given the bilateral data $\left\{ X_{k\ell}\right\} _{k\neq\ell}$
for a well-motivated model and prior. In addition to this finite-sample
Bayesian motivation, the key takeaway from Section \ref{sec:Theory_Asympotics}
is that the bootstrap procedure is also asymptotically valid in a
frequentist sense. 
\begin{example*}[\citealp{waugh2010international}]
The productivity parameter in \citet{waugh2010international} is
estimated for the full sample and for the two subsets of OECD countries
and non-OECD countries. Using Algorithm \ref{alg:BB}, we can obtain
draws from the posterior distribution of the structural parameter
given the bilateral data. Specifically, for each bootstrap iteration
I compute
\begin{align*}
\hat{\theta}^{*,\left(b\right)} & =-T_{\mathrm{Waugh}}\left(\sum_{k\neq\ell}\frac{V_{k}^{\left(b\right)}\cdot V_{\ell}^{\left(b\right)}}{\sum_{s\neq t}V_{s}^{\left(b\right)}\cdot V_{t}^{\left(b\right)}}\cdot\delta_{X_{k\ell}}\right)\\
 & =-\underset{\vartheta\in\Theta}{\arg\min}\sum_{k\neq\ell}V_{k}^{\left(b\right)}\cdot V_{\ell}^{\left(b\right)}\cdot\left(\log\frac{\lambda_{k\ell}}{\lambda_{kk}}-\vartheta\log\left(\tau_{k\ell}\frac{p_{k}}{p_{\ell}}\right)\right)^{2}.
\end{align*}
It turns out that in the case of simple OLS, for each bootstrap draw
we can just re-weight both the dependent and independent variables
by $\sqrt{V_{k}^{\left(b\right)}\cdot V_{\ell}^{\left(b\right)}}$
for all $k\neq\ell$, and compute the corresponding OLS coefficient.
Because the bootstrap distribution corresponds to a limiting posterior
distribution, we can interpret $\left\{ \hat{\theta}^{*,\left(1\right)},...,\hat{\theta}^{*,\left(B\right)}\right\} $
as posterior draws. From these posterior draws, we can obtain a $100\left(1-\alpha\right)\%$
\textit{credible interval} by computing the $\alpha/2$ and $1-\alpha/2$
quantiles. The 95\% credible intervals are reported in Table \ref{tab:W}
and the posterior distributions are plotted in Figure \ref{fig:W}. 

In \citet{waugh2010international}, no uncertainty quantification
is discussed for $\hat{\theta}$. In the accompanying code, the author
computes the dyad-level heteroskedastic-robust standard error $\sqrt{\hat{\Sigma}_{\theta}}$.
In Table \ref{tab:W} I add the corresponding 95\% confidence intervals,
computed using the familiar $\left[\hat{\theta}\pm1.96\cdot\sqrt{\hat{\Sigma}_{\theta}}\right]$.
In Figure \ref{fig:W}, I also plot a normal distribution with mean
$\hat{\theta}$ and standard error $\sqrt{\hat{\Sigma}_{\theta}}$,
since the standard confidence intervals rely on $\hat{\theta}$ to
be approximately normal centered at $\theta$ with variance $\hat{\Sigma}_{\theta}$.\footnote{Formally, this normality could follow from assumptions on the underlying
data generating process such that a Bernstein-von Mises type result
holds \citep{van2000asymptotic}. In that case the influence of the
prior distribution on $\theta$ becomes negligible and the posterior
distribution approximately equals a normal distribution centered at
the maximum likelihood estimator.} We observe that the posterior is approximately normal but has larger
variance than reported in the accompanying code of the paper, which
suggests that considering dyadic dependence is important. 
\begin{table}[h]
\begin{centering}
\begin{tabular}{|c|c|c|c|}
\hline 
 & Point estimate & As in paper & Bayesian bootstrap\tabularnewline
\hline 
\hline 
All countries, $n=43$ & 5.55 & {[}5.39, 5.71{]} & {[}5.12, 6.02{]}\tabularnewline
\hline 
Only OECD, $n=19$ & 7.91 & {[}7.46, 8.37{]} & {[}6.91, 9.21{]}\tabularnewline
\hline 
Only non-OECD, $n=24$ & 5.45 & {[}5.06, 5.84{]} & {[}4.42, 6.65{]}\tabularnewline
\hline 
\end{tabular}
\par\end{centering}
\caption{\protect\label{tab:W}Uncertainty quantification for productivity
parameters as in \citet{waugh2010international}.}
\end{table}
\begin{figure}[h]
\centering{}\includegraphics[scale=0.16]{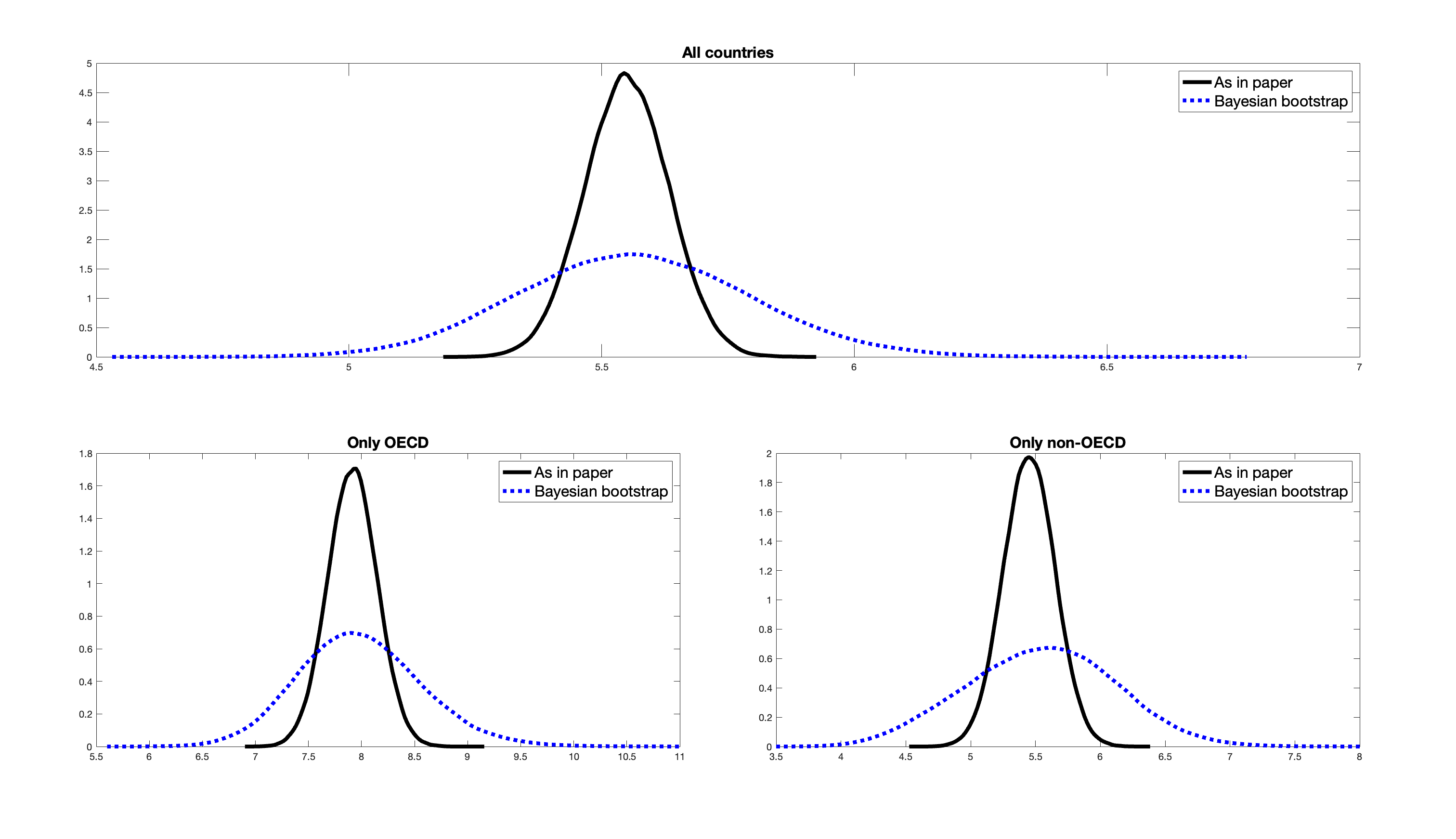}\caption{\protect\label{fig:W}Distributions for productivity parameters as
in \citet{waugh2010international}.}
\end{figure}

Table \ref{tab:W} shows that the confidence intervals and credible
intervals differ substantially. To better understand this discrepancy,
Appendix \ref{subsec:PB_bootstrap_DGP} presents a data-calibrated
simulation exercise based on the pigeonhole bootstrap—a method introduced
in Section \ref{subsec:Pigeonhole}. This setup enables a direct evaluation
of the coverage performance of various uncertainty quantification
methods. Table \ref{tab:PB_bootstrap_DGP_W} shows that using heteroskedasticity-robust
standard errors leads to under-coverage. $\triangle$ 
\begin{table}[h]
\begin{centering}
\begin{tabular}{|c|c|c|}
\hline 
 & \begin{cellvarwidth}[t]
\centering
As in

paper
\end{cellvarwidth} & \begin{cellvarwidth}[t]
\centering
Bayesian

bootstrap
\end{cellvarwidth}\tabularnewline
\hline 
\hline 
All countries, $n=43$ & 0.498 & 0.979\tabularnewline
\hline 
Only OECD, $n=19$ & 0.533 & 0.954\tabularnewline
\hline 
Only non-OECD, $n=24$ & 0.416 & 0.913\tabularnewline
\hline 
\end{tabular}
\par\end{centering}
\caption{\protect\label{tab:PB_bootstrap_DGP_W}Coverage for the approach used
in \citet{waugh2010international} and the Bayesian bootstrap using
the pigeonhole bootstrap DGP as described in Appendix \ref{subsec:PB_bootstrap_DGP}.}
\end{table}
\end{example*}

\subsubsection{\protect\label{subsec:MRGMM}Special Case: Misspecification-Robust
Uncertainty Quantification for GMM}

Often, researchers are interested in over-identified GMM estimators
of the form
\begin{align*}
\hat{\theta} & =\underset{\vartheta\in\Theta}{\arg\min}\ \mathbb{E}_{\mathbb{P}_{n,X_{ij}}}\left[\psi\left(X_{ij};\vartheta\right)\right]^{'}\hat{\Omega}\mathbb{E}_{\mathbb{P}_{n,X_{ij}}}\left[\psi\left(X_{ij};\vartheta\right)\right],
\end{align*}
where $\psi:\mathcal{X}\rightarrow\mathbb{R}^{L}$ with $L\geq K$,
$X_{ij}\in\mathcal{X}$, $\theta\in\Theta\subseteq\mathbb{R}^{K}$
and $\hat{\Omega}$ is an estimated weight matrix. For example, the
PPML estimator in \citet{silva2006log} corresponds to the moment
function
\begin{equation}
\psi\left(F_{ij},R_{ij};\vartheta\right)=\left(F_{ij}-\exp\left\{ R_{ij}^{'}\vartheta\right\} \right)R_{ij},\label{eq:ppml}
\end{equation}
where $F_{ij}\in\mathbb{R}_{+}$ is the dependent variable, $\vartheta\in\mathbb{R}^{d_{\theta}}$
is a vector of parameters and $R_{ij}\in\mathbb{R}^{d_{\theta}}$
is a vector of regressors. 

We know that the optimal weight matrix is the inverse of the variance-covariance
matrix of the moments at $\theta$ \citep{hansen1982large,chamberlain1987asymptotic}.
In practice, since we require an estimate of this optimal weight matrix,
researchers often use a two-step procedure. In the first step the
identity matrix is used as a weight matrix:
\begin{align}
\psi_{n}\left(\vartheta\right) & =\mathbb{E}_{\mathbb{P}_{n,X_{ij}}}\left[\psi\left(X_{ij};\vartheta\right)\right]\label{eq:GMM_1}\\
\hat{\theta}^{\mathrm{1-GMM}} & =\underset{\vartheta\in\Theta}{\arg\min}\ \psi_{n}\left(\vartheta\right)^{'}\psi_{n}\left(\vartheta\right).\label{eq:GMM_2}
\end{align}
The resulting estimator is plugged in to find an estimator of the
optimal weight matrix, which is then used to find the two-step GMM
estimator:\footnote{I follow \citet{lee2014asymptotic} and use the centered weight matrix,
rather than the uncentered version $\hat{\Omega}^{\mathrm{uncentered}}\left(\vartheta\right)=\left(\mathbb{E}_{\mathbb{P}_{n,X_{ij}}}\left[\psi\left(X_{k\ell};\vartheta\right)\psi\left(X_{k\ell};\vartheta\right)^{'}\right]\right)^{-1}$.
The choice of weight matrix affects the resulting pseudo-true value.
As outlined in \citet{hall2000covariance}, the uncentered version
includes bias terms of the moment function, which makes the centered
version better behaved under misspecification.} 
\begin{align}
\hat{\Omega}\left(\vartheta\right) & =\left(\mathbb{E}_{\mathbb{P}_{n,X_{ij}}}\left[\left\{ \psi\left(X_{ij};\vartheta\right)-\psi_{n}\left(\vartheta\right)\right\} \left\{ \psi\left(X_{ij};\vartheta\right)-\psi_{n}\left(\vartheta\right)\right\} ^{'}\right]\right)^{-1}\label{eq:GMM_3}\\
\hat{\theta}^{\mathrm{2-GMM}} & =\underset{\vartheta\in\Theta}{\arg\min}\ \psi_{n}\left(\vartheta\right)^{'}\hat{\Omega}\left(\hat{\theta}^{\mathrm{1-GMM}}\right)\psi_{n}\left(\vartheta\right).\label{eq:GMM_4}
\end{align}
The two-step estimator satisfies Assumption \ref{assu:theta_tilde},
which implies that for the purposes of estimation it is without loss
to assume that the elements of $\left\{ X_{k\ell}\right\} _{k\neq\ell}$
have a common marginal distribution $\mathbb{P}_{X_{ij}}$. The relevant
moment conditions then are
\begin{equation}
\mathbb{E}_{\mathbb{P}_{X_{ij}}}\left[\psi\left(X_{ij};\theta\right)\right]=0.\label{eq:moments_over_identified}
\end{equation}
These moments might be misspecified, meaning that there exists no
$\theta\in\Theta$ such that the moment equations in Equation \eqref{eq:moments_over_identified}
hold. In this case, we might still be interested in doing uncertainty
quantification for the probability limit of the two-step GMM estimator
in Equation \eqref{eq:GMM_4}—the pseudo-true parameter. However,
valid uncertainty quantification using the conventional GMM standard
errors hinges on the moments being well-specified \citep{hall2003large,lee2014asymptotic}.

The Bayesian bootstrap procedure from Algorithm \ref{alg:BB} is robust
to misspecification of the two-step GMM estimator. This means that
it yields valid uncertainty quantification in both the finite-sample
Bayesian and asymptotic frequentist senses. I will make the claim
of valid asymptotic frequentist uncertainty quantification precise
in Section \ref{subsec:Theory_Asymptotics_MRGMM}. The resulting Bayesian
bootstrap procedure is summarized in Algorithm \ref{alg:MRGMM-BB}.
Effectively, each empirical distribution $\mathbb{P}_{n,X_{ij}}$
in Equations \eqref{eq:GMM_1}-\eqref{eq:GMM_4} is replaced by its
weighted analog. 
\begin{algorithm}[h]
\caption{\protect\label{alg:MRGMM-BB}Bayesian bootstrap procedure for GMM}

\begin{enumerate}
\item Input: Bilateral data $\left\{ X_{k\ell}\right\} _{k\neq\ell}$, moment
equations $\psi:\mathcal{X}\rightarrow\Theta$.
\item For each bootstrap draw $b=1,...,B$:
\begin{enumerate}
\item Sample $\left(V_{1}^{\left(b\right)},...,V_{n}^{\left(b\right)}\right)\overset{\mathrm{iid}}{\sim}\mathrm{Exp}\left(1\right)$.
\item Construct $\omega_{k\ell}^{\left(b\right)}=V_{k}^{\left(b\right)}\cdot V_{\ell}^{\left(b\right)}/\left(\sum_{s\neq t}V_{s}^{\left(b\right)}\cdot V_{t}^{\left(b\right)}\right)$,
for $k,\ell=1,...,n$. 
\item Solve for $\hat{\theta}^{*,\mathrm{2-GMM},\left(b\right)}$ from
\begin{align*}
\psi_{n}^{\left(b\right)}\left(\vartheta\right) & =\sum_{k\neq\ell}\omega_{k\ell}^{\left(b\right)}\cdot\psi\left(X_{k\ell};\vartheta\right)\\
\hat{\theta}^{*,\mathrm{1-GMM},\left(b\right)} & =\underset{\vartheta\in\Theta}{\arg\min}\ \psi_{n}^{\left(b\right)}\left(\vartheta\right)^{'}\psi_{n}^{\left(b\right)}\left(\vartheta\right)\\
\hat{\Omega}^{\left(b\right)}\left(\vartheta\right) & =\left(\sum_{k\neq\ell}\omega_{k\ell}^{\left(b\right)}\cdot\left\{ \psi\left(X_{k\ell};\vartheta\right)-\psi_{n}^{\left(b\right)}\left(\vartheta\right)\right\} \left\{ \psi\left(X_{k\ell};\vartheta\right)-\psi_{n}^{\left(b\right)}\left(\vartheta\right)\right\} ^{'}\right)^{-1}\\
\hat{\theta}^{*,\mathrm{2-GMM},\left(b\right)} & =\underset{\vartheta\in\Theta}{\arg\min}\ \psi_{n}^{\left(b\right)}\left(\vartheta\right)^{'}\hat{\Omega}^{\left(b\right)}\left(\hat{\theta}^{*,\mathrm{1-GMM},\left(b\right)}\right)\psi_{n}^{\left(b\right)}\left(\vartheta\right).
\end{align*}
\end{enumerate}
\item Report the quantiles of interest of $\left\{ \hat{\theta}^{*,\mathrm{2-GMM},\left(1\right)},...,\hat{\theta}^{*,\mathrm{2-GMM},\left(B\right)}\right\} $.
\end{enumerate}
\end{algorithm}

\begin{example*}[\citealp{waugh2010international}]
 The estimator in Equation \eqref{eq:theta_tilde_Waugh} has corresponding
moment function
\begin{equation}
\psi_{\mathrm{Waugh}}\left(X_{ij};\vartheta\right)=\left(\log\frac{\lambda_{ij}}{\lambda_{ii}}-\left(-\vartheta\right)\log\left(\tau_{ij}\frac{p_{i}}{p_{j}}\right)\right)\log\left(\tau_{ij}\frac{p_{i}}{p_{j}}\right).\label{eq:moment_W}
\end{equation}
In \citet{waugh2010international}, whenever $\lambda_{k\ell}=0$
for countries $k$ and $\ell$, the corresponding observation $X_{k\ell}$
is omitted. This results in removing $433$ out of the possible $43\cdot42=1806$
bilateral observations. To avoid removing these observations, one
could adapt the simple OLS estimator to a PPML estimator as in Equation
\eqref{eq:ppml}, with corresponding sample moment condition
\begin{equation}
\psi_{\mathrm{Waugh,PPML}}\left(X_{ij};\vartheta\right)=\left(\frac{\lambda_{ij}}{\lambda_{ii}}-\exp\left\{ -\vartheta\log\left(\tau_{ij}\frac{p_{i}}{p_{j}}\right)\right\} \right)\log\left(\tau_{ij}\frac{p_{i}}{p_{j}}\right).\label{eq:moment_W_ppml}
\end{equation}
In Appendix \ref{subsec:W_PPML} I compute the point estimates and
posterior distributions while not omitting zeros and using PPML. The
point estimates drop considerably and there is more uncertainty. $\triangle$
\end{example*}

\subsection{\protect\label{subsec:UQ_gamma}Bayesian Uncertainty Quantification
for $\hat{\gamma}$}

Taking a Bayesian perspective on uncertainty quantification, in Section
\ref{sec:Theory_FS_Bayesian} I show that for a specific choice of
model and prior, the posterior for $\theta$ given the bilateral data
$\left\{ X_{k\ell}\right\} _{k\neq\ell}$ converges to the Bayesian
bootstrap distribution as a certain informativeness parameter is taken
to zero. 

Since we are also interested in uncertainty quantification for the
counterfactual prediction, we aim to find the corresponding limiting
posterior for $\gamma$ given the realized bilateral data $\left\{ X_{k\ell}\right\} _{k\neq\ell}$.
Towards this end, note that, conditional on the realized data $\left\{ X_{k\ell}\right\} _{k\neq\ell}$,
the only randomness is coming from the posterior for the structural
parameter. So having obtained draws $\left\{ \hat{\theta}^{*,\left(1\right)},...,\hat{\theta}^{*,\left(B\right)}\right\} $
from the limiting posterior distribution for $\theta$ given the bilateral
data $\left\{ X_{k\ell}\right\} _{k\neq\ell}$ using the Bayesian
bootstrap procedure, we can use Assumption \ref{assu:gamma} to obtain
draws from the limiting posterior distribution for $\gamma$ given
the bilateral data $\left\{ X_{k\ell}\right\} _{k\neq\ell}$,
\begin{equation}
\hat{\gamma}^{*,\left(b\right)}=g\left(\left\{ X_{k\ell}\right\} _{k\neq\ell},\hat{\theta}^{*,\left(b\right)}\right),\quad b=1,...,B.\label{eq:BUQ_gamma}
\end{equation}
To construct Bayesian credible intervals, we can then report the relevant
quantiles of the draws $\left\{ \hat{\gamma}^{*,\left(1\right)},...,\hat{\gamma}^{*,\left(B\right)}\right\} $.
\begin{example*}[\citealp{waugh2010international}]
In Table \ref{tab:W_cfs_bay}, I reproduce Table 4 of \citet{waugh2010international},
but include 95\% Bayesian credible intervals. The resulting intervals
are small, implying there is not much economically meaningful uncertainty
in the counterfactuals. $\triangle$ 
\begin{table}[h]
\begin{centering}
\begin{tabular}{|c|c|c|}
\hline 
 & Baseline & Autarky\tabularnewline
\hline 
 & $\tau_{ij}^{\mathrm{cf}}=\tau_{ij}$ & $\tau_{ij}^{\mathrm{cf}}=\infty\cdot\mathbb{I}\left\{ i\neq j\right\} $\tabularnewline
\hline 
\hline 
Variance of log wages & 1.30 {[}1.28, 1.32{]} & 1.35 {[}1.31, 1.38{]}\tabularnewline
\hline 
90th/10th percentile of wages & 25.7 {[}25.1, 26.2{]} & 23.5 {[}22.6, 24.2{]}\tabularnewline
\hline 
Mean \% change in wages & - & -10.5 {[}-11.4, -9.6{]}\tabularnewline
\hline 
\end{tabular}
\par\end{centering}
\begin{centering}
\begin{tabular}{|c|c|c|}
\hline 
 & Symmetry & Free trade\tabularnewline
\hline 
 & $\tau_{ij}^{\mathrm{cf}}=\min\left\{ \tau_{ij},\tau_{ji}\right\} $ & $\tau_{ij}^{\mathrm{cf}}=1$\tabularnewline
\hline 
\hline 
Variance of log wages & 1.05 {[}1.05, 1.05{]} & 0.76 {[}0.75, 0.78{]}\tabularnewline
\hline 
90th/10th percentile of wages & 17.3 {[}17.2, 17.4{]} & 11.4 {[}11.0, 11.9{]}\tabularnewline
\hline 
Mean \% change in wages & 24.2 {[}22.4, 25.8{]} & 128.0 {[}114.4, 140.7{]}\tabularnewline
\hline 
\end{tabular}
\par\end{centering}
\caption{\protect\label{tab:W_cfs_bay}Bayesian uncertainty quantification
for counterfactual predictions as in \citet{waugh2010international}.}
\end{table}
\end{example*}

\section{\protect\label{sec:Theory_FS_Bayesian}Theory— Finite-Sample Bayesian
Results}

In this section I formally introduce and motivate the model and prior.
I then present the key result of the paper: the bootstrap procedure
in Algorithm \ref{alg:BB} admits a finite-sample Bayesian interpretation.

\subsection{\protect\label{subsec:Model}Model}

We observe a sample of bilateral data $\left\{ X_{k\ell}\right\} _{k\neq\ell}\in\mathcal{X}^{n\left(n-1\right)}$,
with $\mathcal{X}\subseteq\mathbb{R}^{d_{X}}$. I adopt a Bayesian
approach, which requires specifying both a model and a prior. I assume
the following model:
\begin{assumption}[Model]
\label{assu:Model}The data $\left\{ X_{k\ell}\right\} _{k\neq\ell}$
are generated according to
\begin{align}
C_{1},...,C_{n}|h,\mathbb{P}_{C} & \overset{\mathrm{iid}}{\sim}\mathbb{P}_{C}\label{eq:DGP_C}\\
X_{ij} & =h\left(C_{i},C_{j}\right),\quad\mathrm{for\ }C_{i}\neq C_{j},\label{eq:DGP_X}
\end{align}
where the latent variables $\left\{ C_{k}\right\} $ are continuous
and take values in $\mathcal{C}\subseteq\mathbb{R}^{d_{C}}$ for finite
$d_{C}$ and $h:\mathcal{C}^{2}\rightarrow\mathcal{X}$ is some measurable
function.
\end{assumption}
The observation $X_{ij}$ may also depend on general equilibrium effects,
captured by a common random variable $U$. This results in the more
general model:
\begin{align*}
U|\tilde{h},\mathbb{P}_{C} & \sim U\left[0,1\right]\\
C_{1},...,C_{n}|\tilde{h},\mathbb{P}_{C},U & \overset{\mathrm{iid}}{\sim}\mathbb{P}_{C}\\
X_{ij} & =\tilde{h}\left(U,C_{i},C_{j}\right),\quad\mathrm{for\ }C_{i}\neq C_{j}.
\end{align*}
The variable $U$ can be thought of as capturing general equilibrium
effects or system-wide interdependencies. Following \citet{graham2020dyadic},
I condition on $U$ and suppress it going forward. Specifically, I
define
\[
h\left(C_{i},C_{j}\right)\equiv\tilde{h}\left(U,C_{i},C_{j}\right),
\]
so that the model reduces to the one in Assumption \ref{assu:Model}.

\subsubsection{\protect\label{subsec:AH}Theoretical Motivation for Model}

To motivate Assumption \ref{assu:Model}, first note that it implies
joint exchangeability of the data $\left\{ X_{k\ell}\right\} _{k\neq\ell}$,
as discussed in Section \ref{subsec:theta}. Conversely, starting
from joint exchangeability and viewing the realized data as being
sampled from a superpopulation, we can use the Aldous-Hoover representation
\citep{aldous1981representations,hoover1979relations} to motivate
Assumption \ref{assu:Model}.

Specifically, suppose the data $\left\{ X_{k\ell}\right\} _{k\neq\ell}$
are sampled from the infinite random array $\left\{ X_{ij}\right\} _{i,j\in\mathbb{N},i\neq j}.$
That is, we sample $\left\{ 1,...,n\right\} $ from the natural numbers
$\mathbb{N}$ and only keep the corresponding rows and columns. 
\begin{example*}[\citealp{waugh2010international}]
The corresponding thought experiment in the application of \citet{waugh2010international}
is that we observe a draw of $43$ countries from an infinite superpopulation
of countries. Since there are only $195$ countries in the world,
the existence of an infinite superpopulation might feel unnatural.
This is however the underlying thought experiment that is necessary
for existing asymptotic justifications \citep{graham2020dyadic,graham2020network,davezies2021empirical,menzel2021bootstrap}.
$\triangle$
\end{example*}
Since the superpopulation is an infinite random array, we have the
following result from \citet{aldous1981representations}:
\begin{lem}[Theorem 1.4 in \citealp{aldous1981representations}]
\label{lem:AH}If for every permutation $\sigma:\mathbb{N}\rightarrow\mathbb{N}$
we have 
\[
\left\{ X_{ij}\right\} _{i,j\in\mathbb{N},i\neq j}\overset{d}{=}\left\{ X_{\sigma\left(i\right)\sigma\left(j\right)}\right\} _{i,j\in\mathbb{N},i\neq j},
\]
then there exists another array $\left\{ X_{ij}^{*}\right\} _{i,j\in\mathbb{N},i\neq j}$
generated according to
\begin{equation}
X_{ij}^{*}=\tilde{h}^{AH}\left(U,C_{i},C_{j},D_{ij}\right),\label{eq:h_AH_tilde}
\end{equation}
for $U,\left\{ C_{i}\right\} ,\left\{ D_{ij}\right\} \overset{\mathrm{iid}}{\sim}U\left[0,1\right],$
such that 
\[
\left\{ X_{ij}\right\} _{i,j\in\mathbb{N},i\neq j}\overset{d}{=}\left\{ X_{ij}^{*}\right\} _{i,j\in\mathbb{N},i\neq j}.
\]
\end{lem}
Here, $U$ is a common ``mixture variable'' that is unidentifiable
\citep{bickel2009nonparametric,graham2020dyadic}. Conditioning on
this random variable yields $h^{AH}\left(C_{i},C_{j},D_{ij}\right)\equiv\tilde{h}^{AH}\left(U,C_{i},C_{j},D_{ij}\right)$.
The only difference between the models $h\left(C_{i},C_{j}\right)$
and $h^{AH}\left(C_{i},C_{j},D_{ij}\right)$ is then the idiosyncratic
component $D_{ij}$. Although the Aldous-Hoover representation is
more general and can hence generate more distributions for the bilateral
data, given observed data $\left\{ X_{k\ell}\right\} _{k\neq\ell}$
with finite sample size and arbitrarily flexible $h$, the models
are observationally equivalent. That is, we cannot reject $h\left(C_{i},C_{j}\right)$
relative to $h^{AH}\left(C_{i},C_{j},D_{ij}\right)$ observing only
$\left\{ X_{k\ell}\right\} _{k\neq\ell}$.
\begin{example*}[\citealp{waugh2010international}]
Applying Lemma \ref{lem:AH} to the setting in \citet{waugh2010international}
yields the representation
\[
X_{k\ell}=\tilde{h}^{AH}\left(U,C_{k},C_{\ell},D_{k\ell}\right),
\]
for each $k\neq\ell$. Here, $U$ is a latent variable common to all
countries, which can be interpreted as capturing economy-wide spillovers.
$C_{k}$ is a latent variable specific to country $k$, and $D_{k\ell}$
is a latent variable specific to the country pair $\left(k,\ell\right)$.
Conditional on $U$, and given the observed data $\left\{ X_{k\ell}\right\} _{k\neq\ell}$,
we cannot distinguish a model that includes pair-specific latent variables
from one that flexibly combines the country-specific latent variables.
$\triangle$
\end{example*}

\subsection{\protect\label{subsec:DPP_BI}Dirichlet Process Prior and Bayesian
Interpretation}

Having specified the model in Assumption \ref{assu:Model}, I assume
the following prior on $h$ and $\mathbb{P}_{C}$:
\begin{assumption}[Prior]
\label{assu:prior}We have that $h$ and $\mathbb{P}_{C}$ are independently
drawn according to
\begin{equation}
\left(h,\mathbb{P}_{C}\right)\sim\pi\left(h\right)\cdot\pi\left(\mathbb{P}_{C}\right)=\pi\left(h\right)\cdot DP\left(Q,\alpha\right).\label{eq:prior}
\end{equation}
\end{assumption}
Note that $\pi\left(h\right)$ is a distribution over functions, while
$\pi\left(\mathbb{P}_{C}\right)$ is a distribution over distributions.
Here, $DP\left(Q,\alpha\right)$ denotes a Dirichlet process, where
$Q$ is a probability measure on $\mathcal{C}$, referred to as the
\textit{center measure,} and $\alpha>0$ is a scalar known as the
\textit{prior precision}. The Dirichlet process prior implies that
for any partition $\left\{ A_{1},...,A_{R}\right\} $ of $\mathcal{C}$,
we have
\begin{align*}
\left(\mathbb{P}_{C}\left(A_{1}\right),....,\mathbb{P}_{C}\left(A_{R_{}}\right)\right) & \sim\mathrm{Dir}\left(R;\alpha\cdot Q\left(A_{1}\right),...,\alpha\cdot Q\left(A_{R_{}}\right)\right).
\end{align*}
Given the model in Assumption \ref{assu:Model} and the prior in Assumption
\ref{assu:prior}, we are interested in finding the posterior of $\theta$
given the observed data $\left\{ X_{k\ell}\right\} _{k\neq\ell}$.
The key result of this paper is that we can interpret the draws of
the Bayesian bootstrap procedure in Algorithm \ref{alg:BB} as draws
from this posterior in the uninformative limit where $\alpha\downarrow0$: 
\begin{thm}[Finite-sample Bayesian interpretation]
\label{prop:Bayesian_interpretation} Under Assumptions \ref{assu:Model}
and \ref{assu:prior}, in the uninformative limit $\alpha\downarrow0$,
the posterior on $\theta$ given the realized data $\left\{ X_{k\ell}\right\} _{k\neq\ell}$
converges in distribution to the distribution of the draws produced
by the Bayesian bootstrap procedure in Algorithm \ref{alg:BB}.
\end{thm}
All proofs can be found in Appendix \ref{sec:Proofs}. The proof of
Theorem \ref{prop:Bayesian_interpretation} proceeds in five steps.
First, I find the posterior of $\mathbb{P}_{C}$ given the function
$h$ and draws $\left\{ C_{k}\right\} $ for a given center measure
$Q$ and precision parameter $\alpha$, and denote it by $\pi_{\alpha}\left(\mathbb{P}_{C}|h,\left\{ C_{k}\right\} \right)$.
This step combines the model in Equation \eqref{eq:DGP_C} and the
prior in Equation \eqref{eq:prior} and uses the conjugacy of the
Dirichlet process. Second, I find the posterior which corresponds
to the case where $\alpha\downarrow0$, and argue that it is proper.
Importantly, it does not depend on the center measure $Q$. Denoting
with $\pi_{0}$ the probability under the limiting posterior as $\alpha\downarrow0$,
we have
\begin{equation}
\pi_{0}\left(\mathbb{P}_{C}|h,\left\{ C_{k}\right\} \right)=DP\left(\sum_{k=1}^{n}\frac{1}{n}\cdot\delta_{C_{k}},n\right).\label{eq:pi_0_P_C}
\end{equation}
 Third, I use the model and properties of Dirichlet processes to find
an expression for $\pi_{0}\left(\mathbb{P}_{X_{ij}}|h,\left\{ C_{k}\right\} \right)$.

The first three steps all consider the thought experiment where we
observe the latent variables $\left\{ C_{k}\right\} $ and know the
function $h$. However, in practice we do not observe the latent variables
$\left\{ C_{k}\right\} $ and do not know the function $h$; we only
observe $\left\{ X_{k\ell}\right\} _{k\neq\ell}$. In the fourth step
I therefore find an expression for $\pi_{0}\left(\mathbb{P}_{X_{ij}}|\left\{ X_{k\ell}\right\} _{k\neq\ell}\right)$,
which I show corresponds to
\begin{align}
 & \mathbb{P}_{n,X_{ij}}^{*}\sim\pi_{0}\left(\mathbb{P}_{X_{ij}}|\left\{ X_{k\ell}\right\} _{k\neq\ell}\right)\nonumber \\
 & \Rightarrow\mathbb{P}_{n,X_{ij}}^{*}=\sum_{k\neq\ell}\frac{W_{k}\cdot W_{\ell}}{\sum_{s\neq t}W_{s}\cdot W_{t}}\cdot\delta_{X_{k\ell}},\quad\left(W_{1},...,W_{n}\right)\sim\mathrm{Dir}\left(n;1,...,1\right),\label{eq:posterior_P_X}
\end{align}
which is exactly the distribution we saw in Algorithm \ref{alg:BB}.
Lastly, since $\theta=T\left(\mathbb{P}_{X_{ij}}\right)$, this also
implies a limiting posterior on the structural parameter, $\pi_{0}\left(\theta|\left\{ X_{k\ell}\right\} _{k\neq\ell}\right)$,
and the conclusion follows. 

Concerning the Bayesian interpretation of the counterfactual prediction,
note that—conditional on the realized data $\left\{ X_{k\ell}\right\} _{k\neq\ell}$—the
only remaining source of randomness arises from the posterior distribution
for the structural parameter. Then, combining Theorem \ref{prop:Bayesian_interpretation}
and Assumption \ref{assu:gamma}, it follows that $\pi_{0}\left(\gamma|\left\{ X_{k\ell}\right\} _{k\neq\ell}\right)$
converges to the Bayesian bootstrap distribution characterized by
Algorithm \ref{alg:BB} and Equation \eqref{eq:BUQ_gamma}. 

\subsubsection{Theoretical Motivation for Dirichlet Process Prior}

Theorem \ref{prop:Bayesian_interpretation} shows that the choice
of the Dirichlet process prior implies a finite-sample Bayesian interpretation
for Algorithm \ref{alg:BB}. Moreover, by considering the limit as
the prior precision tends to zero, the procedure becomes agnostic
to the choice of center measure $Q$ and prior on $h$. I further
motivate this class of priors by showing it is uninformative in a
specific sense:
\begin{defn}[Smoothing across events]
\textit{Say the posterior $\pi\left(\mathbb{P}_{X_{ij}}|\left\{ X_{k\ell}\right\} _{k\neq\ell}\right)$
does not }smooth across events\textit{ if for every measurable partition
$\left\{ B_{1},...,B_{R}\right\} $ of the support $\mathcal{X}$
and 
\[
\mathbb{P}_{n,X_{ij}}^{*}\sim\pi\left(\mathbb{P}_{X_{ij}}|\left\{ X_{k\ell}\right\} _{k\neq\ell}\right),
\]
the distribution of 
\[
\left(\mathbb{P}_{n,X_{ij}}^{*}\left(B_{1}\right),...,\mathbb{P}_{n,X_{ij}}^{*}\left(B_{R}\right)\right),
\]
only depends on the indicators $1_{k\ell}^{r}=\mathbb{I}\left\{ X_{k\ell}\in B_{r}\right\} $
for $r=1,...,R$.}
\end{defn}
If a posterior does not smooth across events, to calculate the posterior
probability for a given event $B$, we can replace the data $\left\{ X_{k\ell}\right\} _{k\neq\ell}$
with its binarized version $\left\{ 1_{k\ell}\right\} _{k\neq\ell}$
for $1_{k\ell}=\mathbb{I}\left\{ X_{k\ell}\in B\right\} $. 
\begin{example*}[\citealp{waugh2010international}]
In \citet{waugh2010international}, if the posterior does not smooth
across events, to compute the posterior probability that a bilateral
observation drawn from $\mathbb{P}_{X_{ij}}$ lies in a certain subset
$B\subset\mathcal{X}$, we can binarize the observations $\left\{ X_{k\ell}\right\} _{k\neq\ell}$
into those that lie within $B$ and those that do not. For example,
if we would want to predict the probability that a new observation
will have an own-country trade share $\lambda_{kk}$ less than $0.5$,
we can binarize the observations according to
\begin{align*}
1_{k\ell} & =\mathbb{I}\left\{ X_{k\ell}\in B\right\} =\mathbb{I}\left\{ \lambda_{kk}<0.5\right\} \\
 & =\mathbb{I}\left\{ k\in\left\{ \mathrm{Belgium,\ Benin,\ Ireland,\ Mali,\ Sierra\ Leone}\right\} \right\} .
\end{align*}
In particular, for computation of this posterior probability, all
countries with own-country trade shares above $0.5$ are treated identically.
For example, there is no distinction between Denmark ($\lambda_{kk}=0.523$)
and the United States ($\lambda_{kk}=0.897$). $\triangle$
\end{example*}
We have the following theorem:
\begin{thm}[Smoothing across events and Dirichlet process priors]
\label{prop:Theoretical_motivation}Under Assumption \ref{assu:Model}
and the generic priors
\[
\left(h,\mathbb{P}_{C}\right)\sim\pi\left(h\right)\cdot\pi\left(\mathbb{P}_{C}\right),
\]
we have:
\begin{enumerate}
\item \textup{If} $\pi\left(\mathbb{P}_{C}\right)$ is a Dirichlet process
prior and the prior precision $\alpha$ is taken to zero, then the
resulting posterior $\pi_{0}\left(\mathbb{P}_{X_{ij}}|\left\{ X_{k\ell}\right\} _{k\neq\ell}\right)$
does not smooth across events for all $\pi\left(h\right)$.
\item There exists a prior $\pi\left(h\right)$ such that the corresponding
posterior $\pi\left(\mathbb{P}_{X_{ij}}|\left\{ X_{k\ell}\right\} _{k\neq\ell}\right)$
does not smooth across events \textup{if and only if} $\pi\left(\mathbb{P}_{C}\right)$
is a Dirichlet process prior or a trivial process.\footnote{The three trivial processes, as discussed in Section 4.4 of \citet{ghosal2017fundamentals},
are: (1) $\pi\left(\mathbb{P}_{C}\right)=\rho$ a.s., for a deterministic
probability measure $\rho$, (2) $\pi\left(\mathbb{P}_{C}\right)=\delta_{Y}$,
for a random variable $Y\sim\rho$, (3) $\pi\left(\mathbb{P}_{C}\right)=Z\delta_{a}+\left(1-Z\right)\delta_{b}$,
for deterministic $a,b\in\mathcal{C}$ an arbitrary random variable
$Z$ with values in $\left[0,1\right]$.} 
\end{enumerate}
\end{thm}
So if we want a prior for $\mathbb{P}_{C}$ that ensures the posterior
probability assigned to a set depends only on the data observed within
that set, then this mechanically leads us to use a Dirichlet process
prior. Such a prior reflects a situation where we have no prior reason
to smooth across regions of $\mathcal{X}$: posterior beliefs about
a region of the sample space are updated solely based on whether observed
data fall inside that region. 

\subsubsection{Limiting Marginal Prior for $\theta$}

I am taking a Bayesian approach by specifying a prior in Assumption
\ref{assu:prior}. One might wonder how informative Dirichlet process
priors are. Specifically, it is of interest to plot the implied limiting
marginal prior $\pi\left(\theta\right)$ and compare it to the limiting
posterior $\pi_{0}\left(\theta|\left\{ X_{k\ell}\right\} _{k\neq\ell}\right)$.
By comparing these two distributions, we can see how much information
is drawn from the prior.

However, Theorem \ref{prop:Bayesian_interpretation} shows that the
relevant posterior corresponds to an uninformative limit of posteriors
for any choice of $Q$, which implies that there is not a unique well-defined
implied limiting marginal prior for $\theta$. In this subsection,
I consider a specific choice for the center measure $Q$ and a class
of estimators for which we can plot the limiting distribution of $\pi\left(\theta\right)$.

Concretely, I constrain the Dirichlet process prior in Equation \eqref{eq:prior}
to
\begin{equation}
\mathbb{P}_{C}\sim DP\left(\sum_{k=1}^{n}\frac{1}{n}\cdot\delta_{C_{k}},\alpha\right).\label{eq:DP_C_marg}
\end{equation}
This specific choice of the center measure $Q$ implies that mass
is supported only on the latent variables $\left\{ C_{k}\right\} $.\footnote{One can generalize this to a center measure of $\sum_{k=1}^{n}\omega_{k}\delta_{C_{k}}$
for weights $\left\{ \omega_{k}\right\} $ that sum up to 1. Then
the assumption that mass is supported only on $\left\{ C_{k}\right\} $
becomes less restrictive as the number of units $n$ grows large.
In particular, \citet{andrews2024bootstrap} shows that if $\mathcal{C}$
is a Polish space and $\mathbb{P}_{C}$ has full support, then for
every $\mathbb{P}\in\Delta\left(\mathcal{C}\right)$ and almost every
sequence of draws $\left\{ C_{1},C_{2},...\right\} $ from $\mathbb{P}_{C}$
there exists a sequence of weights $\left\{ \omega_{k}^{n}\right\} $
such that $\sum_{k=1}^{n}\omega_{k}^{n}\delta_{C_{k}}$ converges
weakly to $\mathbb{P}$ as $n\rightarrow\infty$.} This is also the case for the posterior in Equation \eqref{eq:pi_0_P_C},
which can be recovered by setting $\alpha=n$. We then have the following
result:
\begin{thm}[Limiting marginal prior]
\label{cor:Characterization}Under Assumptions \ref{assu:Model}
and using the Dirichlet process prior as in Equation \eqref{eq:DP_C_marg},
if $\hat{\theta}$ is of the form 
\[
\hat{\theta}=T\left(\mathbb{P}_{n,X_{ij}}\right)=\chi\left(\mathbb{E}_{\mathbb{P}_{n,X_{ij}}}\left[\varrho\left(X_{ij}\right)\right]\right),
\]
for known functions $\varrho:\mathcal{X}\rightarrow\mathcal{R}$ and
$\chi:\mathcal{R}\rightarrow\Theta$, and $\chi\left(\cdot\right)$
is continuous at $\varrho\left(X_{k\ell}\right)$ for all $k\neq\ell$,
then, as $\alpha\downarrow0$, the implied marginal prior $\pi\left(\theta\right)$
converges weakly to
\[
\pi^{\infty}\left(\theta\right)=\frac{2}{n\left(n-1\right)}\sum_{k>\ell}\delta_{\frac{\chi\left(\varrho\left(X_{k\ell}\right)\right)+\chi\left(\varrho\left(X_{\ell k}\right)\right)}{2}}.
\]
\end{thm}
Theorem \ref{cor:Characterization} shows how to characterize the
limiting object for the class of estimators that can be written as
functions of means. For example for the case of simple OLS without
an intercept as in the running example and the application in Section
\ref{subsec:Applications_CP}, continuity of $\chi$ is satisfied.
For estimators that can not be written in this way, Appendix \ref{sec:Extra_results_Marginal_prior}
presents an algorithm for plotting proper priors along the limit sequence. 
\begin{example*}[\citealp{waugh2010international}]
In Figure \ref{fig:W_marg}, I plot the bootstrap posterior and the
limiting marginal prior using Theorem \ref{cor:Characterization},
where we have
\[
\varrho\left(X_{ij}\right)=\left(\begin{array}{c}
\log\left(\tau_{ij}\frac{p_{j}}{p_{i}}\right)^{2}\\
-\log\left(\tau_{ij}\frac{p_{j}}{p_{i}}\right)\cdot\log\frac{\lambda_{ij}}{\lambda_{ii}}
\end{array}\right),\ \chi\left(\left(\begin{array}{c}
a_{1}\\
a_{2}
\end{array}\right)\right)=\frac{a_{2}}{a_{1}},
\]
and continuity of $\chi$ is satisfied. We observe that the limiting
marginal prior is much flatter than the bootstrap posterior. Its diffuse
shape reflects weak prior information, allowing for a wide range of
plausible values for the productivity parameter. $\triangle$ 
\begin{figure}[h]
\centering{}\includegraphics[scale=0.16]{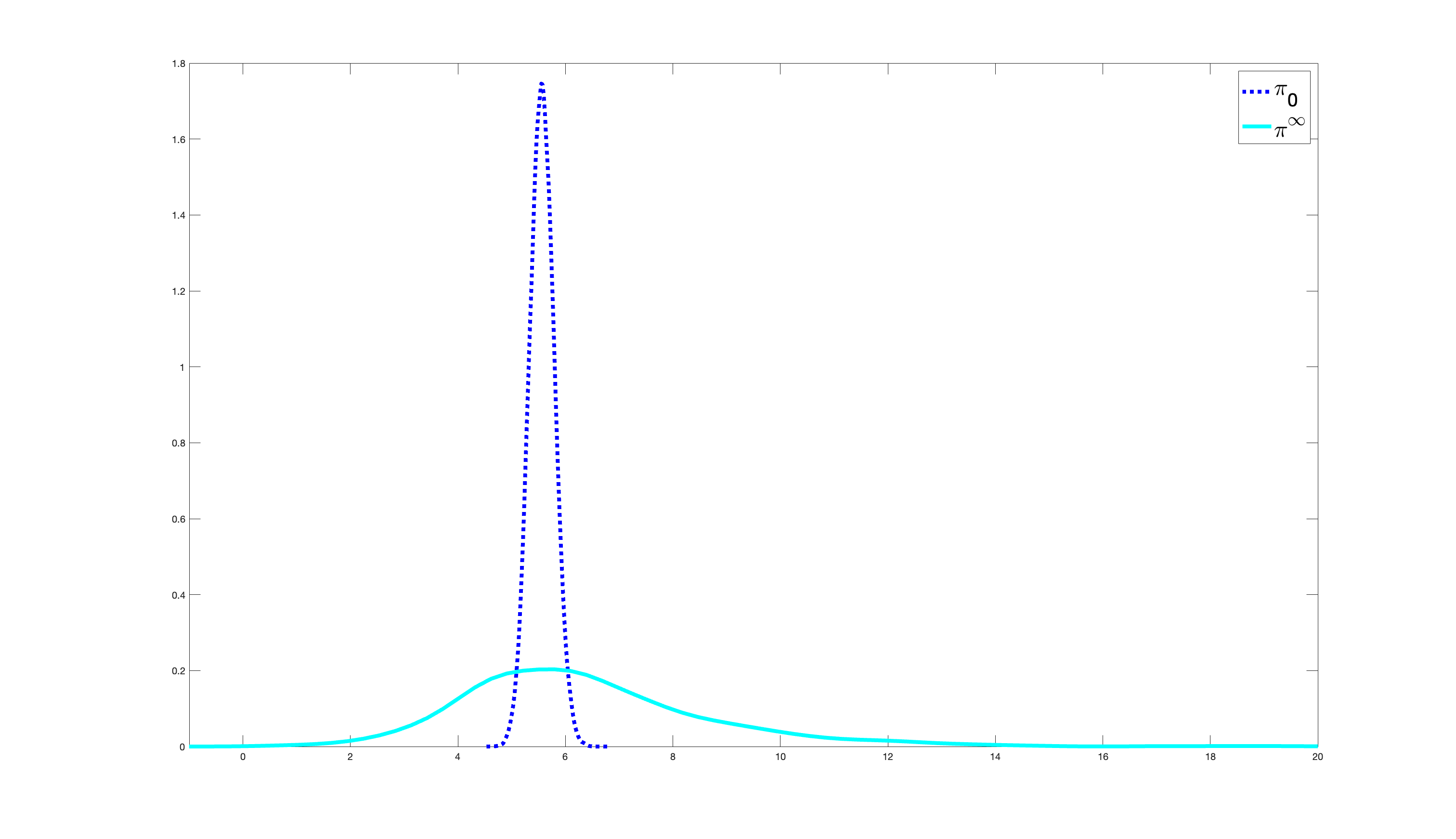}\caption{\protect\label{fig:W_marg}Limiting marginal prior for productivity
parameter using the full sample as in \citet{waugh2010international}.}
\end{figure}
\end{example*}

\section{\protect\label{sec:Theory_Asympotics}Theory— Asymptotic Results}

In this section I provide conditions on $\hat{\theta}$ that guarantee
asymptotic validity of the proposed Bayesian bootstrap procedure.
I again consider misspecification-robust uncertainty quantification
for over-identified GMM as a special case.

\subsection{Frequentist Uncertainty Quantification for $\hat{\theta}$ }

\subsubsection{Sampling Thought Experiment}

In Section \ref{subsec:AH} I introduced the thought experiment that
the data $\left\{ X_{k\ell}\right\} _{k\neq\ell}$ are sampled from
a superpopulation. In this section I will state this as an assumption:
\begin{assumption}[Sampling thought experiment]
\label{assu:Sampling_thought_experiment}The infinite random array
$\left\{ X_{ij}\right\} _{i,j\in\mathbb{N},i\neq j}$ is jointly exchangeable,
so that for every permutation $\sigma:\mathbb{N}\rightarrow\mathbb{N}$
we have 
\[
\left\{ X_{ij}\right\} _{i,j\in\mathbb{N},i\neq j}\overset{d}{=}\left\{ X_{\sigma\left(i\right)\sigma\left(j\right)}\right\} _{i,j\in\mathbb{N},i\neq j}.
\]
The data $\left\{ X_{k\ell}\right\} _{k\neq\ell}$ are generated by
sampling $\left\{ 1,...,n\right\} $ from the natural numbers $\mathbb{N}$
and only keeping the corresponding rows and columns.
\end{assumption}
Assumption \ref{assu:Sampling_thought_experiment} implies that as
we sample more observations from this superpopulation, the resulting
data $\left\{ X_{k\ell}\right\} _{k\neq\ell}$ always will be jointly
exchangeable, and that all observations will have the same marginal
distribution, denoted by $\mathbb{P}_{X_{ij}}$.

\subsubsection{Asymptotic Bootstrap Validity}

The goal of this section is to prove asymptotic validity of the bootstrap
procedure in Algorithm \ref{alg:BB} for a given estimator
\begin{equation}
\hat{\theta}=T\left(\mathbb{P}_{n,X_{ij}}\right)=T\left(\sum_{k\neq\ell}\frac{1}{n\left(n-1\right)}\cdot\delta_{X_{k\ell}}\right).\label{eq:theta_tilde_asymp}
\end{equation}
Going forward, let $\mathbb{P}_{n,X_{ij}}^{*}$ be a given drawn distribution
from $\pi_{0}\left(\mathbb{P}_{X_{ij}}|\left\{ X_{k\ell}\right\} _{k\neq\ell}\right)$,
so that
\[
\mathbb{P}_{n,X_{ij}}^{*}=\sum_{k\neq\ell}\frac{W_{k}\cdot W_{\ell}}{\sum_{s\neq t}W_{s}\cdot W_{t}}\cdot\delta_{X_{k\ell}},\quad\left(W_{1},...,W_{n}\right)\sim\mathrm{Dir}\left(n;1,...,1\right).
\]

\begin{defn}[Asymptotic bootstrap validity]
\textit{The bootstrap procedure is }asymptotically valid\textit{
for the estimator $\hat{\theta}$ as defined in Equation \eqref{eq:theta_tilde_asymp}
if, conditional on the data $\left\{ X_{k\ell}\right\} _{k\neq\ell}$
and almost surely, $\sqrt{n}\left(T\left(\mathbb{P}_{n,X_{ij}}^{*}\right)-T\left(\mathbb{P}_{n,X_{ij}}\right)\right)$
and $\sqrt{n}\left(T\left(\mathbb{P}_{n,X_{ij}}\right)-T\left(\mathbb{P}_{X_{ij}}\right)\right)$
converge in distribution to the same mean zero normal random variable.}
\end{defn}
The main appeal of bootstrap validity for $\hat{\theta}$ is that
it implies asymptotic validity of confidence intervals based on the
bootstrap, because if $n$ grows large, we can approximate the normal
distribution to which $\sqrt{n}\left(\hat{\theta}-\theta\right)$
converges in distribution sufficiently well. 

To show asymptotic validity of the bootstrap for a structural estimator,
I will take a two-step approach. First I show convergence of the empirical
process, and then use the functional delta method to argue validity
of the bootstrap for certain classes of estimators. 

The relevant empirical processes, defined on a class of real-valued
functions $\mathcal{F}$, are
\begin{align*}
\mathbb{G}_{n}f & =\sqrt{n}\left\{ \mathbb{P}_{n,X_{ij}}f-\mathbb{P}_{X_{ij}}f\right\} \\
\mathbb{G}_{n}^{*}f & =\sqrt{n}\left\{ \mathbb{P}_{n,X_{ij}}^{*}f-\mathbb{P}_{n,X_{ij}}f\right\} ,
\end{align*}
for $f\in\mathcal{F}$. Here, $\mathbb{P}_{X_{ij}}f$ denotes $\mathbb{E}_{\mathbb{P}_{X_{ij}}}\left[f\left(X_{ij}\right)\right]$,
and $\mathbb{P}_{n,X_{ij}}f$ and $\mathbb{P}_{n,X_{ij}}^{*}f$ are
defined analogously. We want to show weak convergence over $\ell^{\infty}\left(\mathcal{F}\right)$
of both $\mathbb{G}_{n}$ and $\mathbb{G}_{n}^{*}$ to the same centered
Gaussian process $\mathbb{G}$, where the convergence of $\mathbb{G}_{n}^{*}$
holds conditional on the data $\left\{ X_{k\ell}\right\} _{k\neq\ell}$
and outer almost surely, for $\ell^{\infty}\left(\mathcal{F}\right)$
the set of bounded functions on $\mathcal{F}$. A formal definition
of weak convergence is given in Definition 1.3.3 in \citet{van1996weak}.
To ensure this convergence, we require some regularity conditions
on the function class $\mathcal{F}$.
\begin{assumption}[Regularity conditions on $\mathcal{F}$]
\label{assu:Complexity}Let $\mathcal{F}\subseteq\mathcal{X}^{\mathbb{R}}$
be a measurable class of functions such that:
\begin{enumerate}
\item[(i)] $\mathcal{F}$ is permissible (see page 196 in \citealp{pollard1984convergence})
and admits a positive envelope $F$ with $\mathbb{P}_{X_{ij}}F^{2}<\infty$.
\item[(ii)] We have non-degeneracy, meaning that the covariance kernel is positive
for all elements of $\mathcal{F}$:
\[
K\left(f_{1},f_{2}\right)=\Cov\left(f_{1}\left(X_{12}\right)+f_{1}\left(X_{21}\right),f_{2}\left(X_{12^{'}}\right)+f_{2}\left(X_{2^{'}1}\right)\right)>0\ \forall f_{1},f_{2}\in\mathcal{F}.
\]
\item[(iii)] There exist $0<c,v<\infty$ such that for every $\epsilon>0$ and
probability measure $Q$ with $QF^{2}<\infty$, we have 
\[
N\left(\epsilon\left\Vert F\right\Vert _{L_{2}\left(Q\right)},\mathcal{F},\left\Vert \cdot\right\Vert _{L_{2}\left(Q\right)}\right)\leq c\epsilon^{-v}.
\]
\end{enumerate}
\end{assumption}
Condition (i) captures regularity condition on the function class.
Permissibility is a mild measure-theoretic regularity condition that
ensures function classes meet minimal requirements for measurability
and integration, making them suitable for empirical process analysis.
The existence of an envelope function $F$ for a class $\mathcal{F}$
means that $\left|f\left(x\right)\right|\leq F\left(x\right)$ for
all $f\in\mathcal{F}$ and all $x\in\mathcal{X}$. 

Non-degeneracy in condition (ii) ensures that the limiting processes
of $\mathbb{G}_{n}$ and $\mathbb{G}_{n}^{*}$ are Gaussian with non-zero
variance. Degeneracy may arise, for instance, if the data $\left\{ X_{k\ell}\right\} _{k\neq\ell}$
are in fact i.i.d, in which case the limiting process of $\mathbb{G}_{n}$
is a Gaussian chaos process. In such settings, $\mathbb{G}_{n}^{*}$
will not converge to the correct (non-Gaussian) limit under standard
bootstrap procedures. Alternative bootstrap methods have been developed
to handle degeneracy, including those proposed by \citet{huvskova1993consistency},
\citet{menzel2021bootstrap} and \citet{han2022multiplier}.

Condition (iii) bounds the complexity of $\mathcal{F}$. Here, the
covering number $N\left(\epsilon,\mathcal{F},\left\Vert \cdot\right\Vert _{L_{2}\left(Q\right)}\right)$
is the minimal number of $L_{2}\left(Q\right)$-balls of radius $\varepsilon$
needed to cover $\mathcal{F}$. This condition is for example satisfied
for VC classes of functions by Lemma 4.4 in \citet{alexander1987central}.\footnote{Alternatively, one could assume that $\mathcal{F}$ has polynomial
discrimination, defined on page 17 of \citet{pollard1984convergence}.
By Lemma II.25 in \citet{pollard1984convergence}, this is a sufficient
condition for condition (iii). Also, finite VC-dimension implies polynomial
discrimination due to the Sauer-Shelah lemma, see page 275 in \citet{van2000asymptotic}.} 
\begin{example*}[Smooth functionals of empirical cdf]
Consider the class of estimators that are smooth functionals of the
empirical cdf $F_{n,X_{ij}}$ and suppose for exposition that $X_{ij}$
is a scalar. For some function $\varphi$, we have $\hat{\theta}=\varphi\left(F_{n,X_{ij}}\right)$,
$\theta=\varphi\left(F_{X_{ij}}\right)$ and the relevant function
class is
\[
\mathcal{F}_{\mathrm{cdf}}\equiv\left\{ u\mapsto\mathbb{I}\left\{ u\leq x\right\} :x\in\mathbb{R}\right\} .
\]
As an envelope function we can take the constant function $F_{\mathrm{cdf}}\equiv1$.
The covariance kernel is
\begin{equation}
K_{\mathrm{cdf}}\left(x,y\right)=\Cov\left(\mathbb{I}\left\{ X_{12}\leq x\right\} +\mathbb{I}\left\{ X_{21}\leq x\right\} ,\mathbb{I}\left\{ X_{12^{'}}\leq y\right\} +\mathbb{I}\left\{ X_{2^{'}1}\leq y\right\} \right),\label{eq:K_cdf}
\end{equation}
which we require to be non-zero for all $x,y\in\mathbb{R}$. Lastly,
we know $\mathcal{F}_{\mathrm{cdf}}$ satisfies condition (iii) in
Assumption \ref{assu:Complexity} from Example 19.16 in \citet{van2000asymptotic}.
\end{example*}
We have the following result for the empirical processes:
\begin{thm}[Weak convergence of empirical processes]
\label{thm:empirical_process_conv}If $\mathcal{F}$ satisfies Assumption
\ref{assu:Complexity}, then we have weak convergence over $\ell^{\infty}\left(\mathcal{F}\right)$
of both $\mathbb{G}_{n}$ and $\mathbb{G}_{n}^{*}$ to the same centered
Gaussian process $\mathbb{G}$, where the convergence of $\mathbb{G}_{n}^{*}$
holds conditional on the data $\left\{ X_{k\ell}\right\} _{k\neq\ell}$
and outer almost surely.
\end{thm}
Note that the convergence rate is $\sqrt{n}$ despite having a sample
size of $n\left(n-1\right)$, as is also the case for non-degenerate
U-statistics. The proof of Theorem \ref{thm:empirical_process_conv}
builds on results from \citet{arcones1993limit} and \citet{zhang2001bayesian},
which present a uniform CLT for U-processes and a bootstrap uniform
CLT for U-processes, respectively. 

Once we have established convergence of the empirical process, we
can appeal to the functional delta method for the bootstrap to argue
asymptotic validity of the bootstrap for a given estimator. We require
the estimator to be sufficiently smooth:
\begin{assumption}[Smoothness]
\label{assu:smoothness}Suppose $\hat{\theta}$ is of the form $T\left(\mathbb{P}_{n,X_{ij}}\right)=\varphi\left(\mathbb{P}_{n,X_{ij}}f\right)$
for $f\in\mathcal{F}$, where $\varphi:\ell^{\infty}\left(\mathcal{F}\right)\mapsto\Theta$
with derivative $\varphi^{'}$. The function $\varphi$ is Hadamard
differentiable at $\mathbb{P}_{X_{ij}}f$ tangentially to a subspace
$\ell_{0}^{\infty}\left(\mathcal{F}\right)\subset\ell^{\infty}\left(\mathcal{F}\right)$.
\end{assumption}
The precise definition of Hadamard differentiability is given in Section
20.2 of \citet{van2000asymptotic}. Section 20.3 of \citet{van2000asymptotic}
give examples of Hadamard differentiable functions. 
\begin{example*}[Smooth functionals of empirical cdf]
Consider again the class of estimators that are smooth functionals
of the empirical cdf, so that $\hat{\theta}=\varphi\left(F_{n,X_{ij}}\right)$.
For Assumption \ref{assu:smoothness} to hold we require $\varphi$
to be Hadamard differentiable tangentially to a subspace $\ell_{0}^{\infty}\left(\mathcal{F}_{\mathrm{cdf}}\right)$.
For example, from Lemma 21.3 in \citet{van2000asymptotic} we know
this is the case for the empirical quantiles under mild differentiability
conditions on $F_{X_{ij}}$. 
\end{example*}
Application of the functional delta method for the bootstrap (Theorem
23.9 in \citealp{van2000asymptotic}) then yields the following theorem:
\begin{thm}[Bootstrap validity]
\label{prop:estimator_conv} Under Assumption \ref{assu:Sampling_thought_experiment},
if $\mathcal{F}$ and $\hat{\theta}$ satisfy Assumptions \ref{assu:Complexity}
and \ref{assu:smoothness}, then the bootstrap procedure in Algorithm
\ref{alg:BB} is asymptotically valid for $\hat{\theta}$.
\end{thm}
From the examples throughout this section, we then have the following
corollary:
\begin{cor}[Asymptotic bootstrap validity for smooth functionals of empirical
cdf]
\label{cor:functionals_of_cdf}The bootstrap procedure in Algorithm
\ref{alg:BB} is asymptotically valid for estimators of the form $\hat{\theta}=\varphi\left(F_{n,X_{ij}}\right)$
if $K_{\mathrm{cdf}}\left(x,y\right)$ in Equation \eqref{eq:K_cdf}
is positive for all $x,y\in\mathbb{R}$ and $\varphi$ is Hadamard
differentiable tangentially to a subspace $\ell_{0}^{\infty}\left(\mathcal{F}_{\mathrm{cdf}}\right)$. 
\end{cor}
It will be useful to gather sufficient conditions for Assumptions
\ref{assu:Complexity} and \ref{assu:smoothness} for Z-estimators
in a corollary.
\begin{cor}[Asymptotic bootstrap validity for Z-estimators]
\label{cor:Z-estimators}Suppose $\hat{\theta}$ and $\theta$ solve
\begin{align*}
0 & =\Psi_{n}\left(\vartheta\right)\equiv\underset{\eta\in\mathcal{H}}{\sup}\left|\Psi_{n}\left(\vartheta\right)\left(\eta\right)\right|=\underset{\eta\in\mathcal{H}}{\sup}\left|\mathbb{P}_{n,X_{ij}}\nu_{\vartheta,\eta}\right|\\
0 & =\Psi\left(\vartheta\right)\equiv\underset{\eta\in\mathcal{H}}{\sup}\left|\Psi\left(\vartheta\right)\left(\eta\right)\right|=\underset{\eta\in\mathcal{H}}{\sup}\left|\mathbb{P}_{X_{ij}}\nu_{\vartheta,\eta}\right|,
\end{align*}
and uppose the following conditions hold:
\begin{enumerate}
\item[(i)] $\Psi:\Theta\mapsto\mathbb{R}^{L}$ is uniformly norm-bounded over
$\Theta$, and satisfies $\Psi\left(\theta\right)=0$.
\item[(ii)] $\Psi$ is Fréchet differentiable at $\theta$ with continuously
invertible derivative $\dot{\Psi}_{\theta}$.
\item[(iii)] $\underset{\eta\in\mathcal{H}}{\sup}\left|\Psi\left(\theta_{w}\right)\right|\rightarrow0$
implies $\left\Vert \theta_{w}-\theta_{}\right\Vert \rightarrow0$
for every sequence $\left\{ \theta_{w}\right\} $ in $\Theta$.
\item[(iv)] $\Psi_{n}$ has at least one zero for all $n$ large enough, outer
almost surely (see Section 18.2 in \citet{van2000asymptotic} for
a formal definition).
\item[(v)] The limit of $\vartheta\mapsto\sqrt{n}\left(\Psi_{n}\left(\vartheta\right)-\Psi\left(\vartheta\right)\right)$
is almost surely continuous at $\theta$.
\item[(vi)] The function class $\mathcal{F}_{Z}\equiv\left\{ \nu_{\vartheta,\eta}:\left(\vartheta,\eta\right)\in\Theta\times\mathcal{H}\right\} $
satisfies Assumption \ref{assu:Complexity}.
\end{enumerate}
Then the bootstrap procedure in Algorithm \ref{alg:BB} is asymptotically
valid for $\hat{\theta}$.
\end{cor}

\subsubsection{\protect\label{subsec:Theory_Asymptotics_MRGMM}Special Case: Misspecification-Robust
Uncertainty Quantification for GMM}

Following \citet{imbens1997one}, the two estimation steps of the
two-step GMM estimator from Section \ref{subsec:MRGMM} can be combined
into a single just-identified system,
\begin{align*}
 & \phi\left(X_{ij};\theta_{}^{\mathrm{1-GMM}},\theta_{}^{\mathrm{2-GMM}},m,\mathrm{vec}\left\{ \Omega\right\} ,\mathrm{vec}\left\{ G_{1}\right\} ,\mathrm{vec}\left\{ G_{2}\right\} \right)\\
 & =\left(\begin{array}{c}
\mathrm{vec}\left\{ G_{1}-\frac{\partial}{\partial\theta}\psi\left(X_{ij};\theta_{}^{\mathrm{1-GMM}}\right)\right\} \\
G_{1}^{'}\psi\left(X_{ij};\theta_{}^{\mathrm{1-GMM}}\right)\\
\psi\left(X_{ij};\theta^{\mathrm{1-GMM}}\right)-m\\
\mathrm{vec}\left\{ \Omega-\left[\psi\left(X_{ij};\theta^{\mathrm{1-GMM}}\right)-m\right]\left[\psi\left(X_{ij};\theta^{\mathrm{1-GMM}}\right)-m\right]^{'}\right\} \\
\mathrm{vec}\left\{ G_{2}-\frac{\partial}{\partial\theta}\psi\left(X_{ij};\theta_{}^{\mathrm{2-GMM}}\right)\right\} \\
G_{2}^{'}\Omega_{}\psi\left(X_{ij};\theta_{}^{\mathrm{2-GMM}}\right)
\end{array}\right),
\end{align*}
where
\begin{equation}
\mathbb{E}_{\mathbb{P}_{X_{ij}}}\left[\phi\left(X_{ij};\theta_{}^{\mathrm{1-GMM}},\theta_{}^{\mathrm{2-GMM}},m,\mathrm{vec}\left\{ \Omega\right\} ,\mathrm{vec}\left\{ G_{1}\right\} ,\mathrm{vec}\left\{ G_{2}\right\} \right)\right]=0.\label{eq:moment_just_identified}
\end{equation}
Note that the moment equations in Equation \eqref{eq:moment_just_identified}
hold regardless of whether the moments equations in Equation \eqref{eq:moments_over_identified}
hold for some $\theta\in\Theta$.\footnote{\citet{imbens1997one} also shows that iterated GMM estimator \citep{hansen2021inference}
can be written as a just-identified GMM estimator, but the continuously
updated GMM estimator cannot.} Importantly, running this just-identified GMM procedure is numerically
equivalent to running the two-step GMM procedure. Since the just-identified
GMM estimator is a Z-estimator, we can apply Corollary \ref{cor:Z-estimators}
with 
\begin{align*}
\mathcal{H} & =\left\{ 1,...,2LK+2K+L^{2}\right\} \\
\nu_{\vartheta,\eta}\left(X_{ij}\right) & =\phi_{\eta}\left(X_{ij};\vartheta\right),
\end{align*}
and asymptotic validity of the bootstrap in Algorithm \ref{alg:MRGMM-BB}
amounts to checking relevant conditions on the moment functions.
\begin{example*}[\citealp{waugh2010international}]
Given the moment condition in Equation \eqref{eq:moment_W}, we should
check whether 
\[
\psi\left(X_{ij};\vartheta\right)=\left(\log\frac{\lambda_{ij}}{\lambda_{ii}}+\vartheta\log\left(\tau_{ij}\frac{p_{i}}{p_{j}}\right)\right)\log\left(\tau_{ij}\frac{p_{i}}{p_{j}}\right)
\]
is Fréchet differentiable in $\vartheta$. This is trivially the case
because $\psi\left(X_{ij};\cdot\right)$ is linear. The complexity
condition (iii) in Assumption \ref{assu:Complexity} is also satisfied
for this just-identified case with a single linear moment function.
$\triangle$
\end{example*}

\subsection{Frequentist Uncertainty Quantification for $\hat{\gamma}$ }

Recall the estimand $\gamma=g\left(\left\{ X_{k\ell}\right\} _{k\neq\ell},\theta\right)$,
which is random because it depends on the data $\left\{ X_{k\ell}\right\} _{k\neq\ell}$.
Given that $\hat{\theta}$ is approximately asymptotically normally
distributed, we can use a delta method-type result to find a valid
confidence interval:
\begin{thm}[Delta method for random object]
\label{thm:Delta_method_random_data}Suppose we have $\sqrt{n}\left(\hat{\theta}-\theta\right)\overset{d}{\approx}\mathcal{N}\left(0,\Sigma\right)$,
and we can consistently estimate its asymptotic variance by $\hat{\Sigma}$.
Then, for $G\left(\cdot\right)=\nabla_{\theta}g\left(\left\{ X_{k\ell}\right\} _{k\neq\ell},\cdot\right)$,
if we have 
\begin{equation}
\forall c>0,\underset{\tilde{\theta}:\left\Vert \tilde{\theta}-\theta\right\Vert \leq\frac{c}{\sqrt{n}}}{\sup}\left|G\left(\tilde{\theta}\right)-G\left(\theta\right)\right|\overset{p}{\rightarrow}0,\label{eq:uniform_convergence}
\end{equation}
a valid confidence interval for $\gamma$ is given by 
\[
\left[\hat{\gamma}\pm\Phi^{-1}\left(1-\alpha/2\right)\cdot\sqrt{\frac{1}{n}G\left(\hat{\theta}\right)^{2}\hat{\Sigma}}\right].
\]
\end{thm}
This implies that reporting the quantiles of the bootstrap draws in
Equation \eqref{eq:BUQ_gamma} is an asymptotically valid approach
to uncertainty quantification for the counterfactual prediction in
a frequentist sense.

\section{\protect\label{sec:Extensions}Extensions}

The Bayesian bootstrap procedure in Algorithm \ref{alg:BB} can easily
be adapted to accommodate various extensions. In this section I consider
two such extensions and provide the corresponding changes to the bootstrap
procedure, the model and the priors. In Appendix \ref{sec:Other_extensions}
I additionally discuss multiway clustering and conditional exchangeability.

\subsection{Polyadic data}

The data do not necessarily have to be dyadic. For example in Section
\ref{subsec:Applications_CP} we see that the estimation in \citet{caliendo2015estimates}
corresponds to a triadic regression. 

For the general case with polyadic data of order $P$, denote by $\mathbb{K}_{P}$
the set of all $P$-tuples of $\left\{ 1,...,n\right\} $ without
repetition. In this case, we would sample $\left(V_{1}^{\left(b\right)},...,V_{n}^{\left(b\right)}\right)\overset{\mathrm{iid}}{\sim}\mathrm{Exp}\left(1\right)$,
and compute bootstrap draws according to
\[
\hat{\theta}^{*,\left(b\right)}=T\left(\sum_{k\in\mathbb{K}_{P}}\frac{V_{k_{1}}^{\left(b\right)}\cdot...\cdot V_{k_{P}}^{\left(b\right)}}{\sum_{s\in\mathbb{K}_{P}}V_{s_{1}}^{\left(b\right)}\cdot...\cdot V_{s_{P}}^{\left(b\right)}}\cdot\delta_{X_{k}}\right).
\]
The priors from Assumption \ref{assu:prior} do not change, and in
the model from Assumption \ref{assu:Model} only the link function
changes, so that we have
\begin{align*}
C_{1},...,C_{n}|h,\mathbb{P}_{C} & \overset{\mathrm{iid}}{\sim}\mathbb{P}_{C}\\
X_{i} & =h\left(C_{i_{1}},...,C_{i_{P}}\right),\quad\mathrm{for\ }C_{i_{1}}\neq...\neq C_{i_{P}}.
\end{align*}

\subsection{Missing data}

When we observe the full matrix of bilateral observations, we observe
dyads indexed by the elements of some index set $\mathcal{I}_{\mathrm{non-diag}}=\left\{ \left(i,j\right)\in\left\{ 1,...,n\right\} ^{2}:i\neq j\right\} $.
However, sometimes non-diagonal observations are missing. In quantitative
trade and spatial models, the most common reason for these missing
observations is that zero flows are omitted, as is the case for the
running example based on \citet{waugh2010international} and in the
application based on \citet{caliendo2015estimates} in Section \ref{subsec:Applications_CP}.

To illustrate how to adapt the procedure of Algorithm \ref{alg:BB},
suppose that we only observe dyads in the set $\mathcal{I}_{\mathrm{}}\subset\mathcal{I}_{\mathrm{non-diag}}.$
We would then sample $\left(V_{1}^{\left(b\right)},...,V_{n}^{\left(b\right)}\right)\overset{\mathrm{iid}}{\sim}\mathrm{Exp}\left(1\right)$,
and compute bootstrap draws according to
\[
\hat{\theta}^{*,\left(b\right)}=T\left(\sum_{\left(k,\ell\right)\in\mathcal{I}}\frac{V_{k}^{\left(b\right)}\cdot V_{\ell}^{\left(b\right)}}{\sum_{\left(s,t\right)\in\mathcal{I}}V_{s}^{\left(b\right)}\cdot V_{t}^{\left(b\right)}}\cdot\delta_{X_{k\ell}}\right).
\]
The model in Assumption \ref{assu:Model} can be adapted by assuming
that the function $h$ maps to an empty set if $\left(C_{i},C_{j}\right)$
corresponds to a tuple of indices $\left(i,j\right)$ that was not
observed. We then have
\begin{align*}
C_{1},...,C_{n}|h,\mathbb{P}_{C} & \overset{\mathrm{iid}}{\sim}\mathbb{P}_{C}\\
X_{ij} & =h\left(C_{i},C_{j}\right)\in\mathcal{X}\cup\emptyset,\quad\mathrm{for\ }C_{i}\neq C_{i}\ \mathrm{and}\ h\left(C_{i},C_{j}\right)\neq\emptyset.
\end{align*}
The priors from Assumption \ref{assu:prior} do not change.\footnote{There are cases where we would want to model this differently, for
example if we observe a random sample of dyads. In that case we could
view the index set $\mathcal{I}$ as random and consider priors on
$h$ and $\mathbb{P}_{C}$ conditional on this index set, so that
$\mathcal{I}\sim\pi\left(\mathcal{I}\right)$ and $\left(h,\mathbb{P}_{C}\right)|\mathcal{I}\sim\pi\left(h|\mathcal{I}\right)\cdot DP\left(Q_{\mathcal{I}},\alpha\right).$
The model equations then change to $C_{1},...,C_{n}|h,\mathbb{P}_{C},\mathcal{I}\overset{\mathrm{iid}}{\sim}\mathbb{P}_{C}$
and $X_{ij}=h\left(C_{i},C_{j}\right)$, for $\left(i,j\right)\in\mathcal{I}$.
However, the corresponding bootstrap distribution will not change,
so using such a different underlying Bayesian model has no practical
implications.}

\section{\protect\label{sec:Applications}Applications}

In this section I discuss the applications in \citet{caliendo2015estimates}
and \citet{artucc2010trade}. For both, the number of interacting
units is small, which makes the Bayesian bootstrap procedure an appealing
approach for uncertainty quantification. 

\subsection{\protect\label{subsec:Applications_CP} Application 1: \citet{caliendo2015estimates}}

\subsubsection{Parameter Estimation}

\begin{figure}[h]
\centering{}\includegraphics[scale=0.16]{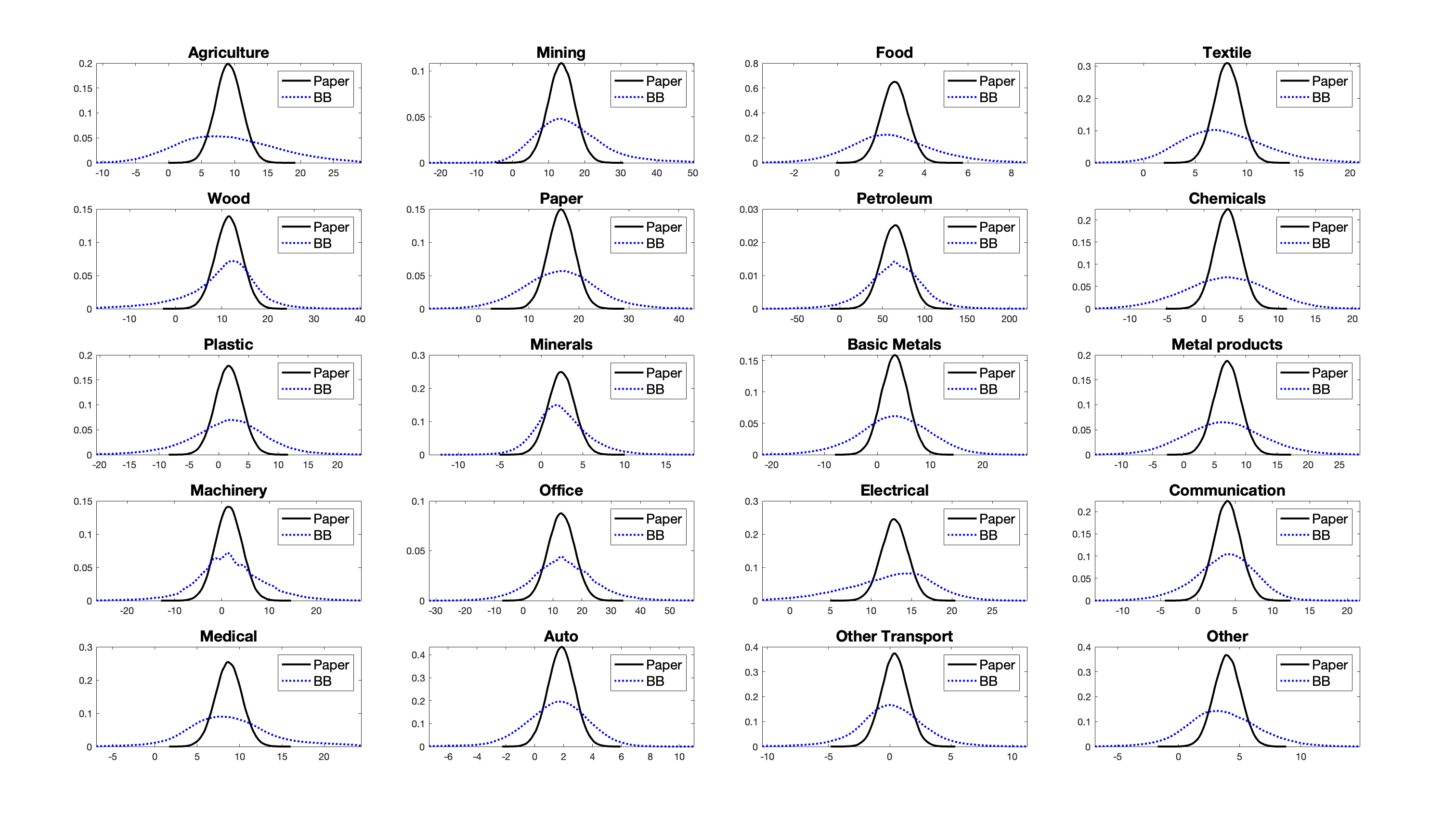}\caption{\protect\label{fig:CP}Distributions of the benchmark estimates (which
remove the countries with the lowest 1\% share of trade for each sector)
in Table 1 of \citet{caliendo2015estimates}. \textquotedblleft Paper\textquotedblright{}
corresponds to the normal approximation as implied by the standard
errors reported in the paper, and \textquotedblleft BB\textquotedblright{}
corresponds to the Bayesian bootstrap posterior.}
\end{figure}

\citet{caliendo2015estimates} introduces a new method to estimate
trade elasticities. Denoting with $F_{ij}^{s}$ and $t_{ij}^{s}$
the trade flow and tariff rate between country $i$ and $j$ in sector
$s$, respectively, the method amounts to running the triadic regressions
\[
\log\left(\frac{F_{ij}^{s}F_{jr}^{s}F_{ri}^{s}}{F_{ji}^{s}F_{rj}^{s}F_{ir}^{s}}\right)=-\theta^{s}\log\left(\frac{t_{ij}^{s}t_{jr}^{s}t_{ri}^{s}}{t_{ji}^{s}t_{rj}^{s}t_{ir}^{s}}\right)+\varepsilon_{ijr}^{s},
\]
with the identification restriction that the random disturbance term
$\varepsilon_{ijr}^{s}$ is orthogonal to the regressor. The number
of interacting units $n$ ranges between $12$ and $15$ across different
sector-specific regressions. Using insights from Section \ref{sec:Extensions},
the bootstrap procedure can easily be adapted to this triadic setting,
where now for each bootstrap draw we compute
\[
\hat{\theta}^{s,*,\left(b\right)}=T^{s}\left(\sum_{\left(k,\ell,m\right)\in\mathcal{I}^{s}}\frac{V_{k}^{\left(b\right)}\cdot V_{\ell}^{\left(b\right)}\cdot V_{m}^{\left(b\right)}}{\sum_{\left(t,u,v\right)\in\mathcal{I}^{s}}V_{t}^{\left(b\right)}\cdot V_{u}^{\left(b\right)}\cdot V_{v}^{\left(b\right)}}\cdot\delta_{X_{k\ell m}^{s}}\right).
\]
Note that I sum over the subset $\mathcal{I}^{s}\subset\left\{ \left(i,j,r\right)\in\left\{ 1,...,n\right\} ^{3}:i\neq j\neq r\right\} $,
because in this application observations with $\frac{F_{k\ell}^{s}F_{\ell m}^{s}F_{mk}^{s}}{F_{\ell k}^{s}F_{m\ell}^{s}F_{km}^{s}}=0$
are dropped. Table \ref{tab:CP} gives the corresponding 95\% Bayesian
credible intervals and 95\% confidence intervals constructed using
the point estimates and heteroskedastic-robust standard errors as
reported in the paper. Figure \ref{fig:CP} plots the corresponding
posterior distributions and implied normal distributions. It is alarming
that many credible intervals include $-1$, which violates the model
assumption that $\theta^{s}>-1$ for all sectors $s$.\footnote{Specifically, the sector-specific Fréchet shape parameter $\theta^{s}$
are assumed to be at least one greater than the within-sector elasticities
of substitution, which is assumed to be strictly positive.} Appendix \ref{subsec:PB_bootstrap_DGP} presents a data-calibrated
simulation exercise, which highlights that using heteroskedastic-robust
standard errors for uncertainty quantification results in under-coverage. 

Figure \ref{fig:CP} highlights that, using the Bayesian bootstrap
procedure, we do not have to ex ante think about which cases will
result in Gaussian posteriors. For example the posterior for the elasticity
for paper looks approximately normal, but the posterior for the elasticity
for mining is skewed with a heavy right tail. 
\begin{table}[h]
\begin{centering}
\begin{tabular}{|c|c|c|c|}
\hline 
 & Point estimate & As in paper & Bayesian bootstrap\tabularnewline
\hline 
\hline 
Agriculture & 9.11 & {[}5.17, 13.05{]} & {[}-4.05, 25.63{]}\tabularnewline
\hline 
Mining & 13.53 & {[}6.34, 20.73{]} & {[}0.69, 42.35{]}\tabularnewline
\hline 
Food & 2.62 & {[}1.43, 3.81{]} & {[}-1.26, 6.83{]}\tabularnewline
\hline 
Textile & 8.10 & {[}5.58, 10.61{]} & {[}0.52, 16.76{]}\tabularnewline
\hline 
Wood & 11.50 & {[}5.87, 17.12{]} & {[}-11.30, 22.88{]}\tabularnewline
\hline 
Paper & 16.52 & {[}11.33, 21.71{]} & {[}1.70, 31.32{]}\tabularnewline
\hline 
Petroleum & 64.44 & {[}33.84, 95.04{]} & {[}-6.41, 128.87{]}\tabularnewline
\hline 
Chemicals & 3.13 & {[}-0.37, 6.62{]} & {[}-8.49, 13.72{]}\tabularnewline
\hline 
Plastic & 1.67 & {[}-2.69, 6.03{]} & {[}-12.65, 14.01{]}\tabularnewline
\hline 
Minerals & 2.41 & {[}-0.72, 5.55{]} & {[}-3.17, 9.47{]}\tabularnewline
\hline 
Basic Metals & 3.28 & {[}-1.64, 8.19{]} & {[}-11.32, 15.91{]}\tabularnewline
\hline 
Metal products & 6.99 & {[}2.82, 11.15{]} & {[}-5.75, 19.46{]}\tabularnewline
\hline 
Machinery & 1.45 & {[}-4.04, 6.93{]} & {[}-12.75, 17.24{]}\tabularnewline
\hline 
Office & 12.95 & {[}4.07, 21.83{]} & {[}-7.71, 36.25{]}\tabularnewline
\hline 
Electrical & 12.91 & {[}9.70, 16.12{]} & {[}0.20, 21.37{]}\tabularnewline
\hline 
Communication & 3.95 & {[}0.48, 7.43{]} & {[}-5.25, 10.98{]}\tabularnewline
\hline 
Medical & 8.71 & {[}5.65, 11.78{]} & {[}-0.66, 26.37{]}\tabularnewline
\hline 
Auto & 1.84 & {[}0.04, 3.64{]} & {[}-3.80, 5.48{]}\tabularnewline
\hline 
Other Transport & 0.39 & {[}-1.73, 2.51{]} & {[}-5.84, 5.67{]}\tabularnewline
\hline 
Other & 3.98 & {[}1.86, 6.11{]} & {[}-2.11, 9.68{]}\tabularnewline
\hline 
\end{tabular}
\par\end{centering}
\caption{\protect\label{tab:CP}Uncertainty quantification for the benchmark
estimates (which remove the countries with the lowest 1\% share of
trade for each sector) in Table 1 of \citet{caliendo2015estimates}.}
\end{table}

\subsubsection{Counterfactual }

\begin{figure}[h]
\centering{}\includegraphics[scale=0.16]{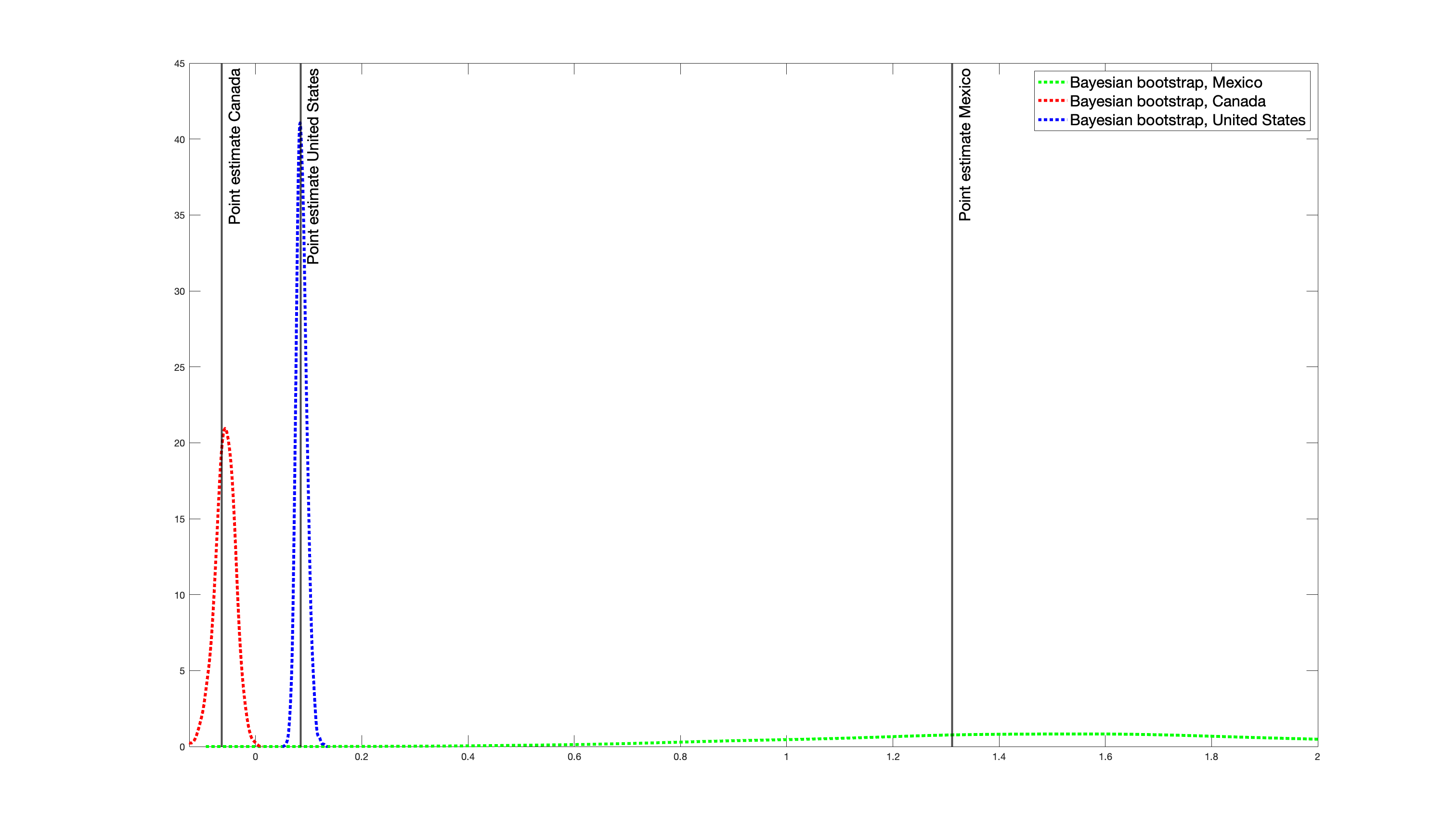}\caption{\protect\label{fig:CP_cfs}Posterior distributions for welfare effects
as in Table 2 of \citet{caliendo2015estimates}}
\end{figure}

The main counterfactual question in \citet{caliendo2015estimates}
concerns the effects of the NAFTA trade agreement on welfare in Mexico,
Canada and the United States. These welfare predictions, which depend
on both the data and the estimated trade elasticities, are reported
in the abstract and in Table 2 of \citet{caliendo2015estimates} without
any uncertainty quantification. In Table \ref{tab:CP_cfs}, I reproduce
these results and include 95\% Bayesian credible intervals. Figure
\ref{fig:CP_cfs} displays the corresponding posterior distributions.
Implementation details and additional results are provided in Appendix
\ref{subsec:Extra_results_CP}.

\begin{table}[h]
\begin{centering}
\begin{tabular}{|c|c|c|}
\hline 
 & Point estimate & Bayesian bootstrap\tabularnewline
\hline 
\hline 
Mexico & 1.31\% & {[}0.65\%, 2.51\%{]}\tabularnewline
\hline 
Canada & -0.06\% & {[}-0.10\%, -0.02\%{]}\tabularnewline
\hline 
U.S. & 0.08\% & {[}0.07\%, 0.11\%{]}\tabularnewline
\hline 
\end{tabular}
\par\end{centering}
\caption{\protect\label{tab:CP_cfs}Bayesian uncertainty quantification for
welfare effects as in Table 2 of \citet{caliendo2015estimates}.}
\end{table}

The credible intervals and posterior distributions show asymmetry
in the distribution of welfare changes, shifting probability mass
away from zero. Furthermore, we observe there is much more uncertainty
around the welfare effect for Mexico than around the welfare effects
for Canada and the United States. However, since none of the credible
intervals include zero, the signs of the effects are robust to uncertainty.
This is also true for the ranking of welfare effects among the three
countries, since for none of the bootstrap draws the ranking is different
from the ranking corresponding to the point estimates. 

\subsection{\protect\label{subsec:Applications_ACM} Application 2: \citet{artucc2010trade}}

\subsubsection{Parameter Estimation}

\citet{artucc2010trade} uses over-identified GMM to estimate the
mean and variance of workers' switching cost, denoted with $\mu$
and $\sigma^{2}$, respectively. The data consists of a panel of dyadic
data across industries. There are $n=6$ industries and $T=23$ years.
Towards uncertainty quantification, \citet{artucc2010trade} ignores
the dependence across years and industries and uses the standard GMM
asymptotic variance formula. Implicitly, this imposes the assumption
that all $690$ ($=n\cdot\left(n-1\right)\cdot T$) observations are
exchangeable. The corresponding moment function is
\begin{equation}
\psi^{\mathrm{industry-year}}\left(X_{ij,t};\theta\right)=\left(Y_{ij,t}-\left(\begin{array}{ccc}
\frac{\zeta-1}{\sigma^{2}}\mu & \frac{\zeta}{\sigma^{2}} & \zeta\end{array}\right)R_{ij,t}\right)Z_{ij,t},\label{eq:ACM_industry_year}
\end{equation}
with
\begin{align*}
Y_{ij,t} & =\log m_{ij,t}-\log m_{ii,t}\\
R_{ij,t} & =\left(\begin{array}{ccc}
1 & w_{j,t+1}-w_{i,t+1} & \log m_{ij,t+1}-\log m_{jj,t+1}\end{array}\right)^{'}\\
Z_{ij,t} & =\left(\begin{array}{ccc}
1 & w_{j,t-1}-w_{i,t-1} & \log m_{ij,t-1}^ {}-\log m_{jj,t-1}\end{array}\right)^{'}.
\end{align*}
Here, $m_{ij,t}$ denotes the fraction of the labor force in industry
$i$ at time $t$ that chooses to move to industry $j$ and $w_{i,t}$
denotes the wage in industry $i$ at time $t$. The parameter $\zeta$
denotes the discount factor which is fixed ex ante. I reproduce the
estimates for $\mu$ and $\sigma^{2}$ in Panel IV of Table 3 in \citet{artucc2010trade},
which fixes $\zeta=0.97$ and corresponds to the authors' preferred
specification.

\begin{figure}[h]
\centering{}\includegraphics[scale=0.16]{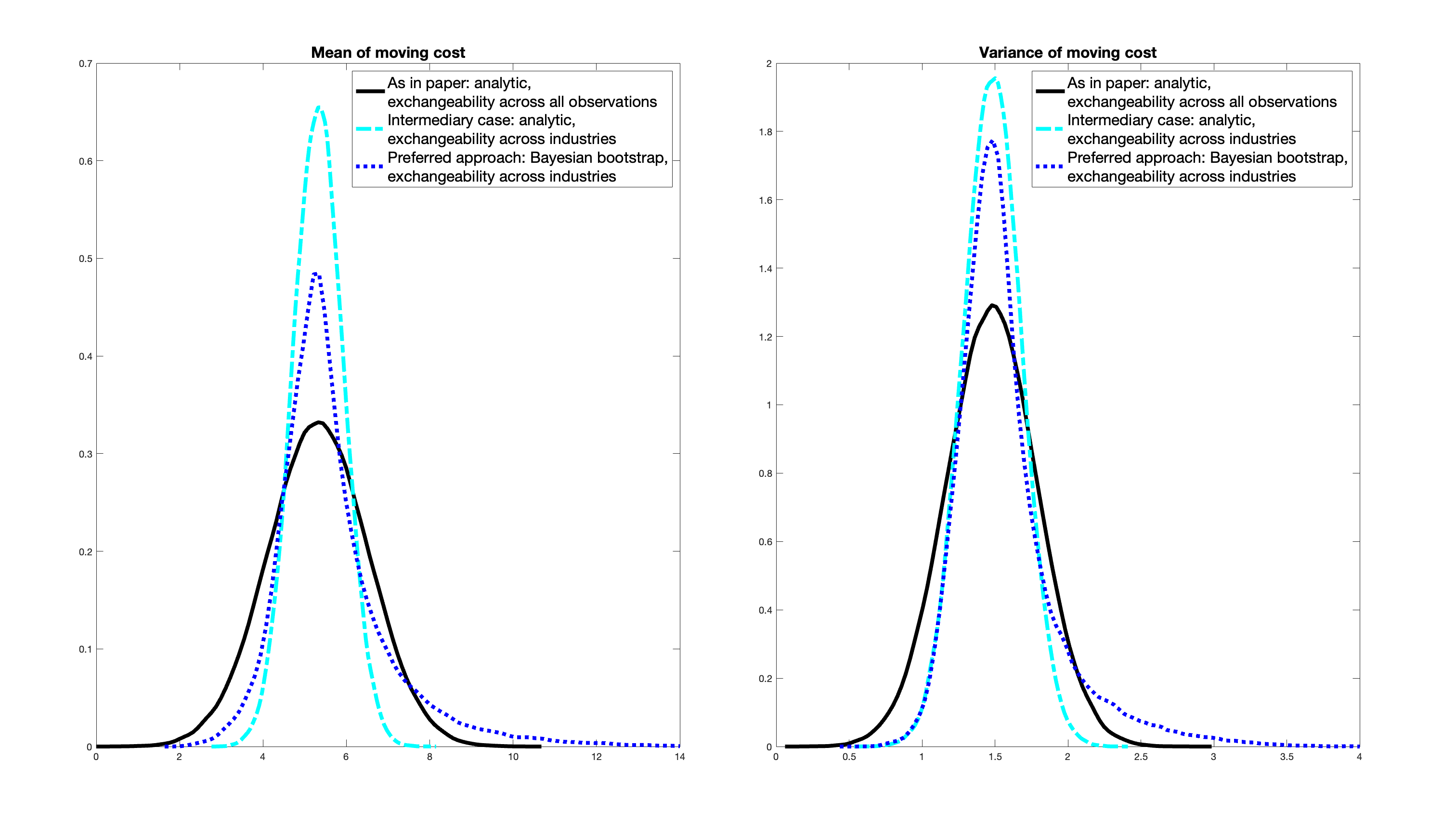}\caption{\protect\label{fig:ACM}Distribution of estimators for Panel IV in
Table 3 in \citet{artucc2010trade} for $\zeta=0.97$.}
\end{figure}

Instead of exchangeability across all observations, one might argue
a more plausible assumption is exchangeability across industries.
The corresponding moment function is
\begin{equation}
\psi^{\mathrm{industry}}\left(X_{ij};\theta\right)=\frac{1}{T}\sum_{t=1}^{T}\left(Y_{ij,t}-\left(\begin{array}{ccc}
\frac{\zeta-1}{\sigma^{2}}\mu & \frac{\zeta}{\sigma^{2}} & \zeta\end{array}\right)R_{ij,t}\right)Z_{ij,t}.\label{eq:ACM_industry}
\end{equation}
Given the moment functions in Equations \eqref{eq:ACM_industry_year}
and \eqref{eq:ACM_industry}, I consider three different approaches
to uncertainty quantification. The first approach follows \citet{artucc2010trade}
and uses Equation \eqref{eq:ACM_industry} to compute the analytic
GMM standard error assuming all observations are exchangeable. The
second approach is an intermediary case; it still computes an analytic
GMM standard error but uses Equation \eqref{eq:ACM_industry} and
hence assumes exchangeability only across industries. The third approach
is my preferred approach, and it uses the Bayesian bootstrap procedure
from Algorithm \ref{alg:MRGMM-BB} and Equation \eqref{eq:ACM_industry}.\footnote{Equivalently to using the moment function in Equation \eqref{eq:ACM_industry}
with weights $\omega_{k\ell}^{\left(b\right)}=V_{k}^{\left(b\right)}\cdot V_{\ell}^{\left(b\right)}/\left(\sum_{u\neq v}V_{u}^{\left(b\right)}\cdot V_{v}^{\left(b\right)}\right)$
for $k,\ell=1,...,n$, one could use the moment function in Equation
\eqref{eq:ACM_industry_year} with weights $\omega_{k\ell,s}^{\left(b\right)}=V_{k}^{\left(b\right)}\cdot V_{\ell}^{\left(b\right)}/\left(T\cdot\sum_{u\neq v}V_{u}^{\left(b\right)}\cdot V_{v}^{\left(b\right)}\right)$
for $k,\ell=1,...,n$ and $s=1,...,T$. } The resulting 95\% confidence intervals and credible intervals are
given in Table \ref{tab:ACM}.\footnote{The point estimates differ slightly from those in \citet{artucc2010trade}.
This is because there the authors use iterated GMM rather than two-step
GMM and they use a different weight matrix. In Appendix \ref{subsec:ACM_iter_GMM}
I consider their exact setup and the conclusions do not change. } The corresponding implied normal distributions and posterior distributions
are plotted in Figure \ref{fig:ACM}. The posterior distributions
for both parameters are non-normal and exhibit heavy right tails,
indicating substantial uncertainty—particularly regarding the possibility
of large switching costs. Implementation details and extra results
can be found in Appendix \ref{subsec:Extra_results_ACM}. Furthermore,
a data-calibrated simulation exercise in Appendix \ref{subsec:PB_bootstrap_DGP}
shows that standard GMM standard errors lead to under-coverage. 
\begin{table}[h]
\begin{centering}
\begin{tabular}{|c|c|c|}
\hline 
 & Mean & Variance\tabularnewline
\hline 
\hline 
Point estimate & 5.33 & 1.48\tabularnewline
\hline 
\begin{cellvarwidth}[m]
\centering
As in paper: analytic  errors,

exchangeability across all observations
\end{cellvarwidth} & {[}2.98, 7.67{]} & {[}0.87, 2.09{]}\tabularnewline
\hline 
\begin{cellvarwidth}[m]
\centering
Intermediary case: analytic  errors,

exchangeability across industries
\end{cellvarwidth} & {[}4.14, 6.51{]} & {[}1.08, 1.88{]}\tabularnewline
\hline 
\begin{cellvarwidth}[m]
\centering
Preferred approach: Bayesian bootstrap,

exchangeability across industries
\end{cellvarwidth} & {[}3.64, 9.39{]} & {[}1.09, 2.57{]}\tabularnewline
\hline 
\end{tabular}
\par\end{centering}
\caption{\protect\label{tab:ACM}Uncertainty quantification for Panel IV in
Table 3 in \citet{artucc2010trade} for $\zeta=0.97$.}
\end{table}

\subsubsection{Counterfactual Prediction}

The estimated mean and variance of the moving cost are then used for
a simulation exercise. The counterfactual scenario of interest is
a sudden liberalization of the manufacturing sector. The main economic
quantities of interest are the pre- and post-employment share of the
manufacturing sector, the pre- and post-wage of the manufacturing
sector, and the expected discounted lifetime utilities before and
after the announcement of liberalization. These counterfactual predictions
are reported in Figures 3, 4 and 5 in \citet{artucc2010trade} without
any uncertainty quantification. The 95\% Bayesian credible intervals
(using my preferred approach, assuming only exchangeability across
industries) for these quantities are given in Table \ref{tab:ACM_cfs}.

\begin{table}[h]
\begin{centering}
\begin{tabular}{|c|c|c|c|}
\hline 
 & Employment share & Wage & Utility\tabularnewline
\hline 
\hline 
Before liberalization & 25.3\% {[}23.3\%, 26.0\%{]} & 1.043 {[}1.028, 1.085{]} & 39.4 {[}36.7, 48.2{]}\tabularnewline
\hline 
After liberalization & 15.7\% {[}15.5\%, 16.1\%{]} & 1.036 {[}1.020, 1.043{]} & 40.8 {[}38.2, 49.3{]}\tabularnewline
\hline 
\end{tabular}
\par\end{centering}
\caption{\protect\label{tab:ACM_cfs}Uncertainty quantification for relevant
economic quantities from Figures 3, 4 and 5 in \citet{artucc2010trade}
for $\zeta=0.97$.}
\end{table}

\begin{figure}[h]
\centering{}\includegraphics[scale=0.16]{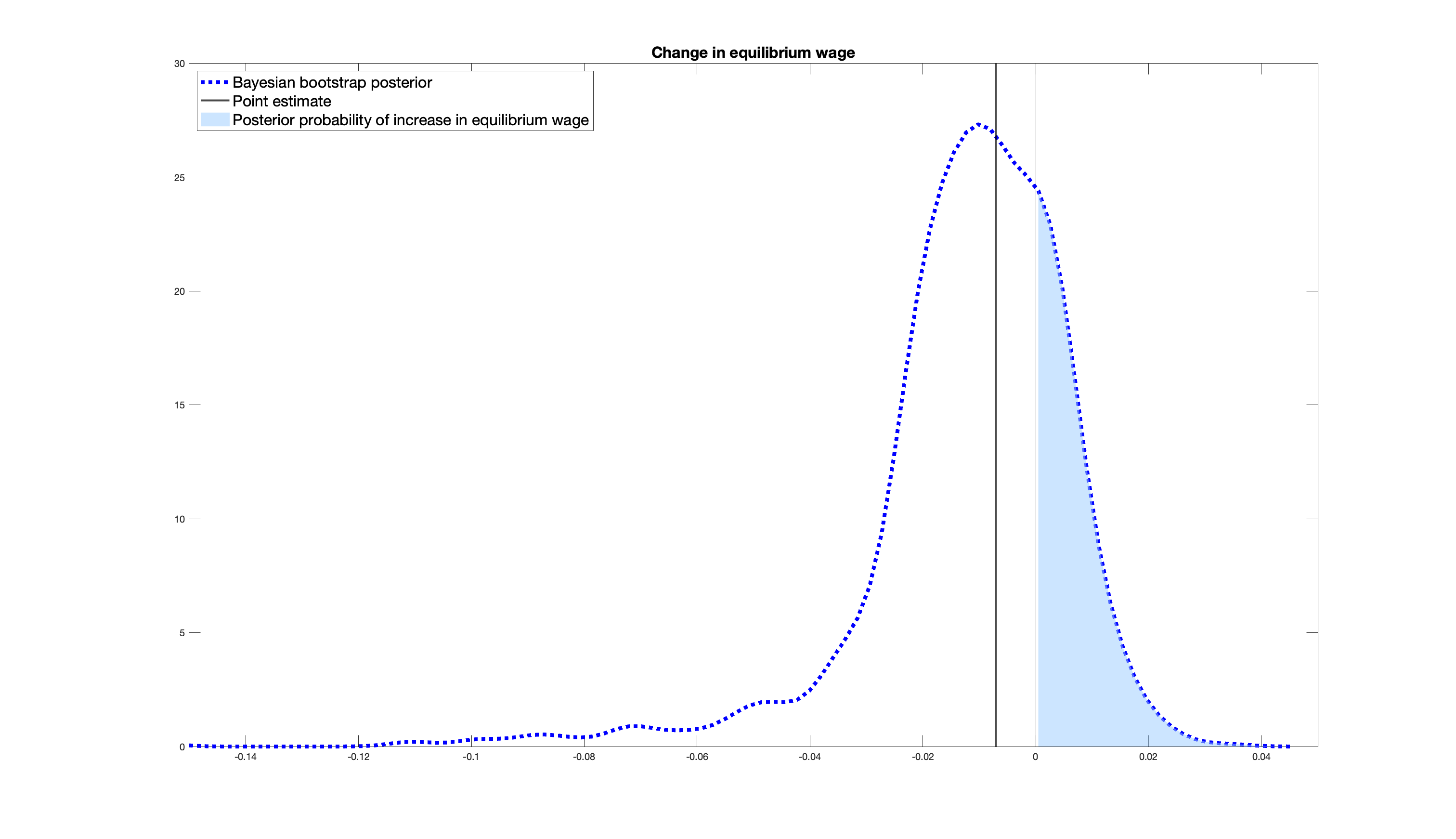}\caption{\protect\label{fig:ACM_cfs}Posterior distribution for the change
in wages based on Figure 4 in \citet{artucc2010trade} for $\zeta=0.97$.}
\end{figure}

The credible intervals are again asymmetric around the point estimates.
In all of the bootstrap draws, the employment share goes down and
the lifetime utility goes up. Notably, in around 25\% of bootstrap
draws, the equilibrium wage after liberalization is higher than the
equilibrium wage before liberalization. To investigate this further,
Figure \ref{fig:ACM_cfs} plots the posterior distribution of the
\textit{difference} between the post- and pre-wage of the manufacturing
sector, which has a heavy left tail but non-negligible mass above
zero. In footnote 26 of \citet{artucc2010trade} it is mentioned that
in principle it could happen that the equilibrium wage rises but ``that
does not happen in this case''. However, when we account for uncertainty
this turns out to be an economically important scenario that should
be taken into consideration—a finding not visible from point estimates
alone. 

\section{\protect\label{sec:Alt_methods}Comparison with Alternative Methods}

As discussed in the introduction, there exist various alternatives
for uncertainty quantification. Here, I discuss an alternative bootstrap
from \citet{davezies2021empirical} based on resampling, and analytic
standard errors based on \citet{graham2020dyadic,graham2020network}. 

\subsection{\protect\label{subsec:Pigeonhole}Pigeonhole Bootstrap}

The closest method for uncertainty quantification for $\hat{\theta}$
that is theoretically grounded is the pigeonhole bootstrap from \citet{davezies2021empirical}.
The method is summarized in Algorithm \ref{alg:PB}. For quantitative
trade and spatial models, the most important disadvantage of the pigeonhole
bootstrap is that its existing theoretical guarantees rely on approximations
that envision large number of units. However, relevant applications
often include a small number of units. 
\begin{algorithm}[h]
\caption{\protect\label{alg:PB}Pigeonhole bootstrap procedure}

\begin{enumerate}
\item Input: Bilateral data $\left\{ X_{k\ell}\right\} _{k\neq\ell}$ and
estimator function $T:\Delta\left(\mathcal{X}\right)\rightarrow\Theta$.
\item For each bootstrap draw $b=1,...,B$:
\begin{enumerate}
\item Sample $n$ units \textit{independently with replacement} from $\left\{ 1,...,n\right\} $
with equal probability. Let $W_{k}^{\mathrm{pb},\left(b\right)}$
denote the number of times that $k$ is sampled.
\item Compute 
\[
\hat{\theta}^{*,\mathrm{pb},\left(b\right)}=T\left(\sum_{k\neq\ell}\frac{W_{k}^{\mathrm{pb},\left(b\right)}\cdot W_{\ell}^{\mathrm{pb},\left(b\right)}}{n\left(n-1\right)}\cdot\delta_{X_{k\ell}}\right).
\]
\end{enumerate}
\item Report the quantiles of interest of $\left\{ \hat{\theta}^{*,\mathrm{pb},\left(1\right)},...,\hat{\theta}^{*,\mathrm{pb},\left(B\right)}\right\} $.
\end{enumerate}
\end{algorithm}
 
\begin{example*}[\citealp{waugh2010international}]
For the application in \citet{waugh2010international}, using the
pigeonhole bootstrap, if we run sufficiently many iterations, we will
eventually draw a world with three copies of Australia and no Belgium.
In contrast, every bootstrap draw in the Bayesian procedure in Algorithm
\ref{alg:BB} will have all $43$ countries, but they are reweighted
using continuous and strictly positive weights. $\triangle$
\end{example*}
Towards uncertainty quantification for the counterfactual prediction
$\hat{\gamma}$, the pigeonhole bootstrap procedure again only delivers
asymptotic frequentist guarantees. If one is confident in the asymptotic
approximation and the validity of the resulting coverage interval
for $\theta$, then uncertainty can be propagated using a delta method
or bootstrap approximation. Specifically, one could compute bootstrap
draws as 
\begin{equation}
\hat{\gamma}^{*,\mathrm{pb},\left(b\right)}=g\left(\left\{ X_{k\ell}\right\} _{k\neq\ell},\hat{\theta}^{*,\mathrm{pb},\left(b\right)}\right),\label{eq:bootstrap_draws_pb}
\end{equation}
 for $b=1,...,B$, and construct a coverage interval for $\gamma$
using these draws. The validity of this approach follows from Theorem
\ref{thm:Delta_method_random_data}. In Appendix \ref{subsec:PB_bootstrap_DGP}
I perform a simulation exercise that for all my applications compares
coverage across methods, assuming the data are generated according
to the pigeonhole bootstrap. 

To illustrate the differences between the Bayesian bootstrap and the
pigeonhole bootstrap, consider the application in \citet{artucc2010trade}
discussed in Section \ref{subsec:Applications_ACM}, where, for my
preferred specification, the number of interacting units is $n=6$.
Table \ref{tab:ACM_pigeon} shows that the credible intervals obtained
from the Bayesian bootstrap are narrower than the coverage intervals
obtained from the pigeonhole bootstrap. Figure \ref{fig:ACM} displays
the corresponding bootstrap distributions, omitting draws outside
of the considered ranges. 

There is a non-negligible probability that the pigeonhole bootstrap
distribution only has bilateral flows between two industries (around
2\% for $n=6$), in which case the optimal weight-matrix is reported
to be near-singular. The pigeonhole bootstrap also produces more extreme
outliers. Specifically, approximately 3\% of the bootstrap draws for
the mean and 1\% for the variance fall more than 10 standard deviations
(as measured by the GMM standard error) away from the point estimate.
For the Bayesian bootstrap, the corresponding rates are 0.5\% and
0.1\%, respectively. In addition, the Bayesian bootstrap draws for
the variance estimator are always nonnegative, whereas about 0.5\%
of the pigeonhole bootstrap draws yield negative values.

\begin{table}[h]
\begin{centering}
\begin{tabular}{|c|c|c|}
\hline 
 & Mean & Variance\tabularnewline
\hline 
\hline 
Point estimate & 5.33 & 1.48\tabularnewline
\hline 
Bayesian bootstrap & {[}3.64, 9.39{]} & {[}1.09, 2.57{]}\tabularnewline
\hline 
Pigeonhole bootstrap & {[}2.76, 11.17{]} & {[}0.59, 3.01{]}\tabularnewline
\hline 
\end{tabular}
\par\end{centering}
\caption{\protect\label{tab:ACM_pigeon}Uncertainty quantification for Panel
IV in Table 3 in \citet{artucc2010trade} for $\zeta=0.97$.}
\end{table}
\begin{figure}[h]
\centering{}\includegraphics[scale=0.16]{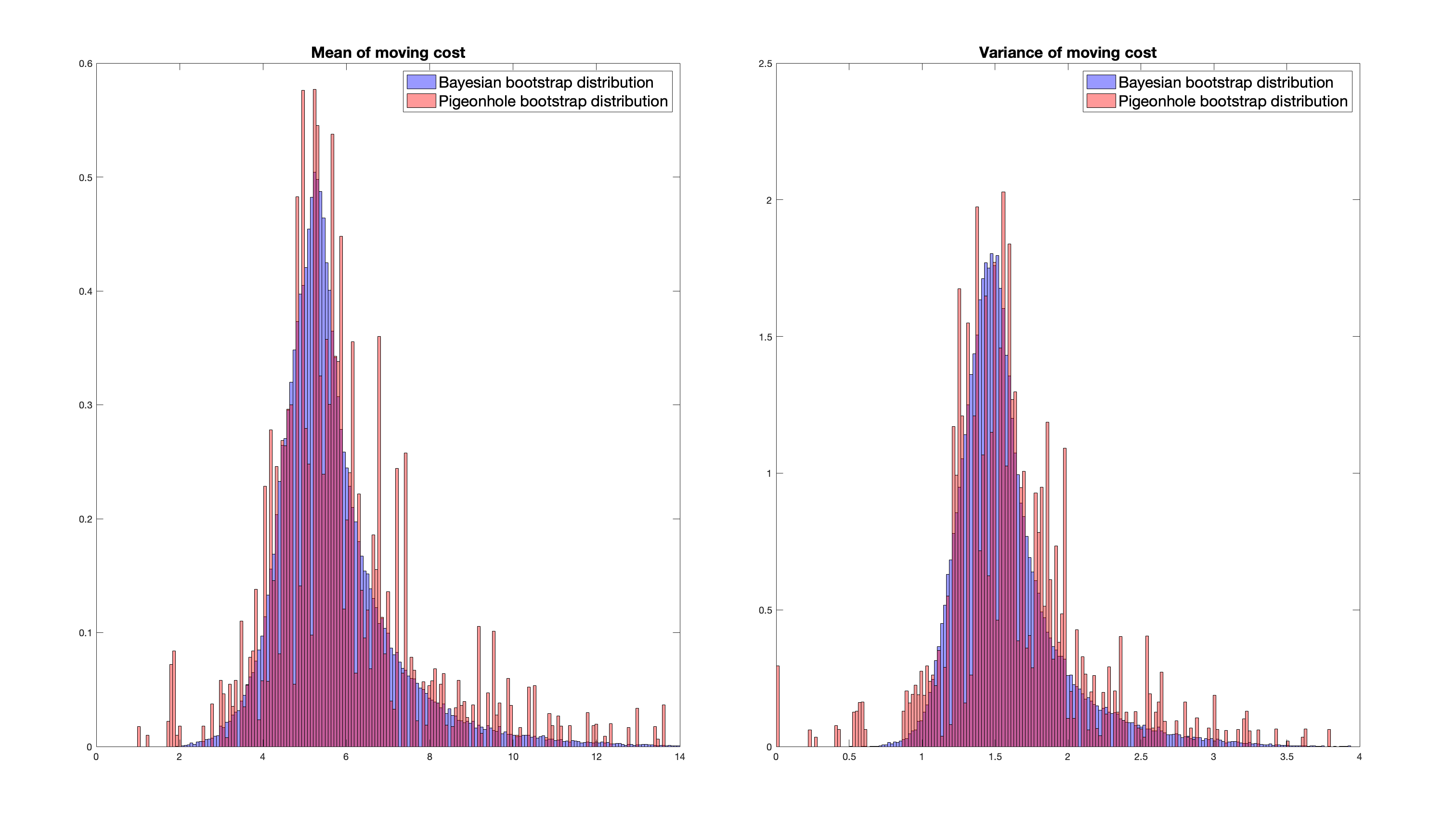}\caption{\protect\label{fig:ACM_pigeon}Bootstrap distributions of estimators
for Panel IV in Table 3 in \citet{artucc2010trade} for $\zeta=0.97$.}
\end{figure}

\subsection{\protect\label{subsec:Analytic}Analytic Standard Errors}

A second alternative approach for uncertainty quantification for $\hat{\theta}$
is to find frequentist standard errors. I adapt the likelihood setting
in \citet{graham2020network} to obtain a new result for Z-estimators:
\begin{prop}[Analytic standard error for Z-estimators]
\label{prop:Bryan_variance}Suppose $\hat{\theta}$ solves $\mathbb{E}_{\mathbb{P}_{n,X_{ij}}}\left[\phi\left(X_{ij};\hat{\theta}\right)\right]=0$
and $\theta$ solves $\mathbb{E}_{\mathbb{P}_{X_{ij}}}\left[\phi\left(X_{ij};\theta\right)\right]=0$.
Then a consistent variance estimator for $\hat{\theta}$ is given
by
\[
\widehat{\Var}_{\mathrm{Graham}}\left(\hat{\theta}\right)=\frac{1}{n}\hat{\Sigma}_{1}^{-1}\left(4\hat{\Sigma}_{2}+\frac{2}{n-1}\left(\hat{\Sigma}_{3}-2\hat{\Sigma}_{2}\right)\right)\left(\hat{\Sigma}_{1}^{-1}\right)^{'}
\]
where 
\begin{align*}
\hat{\Sigma}_{1} & =\frac{1}{n\left(n-1\right)}\sum_{k\ne\ell}\frac{\partial\phi\left(X_{k\ell};\theta\right)}{\partial\theta}|_{\theta=\hat{\theta}}\\
\hat{\Sigma}_{2} & =\left(\begin{array}{c}
n\\
3
\end{array}\right)^{-1}\sum_{k=1}^{n-2}\sum_{\ell=k+1}^{n-1}\sum_{s=\ell+1}^{n}\frac{1}{3}\left\{ \left(\frac{\hat{\phi}_{k\ell}+\hat{\phi}_{\ell k}}{2}\right)\left(\frac{\hat{\phi}_{ks}+\hat{\phi}_{sk}}{2}\right)^{'}\right.\\
 & \quad\left.\left(\frac{\hat{\phi}_{k\ell}+\hat{\phi}_{\ell k}}{2}\right)\left(\frac{\hat{\phi}_{\ell s}+\hat{\phi}_{s\ell}}{2}\right)^{'}+\left(\frac{\hat{\phi}_{ks}+\hat{\phi}_{sk}}{2}\right)\left(\frac{\hat{\phi}_{\ell s}+\hat{\phi}_{s\ell}}{2}\right)^{'}\right\} \\
\hat{\Sigma}_{3} & =\left(\begin{array}{c}
n\\
2
\end{array}\right)^{-1}\sum_{k=1}^{n-1}\sum_{\ell=k+1}^{n}\left(\frac{\hat{\phi}_{k\ell}+\hat{\phi}_{\ell k}}{2}\right)\left(\frac{\hat{\phi}_{k\ell}+\hat{\phi}_{\ell k}}{2}\right)^{'},
\end{align*}
with $\hat{\phi}_{ij}=\phi\left(X_{ij};\hat{\theta}\right).$
\end{prop}
For $\hat{\theta}$ a Z-estimator, one can then report the confidence
interval $\left[\hat{\theta}\pm1.96\cdot\sqrt{\widehat{\Var}_{\mathrm{Graham}}\left(\hat{\theta}\right)}\right]$.
As argued in Theorem \ref{thm:Delta_method_random_data}, under some
regularity conditions we can use the delta method to find a valid
confidence interval for $\gamma$.

There are various reasons why one might prefer using the Bayesian
bootstrap procedure instead of these analytic standard errors. Firstly,
similar to the pigeonhole bootstrap, the validity of the standard
errors relies on asymptotic approximations that envision a large number
of units. Secondly, it is non-trivial how to adjust the analytic approach
to various extensions as discussed in Section \ref{sec:Extensions}
\citep{graham2024sparse}. Lastly, the approach can be difficult to
implement. For example, for over-identified GMM, this approach requires
computing many numerical derivatives.

\subsection{Application 3: \citet{silva2006log}}

That being said, when the sample size is large, the data are dyadic
and there are no missing values, both the pigeonhole bootstrap and
the analytic standard errors result in uncertainty quantification
that is similar to the Bayesian bootstrap procedure for a Z-estimator.
This follows from Corollary \ref{cor:Z-estimators} and Proposition
\ref{prop:Bryan_variance} in the current paper and Theorem 2.4 in
\citet{davezies2021empirical}. To illustrate this, in Appendix \ref{sec:ST}
I revisit the application that was considered in both \citet{graham2020dyadic}
and \citet{davezies2021empirical}, namely a PPML regression based
on data from \citet{silva2006log}. In that setting, despite the Bayesian
bootstrap being the only method with a finite-sample guarantee, all
three methods yield similar uncertainty quantification. 

\section{\protect\label{sec:Conclusion}Conclusion}

This paper considers uncertainty quantification for counterfactual
predictions in polyadic settings. I propose a Bayesian bootstrap procedure
to quantify uncertainty around estimators for structural parameters.
This also implies valid uncertainty quantification for the point estimates
of counterfactual predictions. The method is especially appealing
in applications with a small number of interacting units, as it admits
a finite-sample Bayesian interpretation. At the same time, it provides
frequentist asymptotic guarantees under mild conditions. By revisiting
the applications in \citet{waugh2010international}, \citet{caliendo2015estimates}
and \citet{artucc2010trade}, I illustrate the practical advantages
of the proposed approach.

\end{spacing}

\bibliographystyle{econ-aea}
\bibliography{Abstraction}
\newpage{}

\appendix
\begin{center}
{\huge Appendix}{\huge\par}
\par\end{center}

\section{\protect\label{subsec:Extra_results_Waugh}Extra Results for Running
Example: \citet{waugh2010international}}

\subsection{\protect\label{subsec:W_model_details}Model Details}

In Section \ref{subsec:Counterfactual-Predictions} I introduced the
counterfactual mapping 
\[
\left\{ X_{k\ell}\right\} _{k\neq\ell},\hat{\theta},\left\{ \tau_{k\ell}^{\mathrm{cf}}\right\} \mapsto\left\{ \hat{w}_{k}^{\mathrm{cf}}\right\} .
\]
Here,
\begin{align*}
\left\{ X_{k\ell}\right\} _{k\neq\ell} & =\left\{ \left(\lambda_{k\ell},\lambda_{kk},\tau_{k\ell},p_{k},p_{\ell}\right)\right\} _{k\neq\ell}\\
 & =\left(\left\{ \lambda_{k\ell}\right\} ,\left\{ \tau_{k\ell}\right\} ,\left\{ p_{k}\right\} \right),
\end{align*}
since $\tau_{kk}=1$ for all $k$. Recall that $\lambda_{k\ell}$
denotes country $\ell$'s expenditure share on goods from country
$k$, $\tau_{k\ell}$ denotes estimated iceberg trade costs from country
$k$ to country $\ell$, and $p_{k}$ denotes the aggregate prices
in country $k$.

The equilibrium conditions map rental rates, trade costs, labor endowments,
production parameters and the productivity parameter to aggregated
prices, expenditure shares and wages:
\begin{equation}
\left\{ r_{k}\right\} ,\left\{ \tau_{k\ell}^{\mathrm{}}\right\} ,\left\{ L_{k}\right\} ,\left\{ Q_{k}\right\} ,\alpha,\beta,\theta_{M}\mapsto\left\{ p_{k}\right\} ,\left\{ \lambda_{k\ell}\right\} ,\left\{ w_{k}^{\mathrm{}}\right\} .\label{eq:Waugh_eql}
\end{equation}
Currently, $X_{k\ell}$ only contains the variables that are relevant
for constructing the estimator $\hat{\theta}$. The other variables
that are inputs to the counterfactual analysis are subsumed into $g$.
These are labor endowments $\left\{ L_{k}\right\} $, aggregate capital-labor
ratios $\left\{ K_{k}\right\} $ and the production parameters $\left(\alpha,\beta\right)$.
So the ``data'' that we have in hand are
\[
\left(\left\{ X_{k\ell}\right\} _{k\neq\ell},\left\{ L_{k}\right\} ,\left\{ K_{k}\right\} ,\alpha,\beta\right).
\]
It follows that we require a ``calibration-mapping''
\[
\left\{ X_{k\ell}\right\} _{k\neq\ell},\left\{ L_{k}\right\} ,\left\{ K_{k}\right\} ,\alpha,\beta,\hat{\theta}\mapsto\left\{ \hat{r}_{k}\right\} ,\left\{ \hat{Q}_{k}\right\} .
\]
Such a mapping exists and we can use the equilibrium mapping in Equation
\eqref{eq:Waugh_eql} to arrive at:
\[
\left\{ \hat{r}_{k}\right\} ,\left\{ \tau_{k\ell}^{\mathrm{\mathrm{cf}}}\right\} ,\left\{ L_{k}\right\} ,\left\{ \hat{Q}_{k}\right\} ,\alpha,\beta,\hat{\theta}\mapsto\left\{ \hat{p}_{k}^{\mathrm{cf}}\right\} ,\left\{ \hat{\lambda}_{k\ell}^{\mathrm{cf}}\right\} ,\left\{ \hat{w}_{k}^{\mathrm{cf}}\right\} .
\]
Once we have obtained the counterfactual wage vector $\left\{ \hat{w}_{k}^{\mathrm{cf}}\right\} $,
we can calculate the various inequality statistics. 

\subsection{\protect\label{subsec:W_PPML}Using PPML instead of OLS}

\begin{figure}[h]
\centering{}\includegraphics[scale=0.16]{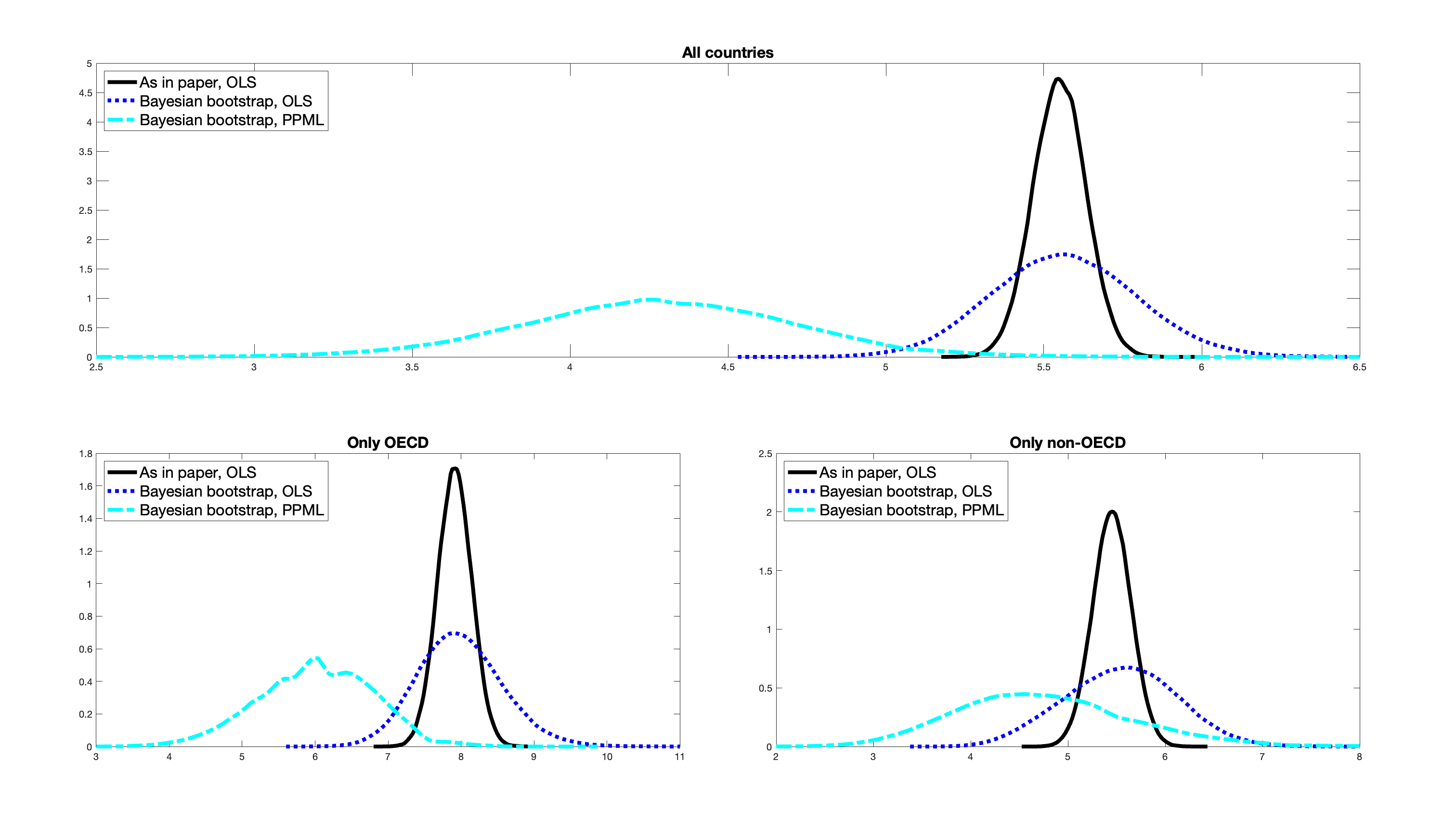}\caption{\protect\label{fig:W_ppml}Distributions for productivity parameters
as in \citet{waugh2010international} using OLS and PPML.}
\end{figure}

Equation \eqref{eq:moment_W_ppml} gives the moment function for the
application in \citet{waugh2010international} when not omitting zeros
and using PPML. Table \ref{tab:W_ppml} and Figure \ref{fig:W_ppml}
add the resulting credible intervals and posterior distributions to
Table \ref{tab:W} and Figure \ref{fig:W}, respectively. The point
estimates drop considerably and there is more uncertainty. 
\begin{table}[h]
\begin{centering}
\begin{tabular}{|c|c|c|c|}
\hline 
 & Point estimate & As in paper & Bayesian bootstrap\tabularnewline
\hline 
\hline 
\textbf{All countries, OLS} & 5.55 & {[}5.39, 5.71{]} & {[}5.12, 6.02{]}\tabularnewline
\hline 
\textbf{Only OECD, OLS} & 7.91 & {[}7.46, 8.37{]} & {[}6.91, 9.21{]}\tabularnewline
\hline 
\textbf{Only non-OECD, OLS} & 5.45 & {[}5.06, 5.84{]} & {[}4.42, 6.65{]}\tabularnewline
\hline 
\textbf{All countries, PPML} & 4.19 & - & {[}3.42, 5.12{]}\tabularnewline
\hline 
\textbf{Only OECD, PPML} & 5.81 & - & {[}4.43, 7.41{]}\tabularnewline
\hline 
\textbf{Only non-OECD, PPML} & 4.49 & - & {[}3.17, 6.68{]}\tabularnewline
\hline 
\end{tabular}
\par\end{centering}
\caption{\protect\label{tab:W_ppml}Uncertainty quantification for productivity
parameters as in \citet{waugh2010international} using OLS and PPML.}
\end{table}

\subsection{Implementation details}

To compute the limiting marginal prior according to Theorem \ref{cor:Characterization},
since there are missing data I use $\delta_{\frac{\chi\left(\varrho\left(X_{k\ell}\right)\right)+\chi\left(\varrho\left(X_{\ell k}\right)\right)}{2}}$
when both $\left(k,\ell\right)$ and $\left(\ell,k\right)$ are observed,
$\delta_{\chi\left(\varrho\left(X_{k\ell}\right)\right)}$ when $\left(k,\ell\right)$
but not $\left(\ell,k\right)$ is observed, and $\delta_{\chi\left(\varrho\left(X_{\ell k}\right)\right)}$
when $\left(\ell,k\right)$ but not $\left(k,\ell\right)$ is observed.

\subsection{Alternative Methods}

\begin{figure}[h]
\centering{}\includegraphics[scale=0.16]{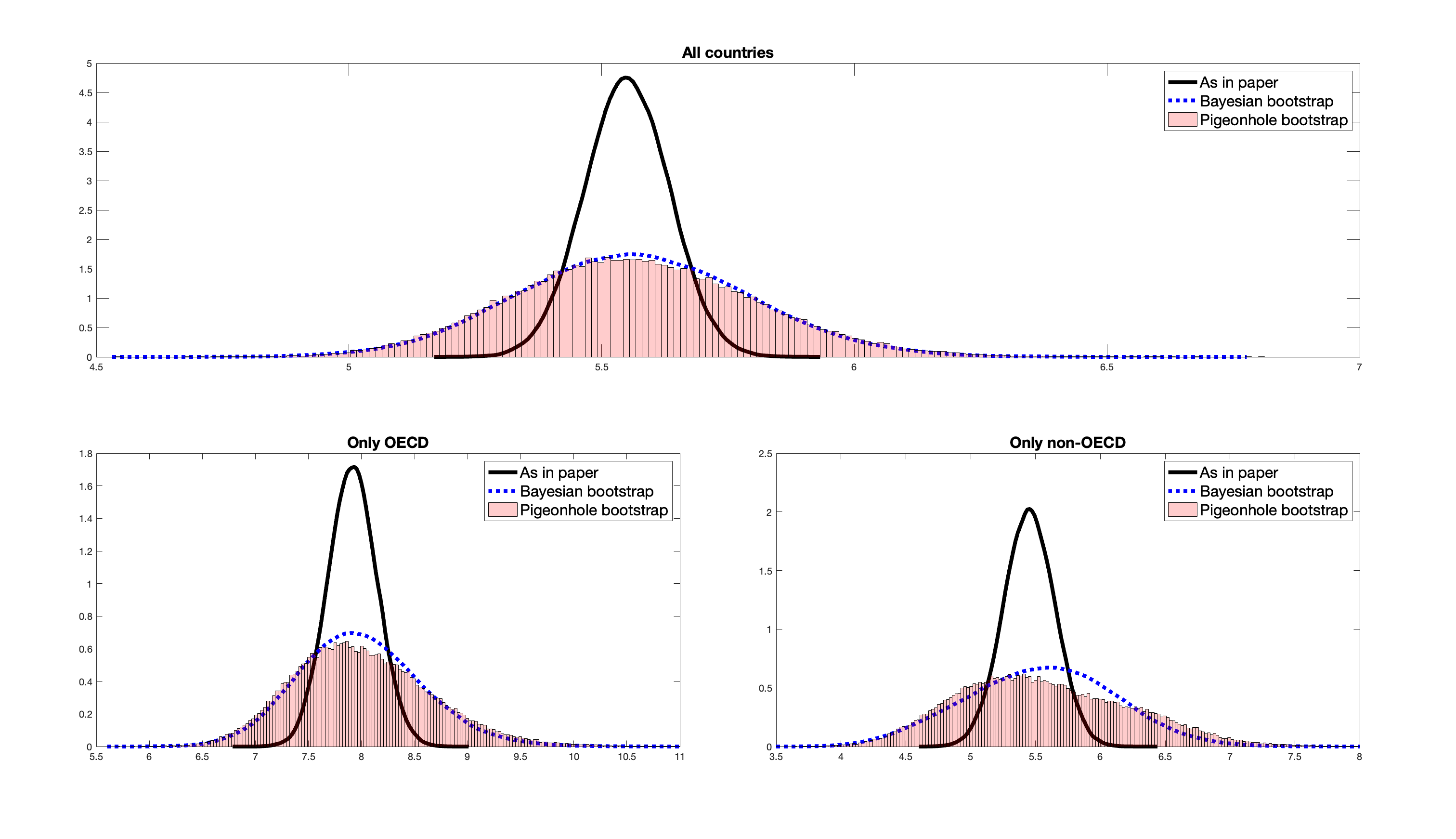}\caption{\protect\label{fig:W_pb}Distributions for productivity parameters
as in \citet{waugh2010international}.}
\end{figure}

Table \ref{tab:W_pb}, Figure \ref{fig:W_pb} and Table \ref{tab:W_cfs_pb}
reproduce Table \ref{tab:W}, Figure \ref{fig:W} and Table \ref{tab:W_cfs_bay},
respectively, but add the results corresponding to the pigeonhole
bootstrap from Section \ref{subsec:Pigeonhole}. There are some small
differences, especially for the non-OECD sample, but overall the economic
conclusions do not change. The approach using analytic standard errors
from Section \ref{subsec:Analytic} cannot be applied here because
a substantial share of the observations are missing. 

\begin{table}[h]
\begin{centering}
\begin{tabular}{|c|c|c|c|c|}
\hline 
 & \begin{cellvarwidth}[t]
\centering
Point 

estimate
\end{cellvarwidth} & \begin{cellvarwidth}[t]
\centering
As in 

paper
\end{cellvarwidth} & \begin{cellvarwidth}[t]
\centering
Bayesian 

bootstrap
\end{cellvarwidth} & \begin{cellvarwidth}[t]
\centering
Pigeonhole

bootstrap
\end{cellvarwidth}\tabularnewline
\hline 
\hline 
All countries, $n=43$ & 5.55 & {[}5.39, 5.71{]} & {[}5.12, 6.02{]} & {[}5.12, 6.05{]}\tabularnewline
\hline 
Only OECD, $n=19$ & 7.91 & {[}7.46, 8.37{]} & {[}6.91, 9.21{]} & {[}6.86, 9.41{]}\tabularnewline
\hline 
Only non-OECD, $n=24$ & 5.45 & {[}5.06, 5.84{]} & {[}4.42, 6.65{]} & {[}4.43, 6.91{]}\tabularnewline
\hline 
\end{tabular}
\par\end{centering}
\caption{\protect\label{tab:W_pb}Uncertainty quantification for productivity
parameters as in \citet{waugh2010international}.}
\end{table}
\begin{table}[h]
\begin{centering}
\begin{tabular}{|c|c|c|c|c|c|c|}
\hline 
 & \multicolumn{3}{c|}{Baseline} & \multicolumn{3}{c|}{Autarky}\tabularnewline
\hline 
 & \multicolumn{3}{c|}{$\tau_{ij}^{\mathrm{cf}}=\tau_{ij}$} & \multicolumn{3}{c|}{$\tau_{ij}^{\mathrm{cf}}=\infty\cdot\mathbb{I}\left\{ i\neq j\right\} $}\tabularnewline
\hline 
 & p.e. & b.b. & p.b. & p.e. & b.b. & p.b.\tabularnewline
\hline 
\hline 
\begin{cellvarwidth}[m]
\centering
Variance 

of log wages
\end{cellvarwidth} & 1.30 & {[}1.28, 1.32{]} & {[}1.28, 1.32{]} & 1.35 & {[}1.31, 1.38{]} & {[}1.31, 1.38{]}\tabularnewline
\hline 
\begin{cellvarwidth}[m]
\centering
90th/10th percentile 

of wages
\end{cellvarwidth} & 25.7 & {[}25.1, 26.2{]} & {[}25.1, 26.2{]} & 23.5 & {[}22.6, 24.2{]} & {[}22.6, 24.2{]}\tabularnewline
\hline 
\begin{cellvarwidth}[m]
\centering
Mean \% change 

in wages
\end{cellvarwidth} & - & - & - & -10.5 & {[}-11.4, -9.6{]} & {[}-11.4, -9.5{]}\tabularnewline
\hline 
\end{tabular}
\par\end{centering}
\begin{centering}
\begin{tabular}{|c|c|c|c|c|c|c|}
\hline 
 & \multicolumn{3}{c|}{Symmetry} & \multicolumn{3}{c|}{Free trade}\tabularnewline
\hline 
 & \multicolumn{3}{c|}{$\tau_{ij}^{\mathrm{cf}}=\min\left\{ \tau_{ij},\tau_{ji}\right\} $} & \multicolumn{3}{c|}{$\tau_{ij}^{\mathrm{cf}}=1$}\tabularnewline
\hline 
 & p.e. & b.b. & p.b. & p.e. & b.b. & p.b.\tabularnewline
\hline 
\hline 
\begin{cellvarwidth}[m]
\centering
Variance 

of log wages
\end{cellvarwidth} & 1.05 & {[}1.05, 1.05{]} & {[}1.05, 1.05{]} & 0.76 & {[}0.75, 0.78{]} & {[}0.75, 0.78{]}\tabularnewline
\hline 
\begin{cellvarwidth}[m]
\centering
90th/10th percentile 

of wages
\end{cellvarwidth} & 17.3 & {[}17.2, 17.4{]} & {[}17.2, 17.4{]} & 11.4 & {[}11.0, 11.9{]} & {[}11.0, 11.9{]}\tabularnewline
\hline 
\begin{cellvarwidth}[m]
\centering
Mean \% change 

in wages
\end{cellvarwidth} & 24.2 & {[}22.4, 25.8{]} & {[}22.3, 25.8{]} & 128.0 & {[}114.4, 140.7{]} & {[}113.5, 140.7{]}\tabularnewline
\hline 
\end{tabular}
\par\end{centering}
\caption{\protect\label{tab:W_cfs_pb}Bayesian uncertainty quantification for
counterfactual predictions as in \citet{waugh2010international}.
Here, \textquotedblleft p.e.\textquotedblright{} denotes point estimate,
\textquotedblleft b.b.\textquotedblright{} denotes Bayesian bootstrap,
and \textquotedblleft p.b.\textquotedblright{} denotes pigeonhole bootstrap.}
\end{table}

\section{\protect\label{subsec:PB_bootstrap_DGP}Comparing Methods using Pigeonhole
Bootstrap DGP}

Recall the model in Assumption \ref{assu:Model}: 
\begin{align*}
C_{1},...,C_{n}|h,\mathbb{P}_{C} & \overset{\mathrm{iid}}{\sim}\mathbb{P}_{C}\\
X_{ij} & =h\left(C_{i},C_{j}\right),\quad\mathrm{for\ }C_{i}\neq C_{j}.
\end{align*}
To test the performance of the various methods discussed in Section
\ref{sec:Alt_methods}, I will use a simulation DGP. Specifically,
consider the thought experiment where we observe the latent variables
$\left\{ C_{k}\right\} $; resample them \textit{with replacement};
and then construct the corresponding data. As summarized in Algorithm
\ref{alg:PB_bootstrap_DGP}, this corresponds exactly to the pigeonhole
bootstrap. After constructing such a dataset, we can use the various
available approaches and check whether the resulting confidence or
credible interval covers the structural estimator $\hat{\theta}$.
By repeating this procedure many times, we can compute the coverage
for each method. 
\begin{algorithm}[h]
\caption{\protect\label{alg:PB_bootstrap_DGP}Pigeonhole bootstrap DGP}

\begin{enumerate}
\item Input: Bilateral data $\left\{ X_{k\ell}\right\} _{k\neq\ell}$.
\item Sample $n$ units \textit{independently with replacement} from $\left\{ 1,...,n\right\} $
with equal probability. Let $W_{k}^{\mathrm{pb},\left(b\right)}$
denote the number of times that $k$ is sampled.
\item Construct a new dataset by replicating the observation $X_{k\ell}$
a specific number of times, namely $W_{k}^{\mathrm{pb},\left(b\right)}\cdot W_{\ell}^{\mathrm{pb},\left(b\right)}$,
for all $k\neq\ell$.
\end{enumerate}
\end{algorithm}
The coverage results for the structural estimators considered in the
body of the paper can be found in Table \ref{tab:PB_bootstrap_DGP}.
\begin{table}[h]
\begin{centering}
\begin{tabular}{|c|c|c|c|}
\hline 
 & \begin{cellvarwidth}[t]
\centering
As in

paper
\end{cellvarwidth} & \begin{cellvarwidth}[t]
\centering
Bayesian

bootstrap
\end{cellvarwidth} & \begin{cellvarwidth}[t]
\centering
Pigeonhole

bootstrap
\end{cellvarwidth}\tabularnewline
\hline 
\hline 
\citet{waugh2010international}, all countries, $n=43$ & 0.498 & 0.979 & 0.986\tabularnewline
\hline 
\citet{waugh2010international}, only OECD, $n=19$ & 0.533 & 0.954 & 0.984\tabularnewline
\hline 
\citet{waugh2010international}, only non-OECD, $n=24$ & 0.416 & 0.913 & 0.941\tabularnewline
\hline 
\citet{caliendo2015estimates}, $\theta^{1}$, $n=15$ & 0.295 & 0.911 & 0.991\tabularnewline
\hline 
\citet{caliendo2015estimates}, $\theta^{2}$, $n=13$ & 0.467 & 0.933 & 0.996\tabularnewline
\hline 
\citet{caliendo2015estimates}, $\theta^{3}$, $n=15$ & 0.360 & 0.950 & 0.999\tabularnewline
\hline 
\citet{caliendo2015estimates}, $\theta^{4}$, $n=14$ & 0.377 & 0.932 & 1.000\tabularnewline
\hline 
\citet{artucc2010trade}, $\mu$, $n=6$ & 0.733 & 0.897 & 0.884\tabularnewline
\hline 
\citet{artucc2010trade}, $\sigma^{2}$, $n=6$ & 0.825 & 0.925 & 0.902\tabularnewline
\hline 
\end{tabular}
\par\end{centering}
\caption{\protect\label{tab:PB_bootstrap_DGP}Coverage for approach used in
paper, Bayesian bootstrap and pigeonhole bootstrap using the pigeonhole
bootstrap DGP from Algorithm \ref{alg:PB_bootstrap_DGP}, using 1000
simulated datasets and $B=1000$. For \citet{waugh2010international}
and \citet{caliendo2015estimates}, I use heteroskedastic-robust standard
errors. For \citet{artucc2010trade} I used the standard GMM variance
formula, assuming only exchangeability across industries.}
\end{table}

\section{\protect\label{sec:Proofs}Proofs}

\subsection{Proof of Theorem \ref{prop:Bayesian_interpretation}}

The proof proceeds in five steps. The first three steps consider the
thought experiment where we observe the latent variables $\left\{ C_{k}\right\} $
and know the function $h$. The fourth and fifth step incorporate
that in practice we only observe $\left\{ X_{k\ell}\right\} _{k\neq\ell}$.

\paragraph{Finding $\pi_{\alpha}\left(\mathbb{P}_{C}|h,\left\{ C_{k}\right\} \right)$.}

Combining Equations \eqref{eq:DGP_C} and \eqref{eq:prior}, we know
from Theorem 4.6 in \citet{ghosal2017fundamentals} that the posterior
for $\mathbb{P}_{C}$ is
\begin{equation}
\pi_{\alpha}\left(\mathbb{P}_{C}|h,\left\{ C_{k}\right\} \right)=DP\left(\frac{\alpha}{\alpha+n}Q+\frac{n}{\alpha+n}\frac{1}{n}\sum_{k=1}^{n}\delta_{C_{k}},\alpha+n\right).\label{eq:posterior}
\end{equation}

\paragraph{Finding $\pi_{0}\left(\mathbb{P}_{C}|h,\left\{ C_{k}\right\} \right)$.}

Applying Theorem 4.16 in \citet{ghosal2017fundamentals} yields the
posterior distribution that sends the precision parameter $\alpha$
to zero:
\[
\pi_{0}\left(\mathbb{P}_{C}|h,\left\{ C_{k}\right\} \right)=\underset{\alpha\downarrow0}{\lim}\pi_{\alpha}\left(\mathbb{P}_{C}|h,\left\{ C_{k}\right\} \right)=DP\left(\sum_{k=1}^{n}\frac{1}{n}\cdot\delta_{C_{k}},n\right).
\]
Note that this posterior distribution is proper. Furthermore, a random
probability distribution $\mathbb{P}_{n,C}^{*}$ drawn from $\pi_{0}\left(\mathbb{P}_{C}|h,\left\{ C_{k}\right\} \right)$
is necessarily supported on the observation points $\left\{ C_{k}\right\} =\left\{ C_{1},...,C_{n}\right\} $.
Hence, by definition of a Dirichlet process, we have
\begin{align*}
\left(\mathbb{P}_{n,C}^{*}\left(C_{1}\right),....,\mathbb{P}_{n,C}^{*}\left(C_{n}\right)\right) & \sim\mathrm{Dir}\left(n;n\cdot\left(\sum_{k=1}^{n}\frac{1}{n}\cdot\delta_{C_{k}}\right)\left(C_{1}\right),...,n\cdot\left(\sum_{k=1}^{n}\frac{1}{n}\cdot\delta_{C_{k}}\right)\left(C_{n}\right)\right)\\
 & \sim\mathrm{Dir}\left(n;1,...,1\right).
\end{align*}
It follows that 
\begin{align}
 & \mathbb{P}_{n,C}^{*}\sim\pi_{0}\left(\mathbb{P}_{C}|h,\left\{ C_{k}\right\} \right)\nonumber \\
 & \Rightarrow\mathbb{P}_{n,C}^{*}=\sum_{k=1}^{n}W_{k}\cdot\delta_{C_{k}},\quad\left(W_{1},...,W_{n}\right)\sim\mathrm{Dir}\left(n;1,...,1\right),\label{eq:posterior_P_C}
\end{align}
and we have
\begin{equation}
Pr_{\pi_{0}}\left\{ C_{i}\in B|h,\left\{ C_{k}\right\} \right\} =\sum_{k=1}^{n}W_{k}\cdot\mathbb{I}\left\{ C_{k}\in B\right\} .\label{eq:posterior_bb_law_C}
\end{equation}

\paragraph{Finding $\pi_{0}\left(\mathbb{P}_{X_{ij}}|h,\left\{ C_{k}\right\} \right)$.}

Next, combining the model from Assumption \ref{assu:Model} and the
Bayesian bootstrap posterior in Equation \eqref{eq:posterior_P_C},
we find
\begin{align*}
Pr_{\pi_{0}}\left\{ X_{ij}\in A|h,\left\{ C_{k}\right\} \right\}  & =Pr_{\pi_{0}}\left\{ h\left(C_{i},C_{j}\right)\in A|C_{i}\neq C_{j},h,\left\{ C_{k}\right\} \right\} \\
 & =\frac{\mathbb{E}_{\pi_{0}}\left[\mathbb{I}\left\{ h\left(C_{i},C_{j}\right)\in A\right\} \cdot\mathbb{I}\left\{ C_{i}\neq C_{j}\right\} |h,\left\{ C_{k}\right\} \right]}{Pr_{\pi_{0}}\left\{ C_{i}\neq C_{j}|h,\left\{ C_{k}\right\} \right\} }\\
 & =\frac{\sum_{k\neq\ell}W_{k}\cdot W_{\ell}\cdot\mathbb{I}\left\{ h\left(C_{k},C_{\ell}\right)\in A\right\} }{1-\sum_{s}W_{s}^{2}}\\
 & =\frac{\sum_{k\neq\ell}W_{k}\cdot W_{\ell}\cdot\mathbb{I}\left\{ X_{k\ell}\in A\right\} }{\sum_{s\neq t}W_{s}\cdot W_{t}}.
\end{align*}
This implies that 
\begin{align}
 & \mathbb{P}_{n,X_{ij}}^{*}\sim\pi_{0}\left(\mathbb{P}_{X_{ij}}|h,\left\{ C_{k}\right\} \right)\nonumber \\
 & \Rightarrow\mathbb{P}_{n,X_{ij}}^{*}=\sum_{k\neq\ell}\frac{W_{k}\cdot W_{\ell}}{\sum_{s\neq t}W_{s}\cdot W_{t}}\cdot\delta_{X_{k\ell}},\quad\left(W_{1},...,W_{n}\right)\sim\mathrm{Dir}\left(n;1,...,1\right).\label{eq:pi_BB_P_Xij_Ck_h}
\end{align}
So we have found an expression for the Bayesian bootstrap posterior
for $\mathbb{P}_{X_{ij}}$ conditional on observing $\left\{ C_{k}\right\} $
and knowing the function $h$. 

\paragraph{Finding $\pi_{0}\left(\mathbb{P}_{X_{ij}}|\left\{ X_{k\ell}\right\} _{k\protect\neq\ell}\right)$.}

However, in practice we do not observe $\left\{ C_{k}\right\} $ and
the function $h$ is unknown. We only observe $\left\{ X_{k\ell}\right\} _{k\neq\ell}$,
so the posterior distribution of interest is $\pi_{0}\left(\mathbb{P}_{X_{ij}}|\left\{ X_{k\ell}\right\} _{k\neq\ell}\right)$.
But using the fact that draws from $\pi_{0}\left(\mathbb{P}_{X_{ij}}|h,\left\{ C_{k}\right\} \right)$
only depend on $\left\{ X_{k\ell}\right\} _{k\neq\ell}$, we can find
an expression for this posterior:
\begin{align*}
\pi_{0}\left(\mathbb{P}_{X_{ij}}|\left\{ X_{k\ell}\right\} _{k\neq\ell}\right) & =\int\pi_{0}\left(\mathbb{P}_{X_{ij}}|h,\left\{ C_{k}\right\} ,\left\{ X_{k\ell}\right\} _{k\neq\ell}\right)d\pi_{0}\left(h,\left\{ C_{k}\right\} |\left\{ X_{k\ell}\right\} _{k\neq\ell}\right)\\
 & =\int\pi_{0}\left(\mathbb{P}_{X_{ij}}|h,\left\{ C_{k}\right\} \right)d\pi_{0}\left(h,\left\{ C_{k}\right\} |\left\{ X_{k\ell}\right\} _{k\neq\ell}\right)\\
 & =\pi_{0}\left(\mathbb{P}_{X_{ij}}|h,\left\{ C_{k}\right\} \right)\int d\pi_{0}\left(h,\left\{ C_{k}\right\} |\left\{ X_{k\ell}\right\} _{k\neq\ell}\right)\\
 & =\pi_{0}\left(\mathbb{P}_{X_{ij}}|h,\left\{ C_{k}\right\} \right).
\end{align*}
The second equality follows from that knowing $\left(h,\left\{ C_{k}\right\} \right)$
implies knowing $\left\{ X_{k\ell}\right\} _{k\neq\ell}$. The third
equality follows from noting that in Equation \eqref{eq:pi_BB_P_Xij_Ck_h},
the posterior $\pi_{0}\left(\mathbb{P}_{X_{ij}}|h,\left\{ C_{k}\right\} \right)$
does not depend on $\left\{ C_{k}\right\} $ or $h$. In conclusion,
we have
\begin{align*}
 & \mathbb{P}_{n,X_{ij}}^{*}\sim\pi_{0}\left(\mathbb{P}_{X_{ij}}|\left\{ X_{k\ell}\right\} _{k\neq\ell}\right)\\
 & \Rightarrow\mathbb{P}_{n,X_{ij}}^{*}=\sum_{k\neq\ell}\frac{W_{k}\cdot W_{\ell}}{\sum_{s\neq t}W_{s}\cdot W_{t}}\cdot\delta_{X_{k\ell}},\quad\left(W_{1},...,W_{n}\right)\sim\mathrm{Dir}\left(n;1,...,1\right).
\end{align*}

\paragraph{Finding $\pi_{0}\left(\theta|\left\{ X_{k\ell}\right\} _{k\protect\neq\ell}\right)$.}

Lastly, since $\theta$ is a function of $\mathbb{P}_{X_{ij}}$, the
limiting posterior $\pi_{0}\left(\mathbb{P}_{X_{ij}}|\left\{ X_{k\ell}\right\} _{k\neq\ell}\right)$
also implies a limiting posterior on structural estimand, $\pi_{0}\left(\theta|\left\{ X_{k\ell}\right\} _{k\neq\ell}\right)$.
So indeed, the procedure from Algorithm \ref{alg:BB} has a Bayesian
interpretation.

\subsection{Proof of Theorem \ref{prop:Theoretical_motivation}}

\paragraph{Statement 1.}

The first part of Theorem \ref{prop:Theoretical_motivation} follows
from the derivations in the proof of Theorem \ref{prop:Bayesian_interpretation},
where we noted that the posterior $\pi_{0}\left(\mathbb{P}_{X_{ij}}|h,\left\{ C_{k}\right\} \right)$
did not depend on $h$ or $Q$, which implies the influences of $\pi\left(h\right)$
and the center measure on the posterior drop out when we take the
prior precision parameter to zero. Furthermore we can write
\begin{align*}
Pr_{\pi_{0}}\left\{ X_{ij}\in B|h,\left\{ C_{k}\right\} \right\}  & =\sum_{k\neq\ell}\frac{W_{k}\cdot W_{\ell}}{\sum_{s\neq t}W_{s}\cdot W_{t}}\cdot\mathbb{I}\left\{ X_{k\ell}\in B\right\} ,
\end{align*}
from which it follows that $\pi_{0}\left(\mathbb{P}_{X_{ij}}|\left\{ X_{k\ell}\right\} _{k\neq\ell}\right)$
does not smooth across events.

\paragraph{Statement 2.}

The second part of Theorem \ref{prop:Theoretical_motivation} builds
on Corollary 4.29 in \citet{ghosal2017fundamentals}, applied to the
prior $\pi\left(\mathbb{P}_{C}\right)$. This result states that if
for every $n$ and every measurable partition $\left\{ A_{1},...,A_{R_{C}}\right\} $
of $\mathcal{C}$, the vector $\left(\mathbb{P}_{n,C}^{*}\left(A_{1}\right),...,\mathbb{P}_{n,C}^{*}\left(A_{R_{C}}\right)\right)$,
where 
\[
\mathbb{P}_{n,C}^{*}\sim\pi\left(\mathbb{P}_{C}|h,\left\{ C_{k}\right\} \right),
\]
depends only on the counts $\left(N_{1}^{C},...,N_{R_{C}}^{C}\right)$,
for $N_{r}^{C}=\sum_{k=1}^{n}\mathbb{I}\left\{ C_{k}\in A_{r}\right\} $,
if and only if the prior $\pi\left(\mathbb{P}_{C}\right)$ is a Dirichlet
process or a trivial process. Consider a prior $\pi\left(h\right)$
that only puts probability mass on the function $h:\mathcal{C}^{2}\rightarrow\mathcal{X}$
defined by $h\left(C_{i},C_{j}\right)=C_{i}$. In that case $\mathcal{X}=\mathcal{C}$
and for a given partition $\left\{ B_{1},...,B_{R_{X}}\right\} $,
we have
\begin{align*}
 & \left(Pr_{\pi}\left\{ X_{ij}\in B_{1}|\left\{ X_{k\ell}\right\} _{k\neq\ell}\right\} ,...,Pr_{\pi}\left\{ X_{ij}\in B_{R_{X}}|\left\{ X_{k\ell}\right\} _{k\neq\ell}\right\} \right)\\
 & =\left(Pr_{\pi}\left\{ C_{i}\in B_{1}|\underbrace{C_{1},...,C_{1}}_{n-1\ \mathrm{times}},C_{2},...,C_{2},...,C_{n},...,C_{n}\right\} ,...,\right.\\
 & \qquad\left.Pr_{\pi}\left\{ C_{i}\in B_{R_{X}}|C_{1},...,C_{1},C_{2},...,C_{2},...,C_{n},...,C_{n}\right\} \right).
\end{align*}
We then know that this vector depends only on the counts $\left(N_{1}^{C},...,N_{R_{X}}^{C}\right)$
that use 
\[
\left\{ C_{1},...,C_{1},C_{2},...,C_{2},...,C_{n},...,C_{n}\right\} ,
\]
which we can relate back to $\left\{ X_{k\ell}\right\} _{k\neq\ell}$:
\begin{align*}
 & \left(N_{1}^{C},...,N_{R_{X}}^{C}\right)\\
 & =\left(\left(n-1\right)\cdot\sum_{k=1}^{n}\mathbb{I}\left\{ C_{k}\in B_{1}\right\} ,...,\left(n-1\right)\cdot\sum_{k=1}^{n}\mathbb{I}\left\{ C_{k}\in B_{R_{X}}\right\} \right)\\
 & =\left(\sum_{k\neq\ell}\mathbb{I}\left\{ X_{k\ell}\in B_{1}\right\} ,...,\sum_{k\neq\ell}\mathbb{I}\left\{ X_{k\ell}\in B_{R_{X}}\right\} \right).
\end{align*}

\subsection{Proof of Theorem \ref{cor:Characterization}}

We first have to check whether the limit exists. We hence require
the function $T:\Delta\left(\mathcal{X}\right)\rightarrow\Theta$
to be well-behaved in some sense. To formalize this, denote the function
that maps a given empirical distribution supported on $\left\{ C_{k}\right\} $
to an element of the parameter space by $T_{C}:\Delta\left(\left\{ C_{k}\right\} \right)\mapsto\Theta$.
I require this function to be well-behaved when evaluated on weighted
empirical distributions where the weights approach a degenerate limit.
\begin{assumption}[Condition for existence of limiting marginal prior]
\label{assu:Marginal_prior_conditions}For any permutation $\sigma:\left\{ 1,...,n\right\} \rightarrow\left\{ 1,...,n\right\} $
and any sequence $\left\{ \omega_{k,w}\right\} _{w}\in\Delta\left(\left\{ 1,..,n\right\} \right)$
with $\omega_{k,w}>0$ and $\lim_{w\rightarrow\infty}\omega_{k+1,w}/\omega_{k,w}=0$
for all $k$, we have that 
\[
\underset{w\rightarrow\infty}{\lim}T_{C}\left(\sum_{k=1}^{n}\omega_{\sigma\left(k\right),w}\delta_{C_{\sigma\left(k\right)}}\right)=\bar{T}_{C}\left(C_{\sigma\left(1\right)},...,C_{\sigma\left(n\right)}\right),
\]
for some limit $\bar{T}_{C}\left(\cdot\right)$.
\end{assumption}
Under Assumption \ref{assu:Marginal_prior_conditions}, the limiting
marginal prior as $\alpha\downarrow0$ takes a simple form:
\begin{lem}[Existence]
\label{prop:Existence} Under Assumptions \ref{assu:Model} and \ref{assu:Marginal_prior_conditions},
and using the Dirichlet process prior 
\[
\mathbb{P}_{C}\sim DP\left(\sum_{k=1}^{n}\frac{1}{n}\cdot\delta_{C_{k}},\alpha\right),
\]
as $\alpha\downarrow0$ the implied marginal prior $\pi\left(\theta\right)$
converges weakly to $\pi^{\infty}\in\Delta\left(\Theta\right)$, where
\[
\pi^{\infty}\left(\theta\right)=\sum_{\sigma\in S}\frac{1}{n!}\mathbb{I}\left\{ \bar{T}_{C}\left(C_{\sigma\left(1\right)},C_{\sigma\left(2\right)},...,C_{\sigma\left(n\right)}\right)=\theta\right\} ,
\]
for $S$ the set of permutations $\sigma:\left\{ 1,...,n\right\} \rightarrow\left\{ 1,...,n\right\} $.
\end{lem}
Lemma \ref{prop:Existence} shows existence of the limiting marginal
prior. For the class of estimators that can be written as functions
of means, we can actually characterize the limiting object. For this
class, we have
\[
T_{C}\left(\mathbb{P}_{C}\right)=\chi\left(\mathbb{E}_{C_{i},C_{j}\sim\mathbb{P}_{C}|C_{i}\neq C_{j}}\left[\varrho\left(h\left(C_{i},C_{j}\right)\right)\right]\right),
\]
and since
\[
T_{C}\left(\sum_{k=1}^{n}\omega_{\sigma\left(k\right),w}\delta_{C_{\sigma\left(k\right)}}\right)=\chi\left(\sum_{k\neq\ell}\frac{\omega_{\sigma\left(k\right),w}\cdot\omega_{\sigma\left(\ell\right),w}}{\sum_{s\neq t}\omega_{\sigma\left(s\right),w}\cdot\omega_{\sigma\left(t\right),w}}\cdot\varrho\left(h\left(C_{\sigma\left(k\right)},C_{\sigma\left(\ell\right)}\right)\right)\right),
\]
Assumption \ref{assu:Marginal_prior_conditions} is satisfied for
\[
\bar{T}\left(c_{1},...,c_{n}\right)=\frac{\chi\left(\varrho\left(h\left(c_{1},c_{2}\right)\right)\right)+\chi\left(\varrho\left(h\left(c_{2},c_{1}\right)\right)\right)}{2}.
\]
This is the case because we have that $\sum_{k\neq\ell}\frac{\omega_{k,w}\cdot\omega_{\ell,w}}{\sum_{s\neq t}\omega_{s,w}\cdot\omega_{t,w}}=1$
and for $k>\ell$ we have 
\begin{align*}
\frac{\omega_{k,w}\cdot\omega_{\ell,w}}{\sum_{s\neq t}\omega_{s,w}\cdot\omega_{t,w}} & =\frac{\omega_{1,w}\cdot\omega_{2,w}}{\sum_{s\neq t}\omega_{s,w}\cdot\omega_{t,w}}\underbrace{\frac{\omega_{k,w}}{\omega_{k-1,w}}\cdots\frac{\omega_{3,w}}{\omega_{2,w}}}_{\rightarrow0}\cdot\underbrace{\frac{\omega_{\ell,w}}{\omega_{\ell-1,w}}\cdots\frac{\omega_{2,w}}{\omega_{1,w}}}_{\rightarrow0}.
\end{align*}
Applying Lemma \ref{prop:Existence} implies that the marginal prior
limit $\pi^{\infty}\left(\theta_{}\right)$ equals 
\[
\frac{2}{n\left(n-1\right)}\sum_{k>\ell}\delta_{\frac{\chi\left(\varrho\left(X_{k\ell}\right)\right)+\chi\left(\varrho\left(X_{\ell k}\right)\right)}{2}}.
\]

\subsection{Proof of Lemma \ref{prop:Existence}}

The proof follows that of Theorem 3 in \citet{andrews2024bootstrap}.
The stick-breaking representation of Dirichlet processes (see e.g.
Theorem 4.12 of \citealp{ghosal2017fundamentals}) implies that we
can write draws from the prior $\pi\left(\mathbb{P}_{C}\right)$ as
\[
\mathbb{P}_{C}=\sum_{m=1}^{\infty}V_{m}\left(\alpha\right)\delta_{\tilde{C}_{m}},
\]
where the random variables $\tilde{C}_{m}$ are drawn i.i.d. from
$\sum_{k=1}^{n}\frac{1}{n}\cdot\delta_{C_{k}}$, and 
\[
V_{m}\left(\alpha\right)=\left(1-U_{m}^{\frac{1}{\alpha}}\right)\prod_{r=1}^{m-1}U_{r}^{\frac{1}{\alpha}}
\]
where the random variables $U_{r}$ are i.i.d. standard uniform. Note
that $Pr\left\{ U_{r}\in\left(0,1\right)\ \mathrm{for\ all}\ r\right\} =1$,
and that conditional on this event $V_{m}\left(\alpha\right)\in\left(0,1\right)$
for all $m$ and all $\alpha>0$, while $V_{m+1}\left(\alpha\right)/V_{m}\left(\alpha\right)\rightarrow0$
as $\alpha\downarrow0$.

Let $\tau\left(1\right)\in\left\{ 1,...,n\right\} $ be the index
for the observation in the original (latent) data with $C_{\tau\left(1\right)}=\tilde{C}_{1}$.
For $r\in\left\{ 2,...,n\right\} $, let $s\left(r\right)$ be the
smallest $s$ such that $\tilde{C}_{s}$ is distinct from $\left\{ C_{\tau\left(1\right)},...,C_{\tau\left(r-1\right)}\right\} $,
and let $C_{\tau\left(r\right)}=\tilde{C}_{s\left(r\right)}$. We
can then equivalently write
\[
\mathbb{P}_{C}=\sum_{k=1}^{n}\omega_{k}\left(\tau,\alpha\right)\delta_{C_{\tau\left(k\right)}},\quad\omega_{k}\left(\tau,\alpha\right)=\sum_{m=1}^{\infty}V_{m}\left(\alpha\right)\mathbb{I}\left\{ \tilde{C}_{m}=C_{\tau\left(k\right)}\right\} .
\]
By construction $\mathbb{P}_{C}\in\Delta\left(\left\{ C_{k}\right\} \right)$,
and $\omega_{k}\left(\tau,\alpha\right)\in\left(0,1\right)$ with
probability one for all $\alpha>0$. Moreover, as $\alpha\downarrow0$
we have $\omega_{k+1}\left(\tau,\alpha\right)/\omega_{k}\left(\tau,\alpha\right)\rightarrow0$
for all $k$, so 
\[
\underset{\alpha\downarrow0}{\lim}T_{C}\left(\mathbb{P}_{C}\right)=\underset{\alpha\downarrow0}{\lim}T_{C}\left(\sum_{k=1}^{n}\omega_{k}\left(\tau,\alpha\right)\delta_{C_{\tau\left(k\right)}}\right)=\bar{T}_{C}\left(C_{\tau\left(1\right)},C_{\tau\left(2\right)},...,C_{\tau\left(n\right)}\right)
\]
by Assumption \ref{assu:Marginal_prior_conditions}. The fact that
we have to multiply by $1/n!$ then follows from the definition of
$\tau$. 

\subsection{Proof of Theorem \ref{thm:empirical_process_conv}}

Recall the definitions of $\mathbb{G}_{n}$ and $\mathbb{G}_{n}^{*}$,
now using Dirichlet draws $\left(W_{1},...,W_{n}\right)\sim\mathrm{Dir}\left(n;1,...,1\right)$
instead of exponential draws:
\begin{align}
\mathbb{G}_{n}f & =\sqrt{n}\left\{ \sum_{k\neq\ell}\frac{1}{n\left(n-1\right)}\cdot f\left(X_{k\ell}\right)-\mathbb{E}_{\mathbb{P}_{X_{ij}}}\left[f\left(X_{ij}\right)\right]\right\} \nonumber \\
\mathbb{G}_{n}^{*}f & =\sqrt{n}\left\{ \sum_{k\neq\ell}\frac{W_{k}\cdot W_{\ell}}{\sum_{s\neq t}W_{s}\cdot W_{t}}\cdot f\left(X_{k\ell}\right)-\sum_{k\neq\ell}\frac{1}{n\left(n-1\right)}\cdot f\left(X_{k\ell}\right)\right\} .\label{eq:G_n*}
\end{align}
We have the following lemma for U-processes based on \citet{arcones1993limit}
and \citet{zhang2001bayesian}: 
\begin{lem}[Weak convergence of empirical processes of U-processes]
\label{lem:U-processes}Let $\tilde{\mathcal{F}}\subseteq\left(\mathcal{C}^{2}\right)^{\mathbb{R}}$
be a measurable class of \textup{symmetric} functions, and let 
\[
C_{1},...,C_{n}\overset{\mathrm{iid}}{\sim}\mathbb{P}_{C}.
\]
The U-process based on $\mathbb{P}_{C}$ and indexed by $\tilde{\mathcal{F}}$
is
\[
U_{n}\left(\tilde{f}\right)=\sum_{k\neq\ell}\frac{1}{n\left(n-1\right)}\cdot\tilde{f}\left(C_{k},C_{\ell}\right).
\]
Suppose that:
\begin{enumerate}
\item[(i)] $\tilde{\mathcal{F}}$ is permissible (see page 196 in \citealp{pollard1984convergence})
and admits a positive envelope $\tilde{F}$ with $\mathbb{P}_{C}\tilde{F}^{2}<\infty$.
\item[(ii)] We have non-degeneracy, meaning that 
\[
\Cov\left(\tilde{f}_{1}\left(C_{1},C_{2}\right),\tilde{f}_{2}\left(C_{1},C_{2^{'}}\right)\right)>0\ \forall\tilde{f}_{1},\tilde{f}_{2}\in\tilde{\mathcal{F}}.
\]
\item[(iii)] There exist $0<c,v<\infty$ such that for every $\epsilon>0$ and
probability measure $\tilde{Q}$ with $\tilde{Q}\tilde{F}^{2}<\infty$,
we have 
\[
N\left(\epsilon\left\Vert \tilde{F}\right\Vert _{L_{2}\left(\tilde{Q}\right)},\tilde{\mathcal{F}},\left\Vert \cdot\right\Vert _{L_{2}\left(\tilde{Q}\right)}\right)\leq c\epsilon^{-v}.
\]
Then, defining the empirical processes
\begin{align*}
\tilde{\mathbb{G}}_{n}\tilde{f} & =\sqrt{n}\left\{ \sum_{k\neq\ell}\frac{1}{n\left(n-1\right)}\cdot\tilde{f}\left(C_{k},C_{\ell}\right)-\mathbb{E}_{\mathbb{P}_{C}}\left[\tilde{f}\left(C_{i},C_{j}\right)\right]\right\} \\
\tilde{\mathbb{G}}_{n}^{*}\tilde{f} & =\sqrt{n}\left\{ \sum_{k\neq\ell}W_{k}\cdot W_{\ell}\cdot\tilde{f}\left(C_{k},C_{\ell}\right)-\sum_{k\neq\ell}\frac{1}{n\left(n-1\right)}\cdot\tilde{f}\left(C_{k},C_{\ell}\right)\right\} ,
\end{align*}
we have weak convergence over $\ell^{\infty}\left(\tilde{\mathcal{F}}\right)$
of both $\tilde{\mathbb{G}}_{n}$ and $\tilde{\mathbb{G}}_{n}^{*}$
to the same centered Gaussian process $\tilde{\mathbb{G}}$ with covariance
kernel
\[
\tilde{K}\left(\tilde{f}_{1},\tilde{f}_{2}\right)=4\Cov\left(\tilde{f}_{1}\left(C_{1},C_{2}\right),\tilde{f}_{2}\left(C_{1},C_{2^{'}}\right)\right),
\]
where the convergence of $\tilde{\mathbb{G}}_{n}^{*}$ holds conditional
on the data $\left\{ C_{k}\right\} $ and outer almost surely. 
\end{enumerate}
\end{lem}
To use this lemma, first note that as $n\rightarrow\infty$, we can
ignore the normalization term in the denominator of Equation \eqref{eq:G_n*},
because
\[
\sum_{s\neq t}W_{s}\cdot W_{t}=1-\sum_{s=1}^{n}W_{s}^{2}\overset{p}{\rightarrow}1.
\]
The convergence in probability follows since $\mathbb{E}\left[\sum_{s=1}^{n}W_{s}^{2}\right]=\frac{2}{n+1}$,
and convergence in mean to zero for a non-negative random variable
implies convergence in probability.

Next, note that we can always ``symmetrize'' a sum of non-symmetric
functions since
\[
\sum_{k\neq\ell}f\left(X_{k\ell}\right)=\sum_{k\neq\ell}\frac{f\left(X_{k\ell}\right)+f\left(X_{\ell k}\right)}{2}.
\]
This symmetrization implies that the relevant covariance kernel is
\[
K\left(f_{1},f_{2}\right)=\Cov\left(f_{1}\left(X_{12}\right)+f_{1}\left(X_{21}\right),f_{2}\left(X_{12^{'}}\right)+f_{2}\left(X_{2^{'}1}\right)\right).
\]
Given these observations, it remains to check that Assumption \ref{assu:Complexity}
implies that we can apply Lemma \ref{lem:U-processes} with $\tilde{\mathcal{F}}=\mathcal{F}\circ h$.

Towards that end, note that $\mathcal{F}$ being permissible implies
$\mathcal{F}\circ h$ is permissible for measurable $h$. The existence
of the positive integrable envelope function $F$ for $\mathcal{F}$
implies the existence of a positive integrable envelope function $F\circ h$
for $\mathcal{F}\circ h$, since for any $\left(f\circ h\right)\in\mathcal{F}\circ h$
and $\left(c_{1},c_{2}\right)\in\mathcal{C}^{2}$,
\begin{align*}
 & \left|\left(f\circ h\right)\left(c_{1},c_{2}\right)\right|=\left|f\left(x_{12}\right)\right|\leq F\left(x_{12}\right)=\left(F\circ h\right)\left(c_{1},c_{2}\right)\\
 & \left(F\circ h\right)\left(c_{1},c_{2}\right)=F\left(x_{12}\right)>0\\
 & \mathbb{P}_{C}\left(F\circ h\right)^{2}=\mathbb{P}_{X_{ij}}F^{2}<\infty.
\end{align*}
Non-degeneracy is satisfied because
\begin{align*}
 & \Cov\left(\left(f_{1}\circ h\right)\left(C_{1},C_{2}\right)+\left(f_{1}\circ h\right)\left(C_{2},C_{1}\right),\left(f_{2}\circ h\right)\left(C_{1},C_{2}^{'}\right)+\left(f_{2}\circ h\right)\left(C_{2}^{'},C_{1}\right)\right)\\
 & =\Cov\left(f_{1}\left(X_{12}\right)+f_{1}\left(X_{21}\right),f_{2}\left(X_{12^{'}}\right)+f_{2}\left(X_{2^{'}1}\right)\right)>0.
\end{align*}
Lastly, for each $\tilde{Q}\in\Delta\left(\mathcal{C}\right)$ such
that $\tilde{Q}\left(F\circ h\right)^{2}<\infty$, there exists a
$Q\in\Delta\left(\mathcal{X}\right)$ such that $QF^{2}<\infty$ and
\[
\left\Vert F\circ h\right\Vert _{L_{2}\left(\tilde{Q}\right)}=\left\Vert F\right\Vert _{L_{2}\left(Q\right)}.
\]
This implies we have
\[
N\left(\epsilon\left\Vert F\circ h\right\Vert _{L_{2}\left(\tilde{Q}\right)},\mathcal{F}\circ h,\left\Vert \cdot\right\Vert _{L_{2}\left(\tilde{Q}\right)}\right)=N\left(\epsilon\left\Vert F\right\Vert _{L_{2}\left(Q\right)},\mathcal{F},\left\Vert \cdot\right\Vert _{L_{2}\left(Q\right)}\right)\leq c\epsilon^{-v}.
\]

\subsection{Proof of Lemma \ref{lem:U-processes}}

The lemma follows almost directly from combining Theorem 4.9 in \citet{arcones1993limit}
and Corollary 1 in \citet{zhang2001bayesian}. Condition (iii) in
Lemma \ref{lem:U-processes} differs from Condition 2 in \citet{zhang2001bayesian},
as the author assumes $\tilde{\mathcal{F}}$ has polynomial discrimination
rather than the condition that the covering numbers are bounded by
a polynomial in $1/\varepsilon$. However, in the proofs of Theorem
2.1 and Corollary 1 of \citet{zhang2001bayesian}, polynomial discrimination
is only used to bounded covering numbers using Lemma II.25 and II.36
in \citet{pollard1984convergence}. So assuming the more familiar
bound on the covering numbers directly is without loss.

\subsection{Proof of Theorem \ref{prop:estimator_conv}}

From Theorem \ref{thm:empirical_process_conv} we know that under
Assumptions \ref{assu:Complexity}, $\mathbb{G}_{n}$ defined by $\mathbb{G}_{n}f=\sqrt{n}\left\{ \mathbb{P}_{n,X_{ij}}f-\mathbb{P}_{X_{ij}}f\right\} $
converges unconditionally in distribution to a tight random element
$\mathbb{G}$, and $\mathbb{G}_{n}^{*}$ defined by $\mathbb{G}_{n}^{*}f=\sqrt{n}\left\{ \mathbb{P}_{n,X_{ij}}^{*}f-\mathbb{P}_{n,X_{ij}}f\right\} $
converges, conditionally given $\left\{ X_{k\ell}\right\} _{k\neq\ell}$
and outer almost surely, to the same random element. This implies
\[
\sup_{\kappa\in\mathrm{BL}_{1}}\left|\mathbb{E}\left[\kappa\left(\mathbb{G}_{n}^{*}\right)|\left\{ X_{k\ell}\right\} _{k\neq\ell}\right]-\mathbb{E}\left[\kappa\left(\mathbb{G}\right)\right]\right|\overset{\mathrm{as}*}{\rightarrow}0,
\]
for $\mathrm{BL}_{1}$ the set of bounded Lipschitz functions from
$\ell^{\infty}\left(\mathcal{F}\right)$ to $\left[0,1\right]$. Since
$\hat{\theta}$ is assumed to be of the form $T\left(\mathbb{P}_{n,X_{ij}}\right)=\varphi\left(\mathbb{P}_{n,X_{ij}}f\right)$
for $f\in\mathcal{F}$, the result then follows by applying the functional
delta method for the bootstrap, Theorem 23.9 in \citet{van2000asymptotic}. 

\subsection{Proof of Corollary \ref{cor:Z-estimators}}

The relevant function class is $\mathcal{F}_{Z}\equiv\left\{ \nu_{\vartheta,\eta}:\left(\vartheta,\eta\right)\in\Theta\times\mathcal{H}\right\} $.
Now, $\varphi:\ell^{\infty}\left(\mathcal{F}_{Z}\right)\mapsto\Theta$
is the map that extracts the zero from the estimating equation, so
that we have, 
\begin{align*}
\theta & =T\left(\mathbb{P}_{X_{ij}}\right)=\varphi\left(\Psi\right)\equiv\varphi\left(\underset{\eta\in\mathcal{H}}{\sup}\left|\mathbb{P}_{X_{ij}}\nu_{\vartheta,\eta}\right|\right)\\
\hat{\theta} & =T\left(\mathbb{P}_{n,X_{ij}}\right)=\varphi\left(\Psi_{n}\right)\equiv\varphi\left(\underset{\eta\in\mathcal{H}}{\sup}\left|\mathbb{P}_{n,X_{ij}}\nu_{\vartheta,\eta}\right|\right)\\
\hat{\theta}^{*} & =T\left(\mathbb{P}_{n,X_{ij}}^{*}\right)=\varphi\left(\Psi_{n}^{*}\right)\equiv\varphi\left(\underset{\eta\in\mathcal{H}}{\sup}\left|\mathbb{P}_{n,X_{ij}}^{*}\nu_{\vartheta,\eta}\right|\right).
\end{align*}
The proof follows Corollary 13.6 in \citet{kosorok2008introduction}.
From Theorem 13.5 in \citet{kosorok2008introduction}, we know that
conditions (i)-(iii) are sufficient conditions for Fréchet differentiability
of $\varphi$. Conditions (iv) and (v) are regularity conditions.
Condition (vi) guarantees convergence weak convergence of the empirical
processes using Theorem \ref{thm:empirical_process_conv}. We can
then apply Theorem \ref{prop:estimator_conv} and the result follows.

\subsection{Proof of Theorem \ref{thm:Delta_method_random_data}}

We can Taylor expand $\hat{\gamma}=\left(\left\{ X_{k\ell}\right\} _{k\neq\ell},\hat{\theta}\right)$
around $\theta$ to find
\[
\sqrt{n}\left(\hat{\gamma}-\gamma\right)=G\left(\bar{\theta}\right)\sqrt{n}\left(\hat{\theta}-\theta\right),
\]
for $G\left(\cdot\right)=\nabla_{\theta}g\left(\left\{ X_{k\ell}\right\} _{k\neq\ell},\cdot\right)$
and $\bar{\theta}$ an intermediate value between $\hat{\theta}$
and $\theta$. The gradient term is random because it depends on the
data $\left\{ X_{k\ell}\right\} _{k\neq\ell}$. However, under the
condition in Equation \eqref{eq:uniform_convergence}, we have that
$G\left(\hat{\theta}\right)=G\left(\bar{\theta}\right)+o_{p}\left(1\right)$.
This leads to the approximation 
\[
\frac{\sqrt{n}\left(\hat{\gamma}-\gamma\right)}{\sqrt{G\left(\hat{\theta}\right)^{2}\hat{\Sigma}}}\overset{d}{\approx}\mathcal{N}\left(0,1\right),
\]
and the result follows.

\section{\protect\label{sec:Extra_results_Marginal_prior}Extra Results for
Limiting Marginal Prior}

Since the Dirichlet process prior in Equation \eqref{eq:DP_C_marg}
is only supported on $\left\{ C_{k}\right\} $, we have
\begin{align*}
\left(\mathbb{P}_{C}^ {}\left(C_{1}\right),....,\mathbb{P}_{C}^ {}\left(C_{n}\right)\right) & \sim\mathrm{Dir}\left(n;\alpha\cdot\sum_{k=1}^{n}\frac{1}{n}\cdot\delta_{C_{k}}\left(C_{1}\right),...,\alpha\cdot\sum_{k=1}^{n}\frac{1}{n}\cdot\delta_{C_{k}}\left(C_{n}\right)\right)\\
 & \sim\mathrm{Dir}\left(n;\frac{\alpha}{n},...,\frac{\alpha}{n}\right),
\end{align*}
which collapses to the Bayesian bootstrap \textit{posterior} in Equation
\eqref{eq:posterior_P_X} by setting $\alpha=n$. From an analogous
argument as in the proof of Theorem \ref{prop:Bayesian_interpretation},
it now follows that for a given choice of $\alpha$ we can sample
from the corresponding marginal prior for $\mathbb{P}_{X_{ij}}$:
\begin{align*}
 & \mathbb{P}_{X_{ij}}\sim\pi\left(\mathbb{P}_{X_{ij}}\right)\\
 & \Rightarrow\mathbb{P}_{X_{ij}}=\sum_{k\neq\ell}\frac{W_{k}\cdot W_{\ell}}{\sum_{s\neq t}W_{s}\cdot W_{t}}\cdot\delta_{X_{k\ell}},\quad\left(W_{1},...,W_{n}\right)\sim\mathrm{Dir}\left(n;\frac{\alpha}{n},...,\frac{\alpha}{n}\right).
\end{align*}
Since $\theta=T\left(\mathbb{P}_{X_{ij}}\right)$, a marginal prior
for $\theta$ is also implied for a given choice of $\alpha$, as
is summarized in Algorithm \ref{alg:Marginal_prior}.
\begin{algorithm}[h]
\caption{\protect\label{alg:Marginal_prior}A marginal prior of $\theta$ along
the uninformative limit sequence}

\begin{enumerate}
\item Input: Bilateral data $\left\{ X_{k\ell}\right\} _{k\neq\ell}$, estimator
function $T:\Delta\left(\mathcal{X}\right)\rightarrow\Theta$ and
prior precision $\alpha$.
\item For each draw $b=1,...,B$:
\begin{enumerate}
\item Sample $\left(V_{1}^{\left(b\right)},...,V_{n}^{\left(b\right)}\right)\overset{\mathrm{iid}}{\sim}\mathrm{Ga}\left(\frac{\alpha}{n},1\right)$.
\item Construct $\omega_{k\ell}^{\left(b\right)}=V_{k}^{\left(b\right)}\cdot V_{\ell}^{\left(b\right)}/\left(\sum_{s\neq t}V_{s}^{\left(b\right)}\cdot V_{t}^{\left(b\right)}\right)$,
for $k,\ell=1,...,n$. 
\item Compute 
\[
\hat{\theta}^{*,\left(b\right)}=T\left(\sum_{k\neq\ell}\omega_{k\ell}^{\left(b\right)}\cdot\delta_{X_{k\ell}}\right).
\]
\end{enumerate}
\item Plot the histogram $\left\{ \hat{\theta}^{*,\left(1\right)},...,\hat{\theta}^{*,\left(B\right)}\right\} $.
\end{enumerate}
\end{algorithm}

\section{\protect\label{sec:Other_extensions}Other Extensions}

\subsection{Multiway Clustering}

One might want to incorporate another dimension of clustering. For
example, in addition to country-heterogeneity one might want to add
sector-heterogeneity or time-heterogeneity. We can accommodate this
by using an additional, separate exchangeability assumption. 

To illustrate, if the observed data $\left\{ X_{k\ell,s}\right\} _{k\neq\ell,s}$
has a time component and we separately want to allow for clustering
across time periods, we would sample
\begin{align*}
\left(W_{1}^{\left(b\right)},...,W_{n}^{\left(b\right)}\right) & \sim\mathrm{Dir}\left(n;1,...,1\right)\\
\left(\check{W}_{1}^{\left(b\right)},...,\check{W}_{T}^{\left(b\right)}\right) & \sim\mathrm{Dir}\left(T;1,...,1\right),
\end{align*}
and compute bootstrap draws according to
\[
\hat{\theta}^{*,\left(b\right)}=T\left(\sum_{k\neq\ell,s}\frac{W_{k}^{\left(b\right)}\cdot W_{\ell}^{\left(b\right)}}{\sum_{u\neq v}W_{u}^{\left(b\right)}\cdot W_{v}^{\left(b\right)}}\cdot\check{W}_{s}^{\left(b\right)}\cdot\delta_{X_{k\ell,s}}\right).
\]
In the model, adding another dimension of heterogeneity corresponds
to independently sampling another set of latent variables. For the
example where we also have time-heterogeneity, we have
\begin{align*}
C_{1},...,C_{n}|h,\mathbb{P}_{C},\mathbb{P}_{\check{C}} & \overset{\mathrm{iid}}{\sim}\mathbb{P}_{C}\\
\check{C}_{1},...,\check{C}_{T}|h,\mathbb{P}_{C},\mathbb{P}_{\check{C}} & \overset{\mathrm{iid}}{\sim}\mathbb{P}_{\check{C}}\\
X_{ij,t} & =h\left(C_{i},C_{j},\check{C}_{t}\right),\quad\mathrm{for\ }C_{i}\neq C_{j}\ \mathrm{for\ each\ }t,
\end{align*}
with corresponding priors 
\[
\left(h,\mathbb{P}_{C},\mathbb{P}_{\check{C}}\right)\sim\pi\left(h\right)\cdot DP\left(Q_{C},\alpha_{C}\right)\cdot DP\left(Q_{\check{C}},\alpha_{\check{C}}\right).
\]

\subsection{Conditional Exchangeability}

The key underlying model assumption, as discussed in Section \ref{subsec:Model},
is that latent characteristics of the units are drawn i.i.d. from
some distribution, or that ``units are exchangeable''. One might
believe this exchangeability assumption only conditional on a set
of covariates. For example one might argue latent characteristics
of countries are only i.i.d. within continent or within trade agreement.
In this case, there exist different ``types'' within which agents
are exchangeable. 

To illustrate, with two types we would independently sample 
\begin{align*}
\left(W_{1}^{\left(b\right)},...,W_{n_{1}}^{\left(b\right)}\right) & \sim\mathrm{Dir}\left(n_{1};1,...,1\right)\\
\left(W_{n_{1}+1}^{\left(b\right)},...,W_{n_{1}+n_{2}}^{\left(b\right)}\right) & \sim\mathrm{Dir}\left(n_{2};1,...,1\right),
\end{align*}
and compute bootstrap draws according to
\[
\hat{\theta}^{*,\left(b\right)}=T\left(\sum_{k=1}^{n_{1}+n_{2}}\sum_{\ell=1,\ell\neq k}^{n_{1}+n_{2}}\frac{W_{k}^{\left(b\right)}\cdot W_{\ell}^{\left(b\right)}}{\sum_{s=1}^{n_{1}+n_{2}}\sum_{t=1,t\neq s}^{n_{1}+n_{2}}W_{s}^{\left(b\right)}\cdot W_{t}^{\left(b\right)}}\cdot\delta_{X_{k\ell}}\right).
\]
The corresponding model is
\begin{align*}
C_{1},...,C_{n_{1}}|h,\mathbb{P}_{C_{1}},\mathbb{P}_{C_{2}} & \overset{\mathrm{iid}}{\sim}\mathbb{P}_{C_{1}}\\
C_{n_{1}+1},...,C_{n_{1}+n_{2}}|h,\mathbb{P}_{C_{1}},\mathbb{P}_{C_{2}} & \overset{\mathrm{iid}}{\sim}\mathbb{P}_{C_{2}}\\
X_{ij} & =h\left(C_{i},C_{j}\right),\quad\mathrm{for\ }C_{i}\neq C_{j},
\end{align*}
and the priors change to
\[
\left(h,\mathbb{P}_{C_{1}},\mathbb{P}_{C_{2}}\right)\sim\pi\left(h\right)\cdot DP\left(Q_{1},\alpha_{1}\right)\cdot DP\left(Q_{2},\alpha_{2}\right).
\]
As in Section \ref{subsec:AH}, we can again motivate the model using
an Aldous-Hoover representation. Now, $\left\{ X_{ij}\right\} _{i,j\in\mathbb{N},i\neq j}$
is assumed to be \textit{relatively exchangeable} with respect to
``types'' $R$, which means that there exist subpopulations within
which agents are exchangeable. Then, 
\[
\left\{ X_{ij}\right\} _{i,j\in\mathbb{N},i\neq j}\overset{d}{=}\left\{ X_{\sigma_{R}\left(i\right)\sigma_{R}\left(j\right)}\right\} _{i,j\in\mathbb{N},i\neq j},
\]
for any within-type relabeling operation $\sigma_{R}:\mathbb{N}\rightarrow\mathbb{N}$.
\citet{graham2020dyadic} uses results from \citet{crane2018relatively}
to show that in this case there exists another array $\left\{ X_{ij}^{*}\right\} _{i,j\in\mathbb{N},i\neq j}$
generated according to
\begin{equation}
X_{ij}^{*}=\tilde{h}^{AH}\left(U,R_{i},R_{j},C_{i},C_{j},D_{ij}\right),\label{eq:h_AH_tilde-1}
\end{equation}
for $U,\left\{ C_{i}\right\} ,\left\{ D_{ij}\right\} \overset{\mathrm{iid}}{\sim}U\left[0,1\right],$
such that 
\[
\left\{ X_{ij}\right\} _{i,j\in\mathbb{N},i\neq j}\overset{d}{=}\left\{ X_{ij}^{*}\right\} _{i,j\in\mathbb{N},i\neq j}.
\]

\section{\protect\label{sec:Extra_results}Extra Results for Applications}

\subsection{\protect\label{subsec:Extra_results_CP}Extra Results for Application
1: \citet{caliendo2015estimates}}

\subsubsection{Details for Implementation}

For the Bayesian bootstrap procedure, for each draw I impose a lower
bound of $\min\left\{ \hat{\theta}^{s}\right\} =1.67$ to ensure the
code runs. I also follow the authors and replace the elasticity for
sectors ``Auto'' and ``Other Transport'' by the average elasticity
of the other sectors.

\subsubsection{Marginal Priors on $\left\{ \theta^{s}\right\} $}

\begin{figure}[h]
\centering{}\includegraphics[scale=0.16]{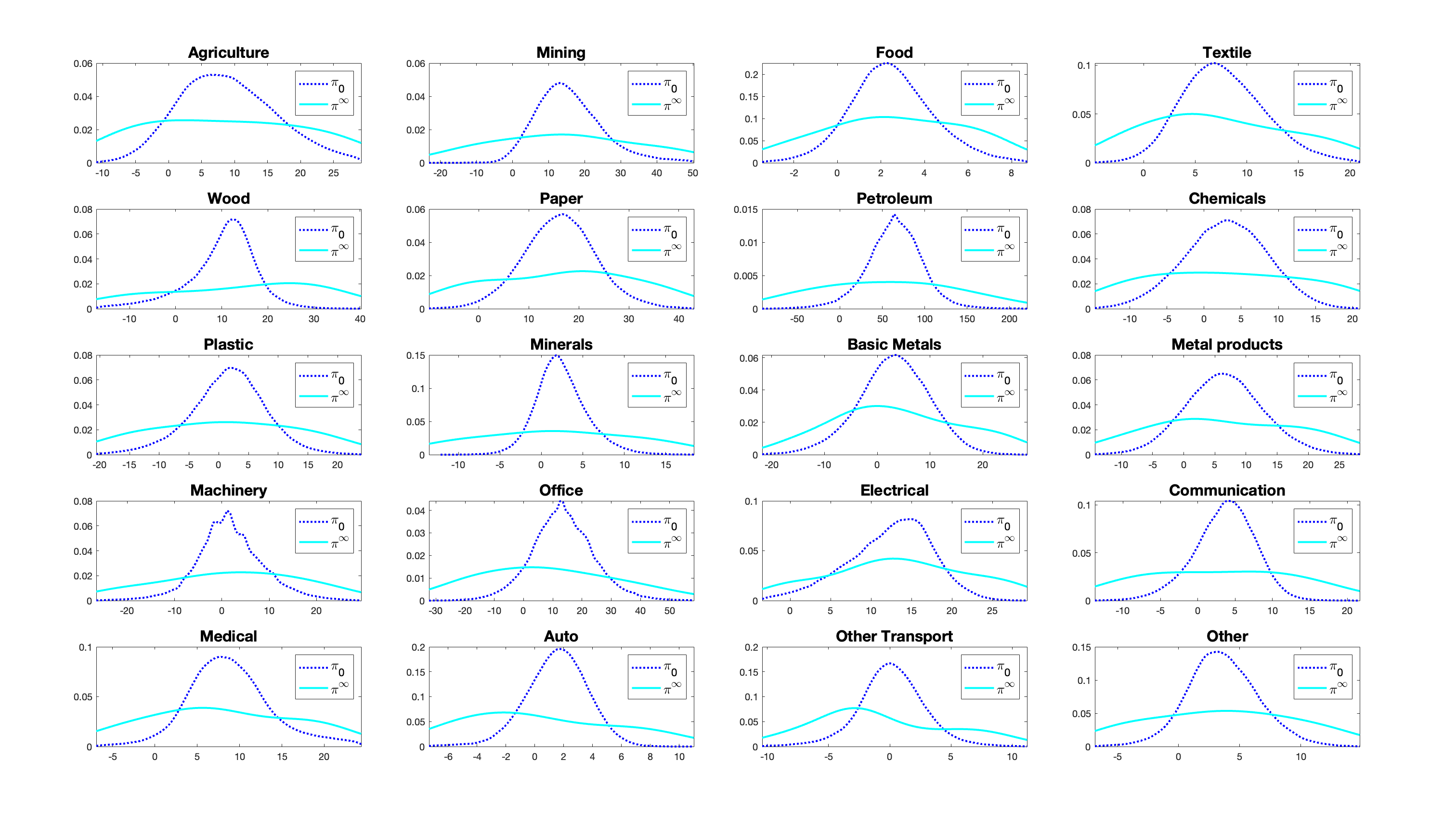}\caption{\protect\label{fig:CP_marg}Limiting marginal prior for trade elasticities
as in \citet{caliendo2015estimates}.}
\end{figure}

We can use Theorem \ref{cor:Characterization} with 
\[
\varrho\left(X_{k\ell m}\right)=\left(\begin{array}{c}
\log\left(\frac{F_{k\ell}^{s}F_{\ell m}^{s}F_{mk}^{s}}{F_{\ell k}^{s}F_{m\ell}^{s}F_{km}^{s}}\right)^{2}\\
-\log\left(\frac{F_{k\ell}^{s}F_{\ell m}^{s}F_{mk}^{s}}{F_{\ell k}^{s}F_{m\ell}^{s}F_{km}^{s}}\right)\cdot\log\left(\frac{t_{k\ell}^{s}t_{\ell m}^{s}t_{mk}^{s}}{t_{\ell k}^{s}t_{m\ell}^{s}t_{km}^{s}}\right)
\end{array}\right),\ \chi\left(\left(\begin{array}{c}
a_{1}\\
a_{2}
\end{array}\right)\right)=\frac{a_{2}}{a_{1}},
\]
and continuity of $\chi$ is satisfied. Figure \ref{fig:CP_marg}
plots the bootstrap posterior and the limiting marginal prior using
Theorem \ref{cor:Characterization}. For almost all cases, the marginal
priors have some outliers in the right tail, so I only consider (normalized)
prior mass within ten standard deviations. We observe that the prior
is much flatter than the bootstrap posterior. 

\subsubsection{Alternative Methods}

\begin{figure}[h]
\centering{}\includegraphics[scale=0.16]{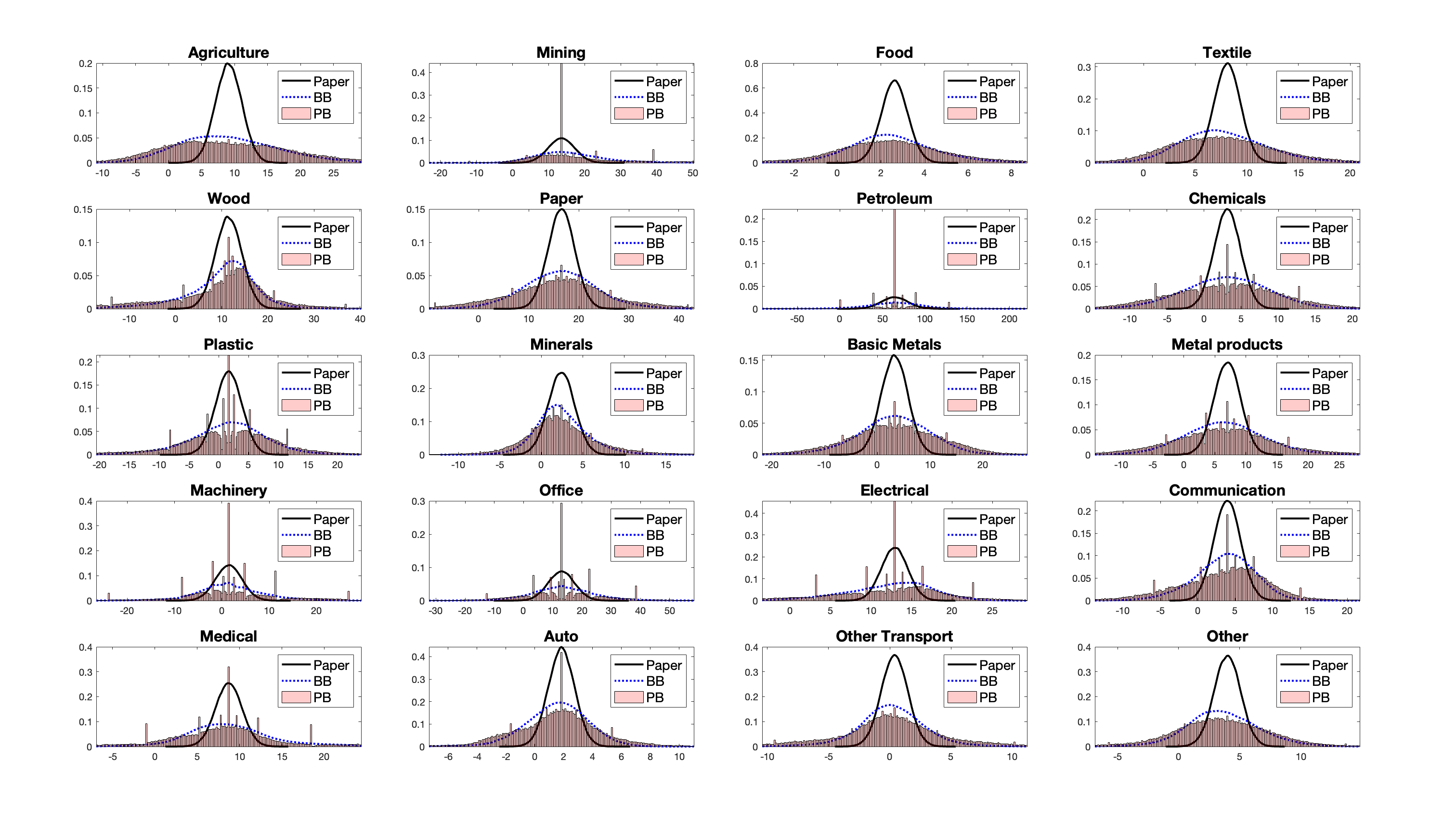}\caption{\protect\label{fig:CP_pb}Distributions of the benchmark estimates
(which remove the countries with the lowest 1\% share of trade for
each sector) in Table 1 of \citet{caliendo2015estimates}. \textquotedblleft Paper\textquotedblright{}
corresponds to the normal approximation as implied by the standard
errors reported in the paper. \textquotedblleft BB\textquotedblright{}
and \textquotedblleft PB\textquotedblright{} correspond to the Bayesian
bootstrap posterior and pigeonhole bootstrap distribution, respectively.}
\end{figure}

Table \ref{tab:CP_pb}, Figure \ref{fig:CP_pb} and Table \ref{tab:CP_cfs_pb}
reproduce Table \ref{tab:CP}, Figure \ref{fig:CP} and Table \ref{tab:CP_cfs},
respectively, but add the results corresponding to the pigeonhole
bootstrap from Section \ref{subsec:Pigeonhole}. The confidence intervals
for the sectoral elasticities constructed using the pigeonhole bootstrap
are consistently larger and all include zero. This results also in
that the confidence intervals for the welfare predictions are somewhat
larger, especially the upper bound on the welfare effect of Mexico.
The approach using analytic standard errors from Section \ref{subsec:Analytic}
cannot be applied here because data are triadic and a small share
of observations are missing. 

\begin{table}[h]
\begin{centering}
\begin{tabular}{|c|c|c|c|c|}
\hline 
 & Point estimate & As in paper & Bayesian bootstrap & Pigeonhole bootstrap\tabularnewline
\hline 
\hline 
Agriculture & 9.11 & {[}5.17, 13.05{]} & {[}-4.05, 25.63{]} & {[}-7.98, 30.19{]}\tabularnewline
\hline 
Mining & 13.53 & {[}6.34, 20.73{]} & {[}0.69, 42.35{]} & {[}-0.89, 71.13{]}\tabularnewline
\hline 
Food & 2.62 & {[}1.43, 3.81{]} & {[}-1.26, 6.83{]} & {[}-3.88, 8.17{]}\tabularnewline
\hline 
Textile & 8.10 & {[}5.58, 10.61{]} & {[}0.52, 16.76{]} & {[}-2.43, 18.95{]}\tabularnewline
\hline 
Wood & 11.50 & {[}5.87, 17.12{]} & {[}-11.30, 22.88{]} & {[}-30.90, 31.29{]}\tabularnewline
\hline 
Paper & 16.52 & {[}11.33, 21.71{]} & {[}1.70, 31.32{]} & {[}-7.18, 39.60{]}\tabularnewline
\hline 
Petroleum & 64.44 & {[}33.84, 95.04{]} & {[}-6.41, 128.87{]} & {[}-30.93, 128.88{]}\tabularnewline
\hline 
Chemicals & 3.13 & {[}-0.37, 6.62{]} & {[}-8.49, 13.72{]} & {[}-11.59, 17.50{]}\tabularnewline
\hline 
Plastic & 1.67 & {[}-2.69, 6.03{]} & {[}-12.65, 14.01{]} & {[}-20.33, 18.74{]}\tabularnewline
\hline 
Minerals & 2.41 & {[}-0.72, 5.55{]} & {[}-3.17, 9.47{]} & {[}-5.99, 13.42{]}\tabularnewline
\hline 
Basic Metals & 3.28 & {[}-1.64, 8.19{]} & {[}-11.32, 15.91{]} & {[}-15.76, 20.16{]}\tabularnewline
\hline 
Metal products & 6.99 & {[}2.82, 11.15{]} & {[}-5.75, 19.46{]} & {[}-11.41, 25.91{]}\tabularnewline
\hline 
Machinery & 1.45 & {[}-4.04, 6.93{]} & {[}-12.75, 17.24{]} & {[}-22.56, 26.82{]}\tabularnewline
\hline 
Office & 12.95 & {[}4.07, 21.83{]} & {[}-7.71, 36.25{]} & {[}-14.35, 45.52{]}\tabularnewline
\hline 
Electrical & 12.91 & {[}9.70, 16.12{]} & {[}0.20, 21.37{]} & {[}-5.35, 25.43{]}\tabularnewline
\hline 
Communication & 3.95 & {[}0.48, 7.43{]} & {[}-5.25, 10.98{]} & {[}-10.96, 14.68{]}\tabularnewline
\hline 
Medical & 8.71 & {[}5.65, 11.78{]} & {[}-0.66, 26.37{]} & {[}-10.96, 14.68{]}\tabularnewline
\hline 
Auto & 1.84 & {[}0.04, 3.64{]} & {[}-3.80, 5.48{]} & {[}-46.82, 10.08{]}\tabularnewline
\hline 
Other Transport & 0.39 & {[}-1.73, 2.51{]} & {[}-5.84, 5.67{]} & {[}-13.30, 10.14{]}\tabularnewline
\hline 
Other & 3.98 & {[}1.86, 6.11{]} & {[}-2.11, 9.68{]} & {[}-6.80, 11.83{]}\tabularnewline
\hline 
\end{tabular}
\par\end{centering}
\caption{\protect\label{tab:CP_pb}Uncertainty quantification for the benchmark
estimates (which remove the countries with the lowest 1\% share of
trade for each sector) in Table 1 of \citet{caliendo2015estimates}.}
\end{table}
\begin{table}[h]
\begin{centering}
\begin{tabular}{|c|c|c|c|}
\hline 
 & Point estimate & Bayesian bootstrap & Bayesian bootstrap\tabularnewline
\hline 
\hline 
Mexico & 1.31\% & {[}0.65\%, 2.51\%{]} & {[}0.68\%, 3.38\%{]}\tabularnewline
\hline 
Canada & -0.06\% & {[}-0.10\%, -0.02\%{]} & {[}-0.10\%, -0.01\%{]}\tabularnewline
\hline 
U.S. & 0.08\% & {[}0.07\%, 0.11\%{]} & {[}0.07\%, 0.13\%{]}\tabularnewline
\hline 
\end{tabular}
\par\end{centering}
\caption{\protect\label{tab:CP_cfs_pb}Bayesian uncertainty quantification
for welfare effects as in Table 2 of \citet{caliendo2015estimates}.}
\end{table}

\subsection{\protect\label{subsec:Extra_results_ACM}Extra Results for Application
2: \citet{artucc2010trade}}

\subsubsection{Details for Implementation}

In a small number of counterfactual draws, the welfare effects are
complex numbers. I omit these draws. 

\subsubsection{\protect\label{subsec:ACM_iter_GMM}Iterated GMM and Different Weight
Matrix}

In \citet{artucc2010trade}, the authors use a different weight matrix,
namely 
\[
\hat{\Omega}^{\mathrm{ACM}}\left(\theta\right)=\left(\left\{ \frac{1}{n\left(n-1\right)T}\sum_{k\neq\ell,s}e_{k\ell,s}\left(\theta\right)^{2}\right\} \left\{ \frac{1}{n\left(n-1\right)T}\sum_{k\neq\ell,s}Z_{k\ell,s}Z_{k\ell,s}^{'}\right\} \right)^{-1},
\]
for the residuals 
\[
e_{ij,t}\left(\theta\right)=\left(Y_{ij,t}-\left(\begin{array}{ccc}
\frac{\zeta-1}{\sigma^{2}}\mu & \frac{\zeta}{\sigma^{2}} & \zeta\end{array}\right)R_{ij,t}\right).
\]
They also use iterated GMM rather than two-step GMM. Using a different
weight matrix and a different GMM estimator results in slightly different
point estimates. However, the iterated GMM estimator satisfies Assumption
\ref{assu:theta_tilde} and, analogous to the discussion in Section
\ref{subsec:Theory_Asymptotics_MRGMM}, \citet{imbens1997one} shows
it can be combined into a single just-identified system, so that the
frequentist guarantees also hold. The resulting Bayesian bootstrap
algorithm that only assumes exchangeability across industries is summarized
in Algorithm \ref{alg:GMM_iter_BB}. 

\begin{figure}[h]
\centering{}\includegraphics[scale=0.16]{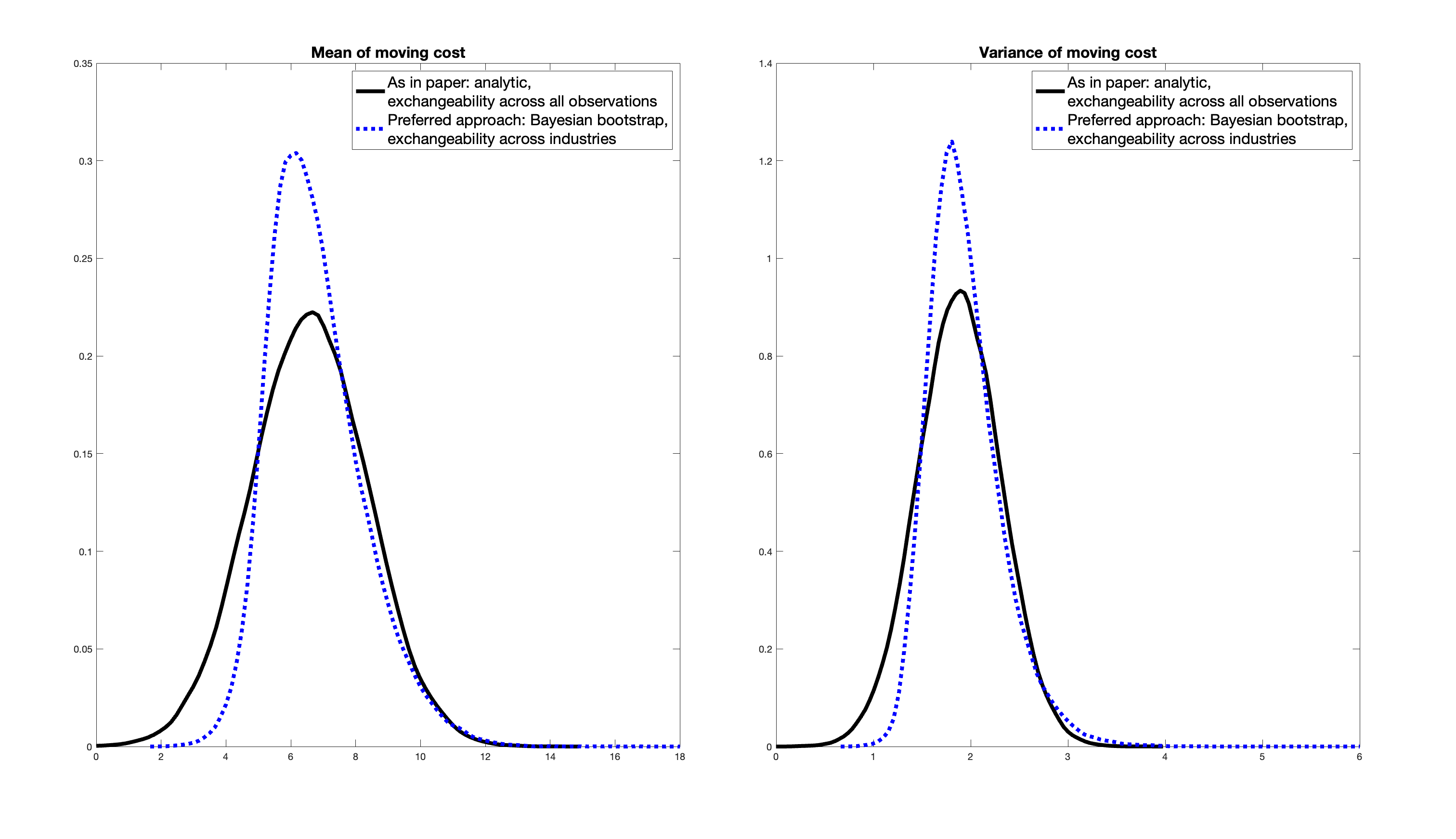}\caption{\protect\label{fig:ACM_iter}Distribution of estimators for Panel
IV in Table 3 in \citet{artucc2010trade} for $\zeta=0.97$.}
\end{figure}

\begin{algorithm}[h]
\caption{\protect\label{alg:GMM_iter_BB}Bayesian bootstrap procedure for iterated
GMM and different weight matrix}

\begin{enumerate}
\item For each bootstrap draw $b=1,...,B$:
\begin{enumerate}
\item Sample $\left(V_{1}^{\left(b\right)},...,V_{6}^{\left(b\right)}\right)\overset{\mathrm{iid}}{\sim}\mathrm{Exp}\left(1\right)$.
\item Compute $\omega_{k\ell,s}^{\left(b\right)}=V_{k}^{\left(b\right)}\cdot V_{\ell}^{\left(b\right)}/\left(\sum_{s=1}^{T}\sum_{u\neq v}V_{u}^{\left(b\right)}\cdot V_{v}^{\left(b\right)}\right)$
for $k,\ell=1,...,n$ and $s=1,...,T$. 
\item Denote
\[
\mathbb{P}_{n,X_{ij,t}}^{*,\left(b\right)}=\sum_{k\neq\ell,s}\omega_{k\ell,s}^{\left(b\right)}\cdot\delta_{X_{k\ell,s}}.
\]
\item Set $\hat{\Omega}_{\left(0\right)}^{\left(b\right)}=I_{3}$.
\item Until convergence, compute
\begin{enumerate}
\item $\hat{\theta}_{\left(w+1\right)}^{\left(b\right)}=\underset{\vartheta\in\Theta}{\arg\min}\mathbb{E}_{\mathbb{P}_{n,X_{ij,t}}^{*,\left(b\right)}}\left[e_{ij,t}\left(\vartheta\right)Z_{ij,t}\right]^{'}\hat{\Omega}_{\left(w\right)}^{\left(b\right)}\mathbb{E}_{\mathbb{P}_{n,X_{ij,t}}^{*,\left(b\right)}}\left[e_{ij,t}\left(\vartheta\right)Z_{ij,t}\right]$
\item $\hat{\Omega}_{\left(w+1\right)}^{\left(b\right)}=\mathbb{E}_{\mathbb{P}_{n,X_{ij,t}}^{*,\left(b\right)}}\left[e_{ij,t}\left(\hat{\theta}_{\left(w+1\right)}^{\left(b\right)}\right)^{2}\right]\mathbb{E}_{\mathbb{P}_{n,X_{ij,t}}^{*,\left(b\right)}}\left[Z_{ij,t}Z_{ij,t}^{'}\right]$.
\end{enumerate}
\end{enumerate}
\item Report the quantiles of interest of $\left\{ \hat{\theta}^{*,\left(1\right)},...,\hat{\theta}^{*,\left(B\right)}\right\} $.
\end{enumerate}
\end{algorithm}
Table \ref{tab:ACM_iter} reports the resulting coverage and credible
intervals. Figure \ref{fig:ACM_iter} plots the implied normal distributions
by the point estimate and t-statistic reported in the paper, and the
posteriors obtained by Algorithm \ref{alg:GMM_iter_BB}. 

\begin{table}[h]
\begin{centering}
\begin{tabular}{|c|c|c|}
\hline 
 & Mean & Variance\tabularnewline
\hline 
\hline 
Point estimate & 6.56 & 1.88\tabularnewline
\hline 
\begin{cellvarwidth}[m]
\centering
As in paper: analytic  errors,

exchangeability across all observations
\end{cellvarwidth} & {[}3.06, 10.07{]} & {[}1.04, 2.72{]}\tabularnewline
\hline 
\begin{cellvarwidth}[m]
\centering
Preferred approach: Bayesian bootstrap,

exchangeability across industries
\end{cellvarwidth} & {[}4.47, 10.09{]} & {[}1.35, 2.83{]}\tabularnewline
\hline 
\end{tabular}
\par\end{centering}
\caption{\protect\label{tab:ACM_iter}Uncertainty quantification for Panel
IV in Table 3 in \citet{artucc2010trade} for $\zeta=0.97$.}
\end{table}

\subsubsection{Alternative Methods}

In principle one could apply the approach using analytic standard
errors from Section \ref{subsec:Analytic} by interpreting the two-step
GMM estimator as a just-identified GMM estimator as per the discussion
in Section \ref{subsec:Theory_Asymptotics_MRGMM}. However, since
this amounts to taking many numerical derivatives, the procedure is
unstable. 

\section{\protect\label{sec:ST}Application 3: \citet{silva2006log}}

\begin{figure}[h]
\centering{}\includegraphics[scale=0.16]{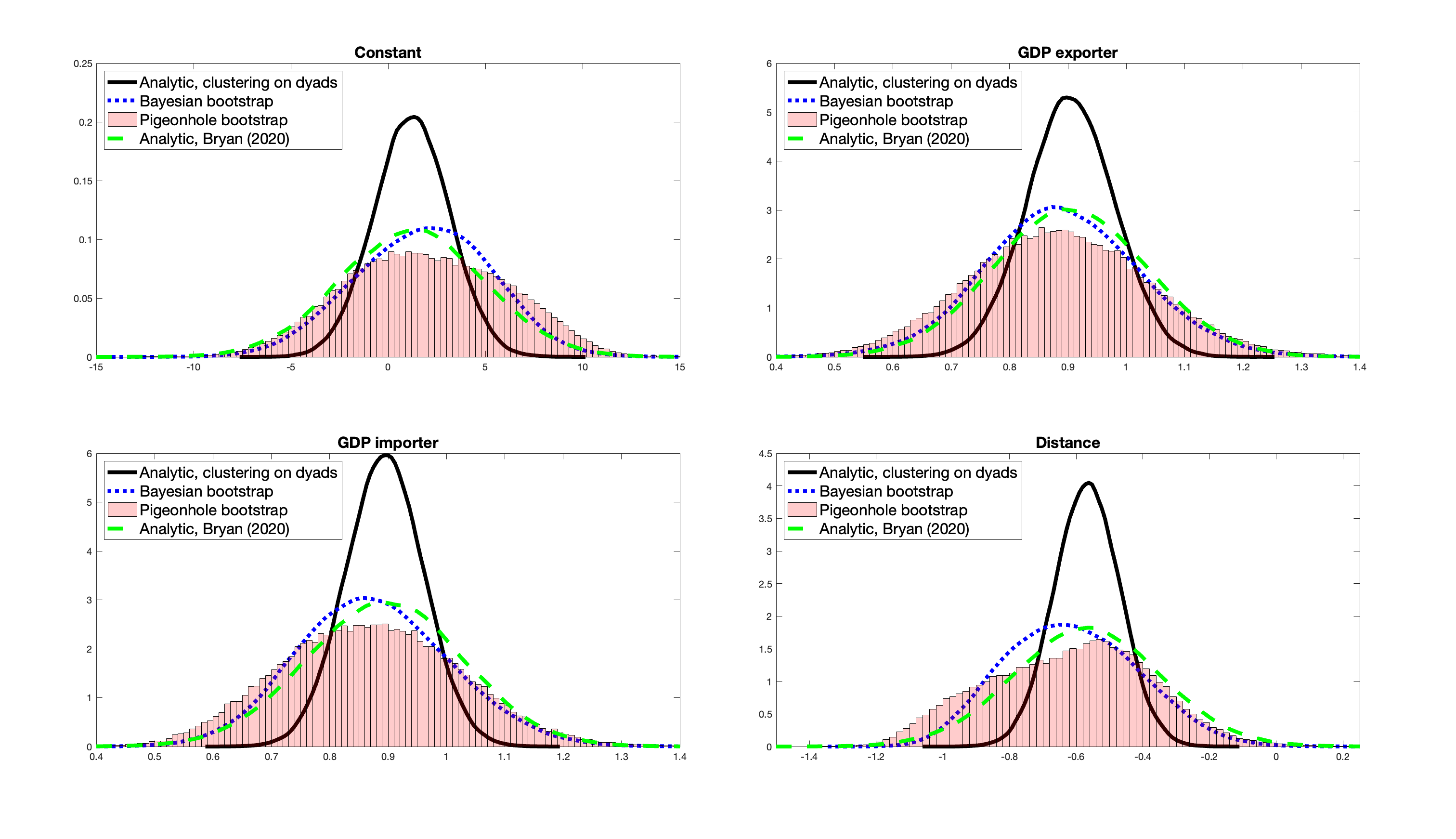}\caption{\protect\label{fig:ST}Distribution of estimators when regressing
bilateral trade flows on a constant, log exporter and importer GDP,
and log distance using PPML.}
\end{figure}

I follow the empirical illustrations in \citet{graham2020dyadic}
and \citet{davezies2021empirical}, which both consider the dyadic
PPML regression from \citet{silva2006log}. Specifically, I consider
the fitted regression function of bilateral trade flows $F_{k\ell}$
on a constant, the exporter's log GDP, the importer's log GDP and
the log distance. By taking the first order condition of the log likelihood,
we can obtain the sample moment condition
\[
\mathbb{E}_{\mathbb{P}_{n,X_{ij}}}\left[\left(F_{ij}-\exp\left\{ \left(\begin{array}{cccc}
1 & \mathrm{GDP}_{i} & \mathrm{GDP}_{j} & \mathrm{dist}_{ij}\end{array}\right)\theta\right\} \right)\left(\begin{array}{cccc}
1 & \mathrm{GDP}_{i} & \mathrm{GDP}_{j} & \mathrm{dist}_{ij}\end{array}\right)^{'}\right]=0.
\]
The basic specification has $n=136$ countries so $n\cdot\left(n-1\right)=18,360$
bilateral trade flows. For each of the four regression coefficients,
I compute a coverage or credible interval in Table \ref{tab:ST} and
plot the resulting posterior or implied distributions in Figure \ref{fig:ST}
using (i) naive analytic standard errors that cluster on dyads; (ii)
the Bayesian bootstrap procedure in Algorithm \ref{alg:BB}; (iii)
the pigeonhole-type bootstrap in Algorithm \ref{alg:PB}; and (iv)
analytic standard errors from Proposition \ref{prop:Bryan_variance}.
Reassuringly, all methods yield comparable results for uncertainty
quantification. 
\begin{table}[h]
\begin{centering}
\begin{tabular}{|c|c|c|c|c|}
\hline 
 & Constant & Exporter GDP & Importer GDP & Distance\tabularnewline
\hline 
\hline 
Point estimate & 1.22 & 0.90 & 0.89 & -0.57\tabularnewline
\hline 
\begin{cellvarwidth}[m]
\centering
Analytic,

clustering on dyads
\end{cellvarwidth} & {[}-2.58, 5.02{]} & {[}0.76, 1.05{]} & {[}0.76, 1.02{]} & {[}-0.76, -0.38{]}\tabularnewline
\hline 
Bayesian bootstrap & {[}-5.12, 8.34{]} & {[}0.63, 1.16{]} & {[}0.63, 1.14{]} & {[}-0.97, -0.21{]}\tabularnewline
\hline 
Pigeonhole bootstrap & {[}-5.77, 9.69{]} & {[}0.59, 1.19{]} & {[}0.58, 1.18{]} & {[}-1.08, -0.17{]}\tabularnewline
\hline 
Analytic, \citet{graham2020dyadic}  & {[}-5.99, 8.43{]} & {[}0.65, 1.16{]} & {[}0.63, 1.16{]} & {[}-1.00, -0.14{]}\tabularnewline
\hline 
\end{tabular}
\par\end{centering}
\caption{\protect\label{tab:ST}Uncertainty quantification when regressing
bilateral trade flows on a constant, log exporter and importer GDP,
and log distance using PPML.}
\end{table}

\end{document}